%% file: Emulating_over_Discontinuites_using_Torn_Embeddings.tex
\documentclass[12pt]{article}

\usepackage{lscape,epsfig,amsmath,setspace}
\usepackage{xypic}
\usepackage{natbib}
\usepackage{acronym}
\usepackage{defs3H}
\usepackage{enumerate}
\usepackage{color}
\usepackage{soul}
\usepackage{amsmath,amsfonts,amssymb}
\usepackage[all]{xy}
\usepackage{psfrag}
\usepackage{fullpage}
\usepackage{psfrag}
\usepackage{epsfig}
\usepackage[all]{xy}
\usepackage{multirow}
\usepackage{bbm}
\usepackage{caption}
\usepackage{subcaption}
\usepackage{ifthen}
\usepackage[colorlinks,allcolors=blue]{hyperref}
\newcommand{\e}[1]{\ensuremath{{\rm E}[#1]}}
\newcommand{\ed}[2]{\ensuremath{{\rm E}_{#1}[#2]}}
\newcommand{\var}[1]{\ensuremath{{\rm Var}[#1]}}
\newcommand{\vard}[2]{\ensuremath{{\rm Var}_{#1}[#2]}}

\newcommand{\cov}[2]{\ensuremath{{\rm Cov}\left[#1,#2\right]}}

\newcommand{\be}{\begin{equation}}
\newcommand{\ee}{\end{equation}}
\newcommand{\ba}{\begin{eqnarray}}
\newcommand{\ea}{\end{eqnarray}}
\newcommand{\bes}{\begin{equation}}
\newcommand{\ees}{\end{equation}}
\newcommand{\bas}{\begin{eqnarray*}}
\newcommand{\eas}{\end{eqnarray*}}
\newcommand{\bi}{\begin{itemize}}
\newcommand{\ei}{\end{itemize}}
\newcommand{\bn}{\begin{enumerate}}
\newcommand{\en}{\end{enumerate}}
\newcommand{\bfi}{\begin{figure}}
\newcommand{\efi}{\end{figure}}

\newcommand{\bsmatrix}{\left( \begin{smallmatrix}}
\newcommand{\esmatrix}{\end{smallmatrix} \right)}

\newcommand{\cm}[1]{\ensuremath{ { #1 }}}

\newcommand{\vv}{\mathbf{v}}    % Define a general embedding function for input parameters in higher dimensions.
\newcommand{\vx}{\mathbf{x}}    % Define a general embedding function for input parameters in higher dimensions.
\newcommand{\vw}{\mathbf{w}}    % Define a general embedding function for input parameters in higher dimensions.
\newcommand{\ve}{\mathbf{e}}    % Define a general embedding function for input parameters in higher dimensions.

\newcommand{\Cov}{\mathrm{Cov}}    %Command for covariance.

\newcommand{\reff}[2]{\ref{#1}}  % use ref as normal

\begin{document}

%\title{Bayesian Emulation using Torn Embeddings for Computer Models with Structured Partial Discontinuities}
\title{Bayesian Emulation for Computer Models with \\Multiple Partial Discontinuities}

\author{Ian Vernon\footnote{i.r.vernon@durham.ac.uk}, Jonathan Owen\footnote{j.owen1@leeds.ac.uk} \ \& Jonathan Carter\footnote{ac8191@coventry.ac.uk} \\
  \hspace{1cm} \\
\small $^*$$^\dagger$Department of Mathematical Sciences, Durham University,  \\
\small Science Laboratories, Durham, DH1 3LE, UK \\
\small $^\ddagger$Research Centre for Fluid and Complex Systems, Coventry University, Coventry, UK
}

\maketitle

\abstract{
Computer models are widely used across a range of scientific disciplines to describe various complex physical systems, 
however to perform 
full uncertainty quantification we often need to employ emulators.
An emulator is a fast statistical construct that mimics the slow to evaluate computer model, and greatly aids the vastly more computationally intensive
uncertainty quantification calculations that an important scientific analysis often requires.
We examine the problem of emulating computer models that possess multiple, partial discontinuities occurring at known non-linear location. We introduce the TENSE framework, based on carefully designed correlation structures that respect the discontinuities while enabling full exploitation of any smoothness/continuity elsewhere. This leads to a single emulator object that can be updated by all runs simultaneously, and also used for efficient design. This approach avoids having to split the input space into multiple subregions. We apply the TENSE framework to the TNO Challenge II, emulating the OLYMPUS reservoir model, which possess multiple such discontinuities. 
}

%\vspace{-2cm}

%%%% setup variable to change plots from arXiv style to BA  style %%%%
\newcounter{plotstyle}
\setcounter{plotstyle}{1}  % 1 for arXive, 2 for BA.

\input{Emulating_over_Discontinuites_using_Torn_Embeddings_core}

\newpage
\section*{Acknowledgments}
I.V. gratefully acknowledges UKRI (EP/W011956/1) and Wellcome (218261/Z/19/Z) funding.
J.O. gratefully acknowledges EPSRC iCase Studentship (Smith Institute) funding.
We thank Rock Flow Dynamics for use of the tNavigator simulator.

\bibliographystyle{asa2}
\bibliography{/Users/ianvernon/Work/Master_Bibliography_and_Related_Files/master_bib.bib}

%%% Appendices %%%

\input{Emulating_over_Discontinuites_using_Torn_Embeddings_appendices}

\end{document}

%% file: Emulating_over_Discontinuites_using_Torn_Embeddings_core.tex
%%%%%%%%%%%%%%%%%%%%%%%%%%%%%%%%%%%%%%%%%%%%%%%%%%%%%
\section{Introduction}
%%%%%%%%%%%%%%%%%%%%%%%%%%%%%%%%%%%%%%%%%%%%%%%%%%%%%

The use of computer models, or {\it simulators}, to describe the dynamics of complex physical systems is now commonplace in a wide variety of scientific disciplines. Often such simulators possess high numbers of input and/or output dimensions and, due to their complexity, take a substantial amount of time to evaluate. This presents an immediate challenge, as the responsible use of a simulator (e.g. for model calibration, prediction, decision support, etc.), usually demands Bayesian uncertainty quantification, to capture all major sources of uncertainty, which typically requires a vast number of simulator evaluations. For complex simulators possessing even a modest runtime, this is utterly infeasible. 
Emulators represent a solution to this problem. An emulator is a statistical construct that seeks to mimic the 
behaviour of the simulator over its input space, but which is several orders of magnitude faster to evaluate.
As the emulator provides both a prediction and an uncertainty statement about the simulator's behaviour at unexplored input locations (an attribute that elevates it above interpolation 
or other proxy modelling approaches), it can naturally be incorporated in a wider Bayesian uncertainty analysis. 

Early uses of Gaussian process emulators for computer models were given by \cite{SWMW89_DACE,Currin91_BayesDACE}. For an early example using multilevel emulation combined with structural discrepancy modelling in a Bayesian history matching context see~\cite{Craig97_Pressure}, and for a fully Bayesian calibration of a complex nuclear radiation model, see~\cite{Kennedy01_Calibration}.
Emulators have now been successfully employed across several scientific disciplines, including 
cosmology~\citep{Vernon10_CS,Vernon10_CS_rej,vernon_astro,PhysRevD.78.063529,Higdon09_Coyote2,Kaufman:2011uz,galf_stat_sci,Vernon:2016aa_short}, climate 
modelling~\citep{Williamson:2013aa,S:2015aa,Holden:aa,Edwards:2019aa,Edwards:2021wk}, engineering~\citep{0543c39b6ec042e4b3f0e3794152ede6}, 
epidemiology~\citep{Yiannis_HIV_1,Yiannis_HIV_3,McKinley:2017aa,McCreesh2017,Vernon_HM_June_2022}, systems biology~\citep{Vernon_sysbio_hm_2016,Jackson1}, oil reservoir 
modelling~\citep{JAC_Handbook,JAC_sma_samp},
environmental science~\citep{asses_mod}, vulcanology~\citep{Bayarri:2009aa,PPGPE, marshall2019ehe} and even to Bayesian analysis itself~\citep{BABA_paper1}.
%\ian{list some more from other authors: OHagan, Gosling, Oakley, Wilkinson, Higdon, Berger etc}.
The development of improved emulation strategies therefore has the potential to benefit multiple scientific areas, allowing more accurate 
analyses with lower computational cost.\nocite{Higdon08a_calibration}
%We develop one such general strategy here.

Most emulator constructions exploit prior judgements about the behaviour of the simulator in terms of its smoothness/differentiability/continuity etc. 
%either directly in the case of a deterministic simulator, or as applied to the attributes of the distributional output of a stochastic simulator. 
In this work, however, we are confronted with a problem 
arising in the TNO Challenge II: a joint industrial and academic challenge posed in the oil industry (see section~\ref{sec_TNO} for details). A key part of this problem requires 
the emulation of simulators that are anticipated to be smooth over much of the input space, but that also possess multiple, partial 
discontinuities of known, non-linear location. We use the term ``partial" in the sense that the location of the discontinuities begin within the input space, typically ending on the boundary, and hence are not closed, nor do they necessarily bisect the space. Examples of the location of these discontinuities are shown in figure~\ref{fig_olympus_model_a} (with toy versions in figures~\ref{fig_toymod1} to \ref{fig_toy3_curved_disc}).

%As we discuss in more detail below, various emulation techniques have been developed for dealing with non-stationary features, however they are not ideally suited for the case we consider here. 

A possible way to incorporate discontinuities is to partition the input space into various subregions, and then fit separate, independent emulators in each subregion. For example, Treed GPs  \citep{doi:10.1198/016214508000000689} which use rectangular, axis aligned subregions, or \cite{Pope:2021wk} who use Voronoi tessellations. Although flexible, these approaches typically require substantial numbers of simulator evaluations, especially in higher dimensions, and critically will not exploit the smoothness around the discontinuity endpoints, which we wish to do here. In addition, many subregions maybe required to handle curved discontinuities (especially for Treed GPs). \cite{BUACPS} use emulators to identify discontinuities caused by tipping points, and then emulate the output separately in each region. This however, is used for discontinuities that bisect the input space, unlike the case here, and 
the identification of the discontinuities is reported to be time-consuming (see also \cite{Ghosh2018GaussianPE}).
Deep GPs (see e.g. \cite{Dunlop:2018uz} and references therein) whereby either the correlation lengths or GP inputs are modelled by a second layer GP with inputs or dependant parameters in turn modelled by the next layer GP etc. have almost unlimited flexibility but this comes at a cost, requiring substantial numbers of runs to train, whereas for our application run numbers will be extremely limited. Deep GPs also typically have non-analytic uncertainty propagation, which poses problems for full UQ \citep{doi:10.1080/00401706.2021.2008505}. More importantly, even a deep GP based on smooth layers may fail diagnostics on closer examination, as the impact of the discontinuity will percolate down the layers and still be evident at each level e.g. mimicking rapid (i.e. discontinuous) change in the simulator on the top layer would require rapid (also discontinuous) change of inputs or correlation lengths on the second layer, and so on. 
\cite{https://doi.org/10.48550/arxiv.1903.02071} attempted to emulate across simple 1D step functions using a variety of interesting covariance structures 
with moderate success, although most structures used were either still essentially continuous and hence couldn't fully represent the discontinuity, or induced additional unwanted features.

We instead introduce the TENSE framework, based around carefully designed covariance structures that respect the discontinuities while fully exploiting any smoothness/continuity elsewhere, leading to a single emulator object that can be updated by all runs simultaneously. 
The layout of the article is as follows. In Section~\ref{sec_2} we construct emulators that exhibit partial discontinuities using torn embeddings, before showing how to correct for various induced warpings in Section~\ref{sec_contr_warping}. In Section~\ref{sec_TNO} we apply the TENSE framework to the TNO Challenge II. Example code to reproduce the plots in 
Sections~\ref{sec_2} and \ref{sec_contr_warping} can be found at \url{https://github.com/ivernon/TENSE.git}.

%%%%%%%%%%%%%%%%%%%%%%%%%%%%%%%%%%%%%%%%%%%%%%%%%%%%%
\section{Emulating Computer Models with Partial Discontinuities using Torn Embeddings.}\label{sec_2}
%%%%%%%%%%%%%%%%%%%%%%%%%%%%%%%%%%%%%%%%%%%%%%%%%%%%%

\subsection{Emulation of Computer Models}

We now summarise the standard emulation of computer models approach. 
We consider a complex computer model represented by a function $f(\vx)$, where $\vx\in\mathcal{X}$ denotes a $d$-dimensional vector containing the computer model's input parameters, and $\mathcal{X}\subset \mathbb{R}^d$ is a pre-specified input parameter space of interest. 
We imagine that due to its complexity, a single evaluation of the computer model will take a substantial amount of time to complete, and due to limited computational resources we will only be able evaluate it at a relatively small 
number of locations across the input space. 
Here we assume $f(\vx)$ is univariate, but the methods we develop should in principle generalise to the multivariate case.
Following the Bayesian paradigm, we represent our beliefs about the unknown $f(\vx)$ at unevaluated input $\vx$ via an emulator. 
A typical approach is to use a pure Gaussian process (GP) for the emulator, such that  
\be
  f(\cdot)|m(\cdot),c(\cdot,\cdot)  \;\; \sim \;\;  GP(m(\cdot),c(\cdot,\cdot)),
\ee
for some mean function $m(\cdot)$ and covariance function $c(\cdot,\cdot)$ \citep{Kennedy01_Calibration}, chosen corresponding to any prior beliefs we hold about the properties of the function $f(\vx)$.
%e.g. whether it is continuous or smooth etc. If we perform a set of runs at locations $\vx_D=(\vx^{(1)},\dots,\vx^{(n)})$ over the input space of interest $\mathcal{X}$, giving computer model outputs 
%as the column vector $D = (f(\vx^{(1)}),\dots,f(\vx^{(n)}))^T$, then we can update our beliefs about the computer model $f(\vx)$ in light of $D$.
While this form of GP emulator has been successfully employed 
in a large number of applications, it is sometimes argued that it is the core second-order structure of the GP that is its most important feature, a structure which aligns more closely with our actual beliefs about the behaviour of $f(\vx)$. The additional distributional assumptions that use of a GP entails, namely that any finite collection of outputs 
$ \{f(\vx^{(1)}),\dots,f(\vx^{(n)}) \} $ 
have specifically a multivariate normal distribution is, in some cases, too strong an assumption, which can have unintended consequences.
%which in addition to requiring more stringent  diagnostics (cite   Bastos), can sometimes have unintended consequences e.g. when employed for model calibration and prediction (  cite Michael Stein's talk?).

Therefore, we often prefer to focus directly on the second-order structure itself, and employ Bayes linear emulators instead of the above GP version. Bayes linear methods follow the foundational work of DeFinetti \citep{DeFinetti1} by treating expectation instead of probability as primitive, and respect the subjectivist Bayesian paradigm, but require only a second-order specification~\citep{Goldstein_99,Goldstein07_BayesLinearBook}. 
In this framework, instead of a GP we represent $f(\vx)$ as a weakly stationary stochastic process. A simple prior specification appropriate for some computer models (see appendix~\reff{app_A}{A} for a more complex version) would be to set $\e{f(\vx)} = m(\vx)$ for some mean function $m(\vx)$, and to specify the covariance structure as
\be \label{eq_cor_struc}
\cov{f(\vx)}{f(\vx')} \;=\; \sigma^2 \, r(\vx-\vx') 
\ee
where $\sigma^2$ represents the prior variance of $f(\vx)$, and $r(\vx-\vx')$ defines a stationary correlation structure, of which there are many possible options (see \cite{GPML}). A popular choice for smooth (i.e. infinitely differentiable) functions being the squared exponential:
\be\label{eq_prodgausscor}
%r(\vx-\vx') \;=\; \exp\{-\|\vx-\vx'\|^2/\theta^2\}   % \;=\;   \prod_{i=1}^d \exp\{-|x_i-x_i'|^2/\theta^2\}
%r(\vx-\vx') \;=\; {\rm exp}\left\{- (\vx-\vx')^T \Sigma^{-1}_{2D}(\vx-\vx') \right\}  
r(\vx-\vx') \;=\; {\rm exp}\left\{- (\vx-\vx')^T \Sigma^{-1}(\vx-\vx') \right\}  
\ee
where $ \Sigma$ is a covariance matrix governing general Mahalanobis distances. Setting 
$  \Sigma = \rm{diag} \{\theta,\dots,\theta\} $, regains the usual isotropic form, where $\theta$ is the standard correlation length.
%where $\theta$ is the correlation length parameter. This can easily be generalised to a non-isotropic version, which we will use below.
Another widely used choice is the Matérn correlation function:
\be\label{eq_cor_matern}
r(\vx-\vx') \;\;=\;\;  \frac{2^{1-\nu}}{\Gamma (\nu)} \left( \frac{\sqrt{2\nu} \|\vx-\vx'\|}{\theta} \right)^{\nu} K_{\nu} \left( \frac{\sqrt{2\nu}\|\vx-\vx'\|}{\theta} \right).
\ee
where $K_{\nu}$ is a modified Bessel function of the second kind and $\theta$ and $\nu$ are parameters to be specified that govern the correlation length 
and the derivatives of the computer model respectively ($\nu$ rounded up to the next integer gives the number of derivatives that exist). 

Given such a second-order specification and a set of model evaluations at locations $\vx^{(1)},\dots,\vx^{(n)}$, yielding simulator outputs $D = (f(\vx^{(1)}),\dots,f(\vx^{(n)}))^T$, we can update our second-order beliefs about $f(\vx)$ at unevaluated location $\vx$ via the Bayes linear adjustment formulae:
%~\citep{Goldstein_99,Goldstein07_BayesLinearBook}:
\ba
\ed{D}{f(\vx)} &=& \e{f(\vx)} + \cov{f(\vx)}{D} \var{D} ^{-1}(D- \e{D}) \label{eq_BLm}\\
\vard{D}{f(\vx)} &=& \var{f(\vx)} - \cov{f(\vx)}{D} \var{D}^{-1}\cov{D}{f(\vx)} \label{eq_BLv}
\ea
where $\ed{D}{f(\vx)}$ and $\vard{D}{f(\vx)}$ are the expectation and variance of $f(\vx)$ adjusted by $D$. See \cite{Goldstein_99,Goldstein07_BayesLinearBook} for details and discussion of the benefits of using a Bayes linear approach, and  \cite{Vernon10_CS,Vernon10_CS_rej,Vernon_sysbio_hm_2016} for the benefits within a computer model setting. 
The fully specified Bayesian GP based calculation, would of course yield similar update formulae for the analogous posterior mean and variance quantities (conditioned upon various hyperparameters in the definitions of $c(\cdot,\cdot)$ and $m(\cdot)$). 
While the results derived in this article apply to both the Bayes linear and the fully specified GP emulator frameworks, we will
most often refer to the Bayes linear case, as the core arguments concern the second order covariance structure itself, and how we adapt it to the presence of discontinuities. 
Additionally, for clarity of exposition, we will mainly focus on the standard emulator specification as given by equations~(\ref{eq_cor_struc}) and (\ref{eq_prodgausscor}), however see appendix~\reff{app_A}{A} for more advanced emulator specifications.

\subsection{Emulation Problems caused by Partial Discontinuities}\label{ssec_EPPD}

It is worth discussing the specific difficulties that partial discontinuities pose for standard emulators of the form described in the previous section.
An example toy computer model that exhibits a partial discontinuity is given by the function:
\be \label{eq_toy1}
f(\vx) \;\equiv\; f(x,y) \;=\; 0.4\sin(5x) + 0.4\cos(5y) + 0.8(x-0.75)^2\sign(y-1)\mathbbm{1}_{\{x>0.75\}}
%f(x,y) \;=\; \frac{2}{5}\sin(5x) + 0.4\cos(5y) + 0.8(x-0.75)^2\sign(y-1)\mathbbm{1}_{\{x>0.75\}}
%0.4*sin(5*x) + 0.4*cos(5*y) + 0.8*(x>0.75)*(x-0.75)^2*(y>1.15) - 0.8*(x>0.75)*(x-0.75)^2*(y<0.85)
%yrift <- c(0.75,1.25)								# y-coordinate of rifts
%xrift <- c(0.6,1)								# x-coordinate of left edge of rift
%sim2d_fun <- function(x,y) 0.4*sin(5*x) + 0.4*cos(5*y) + 
%							1.2*(x>xrift[2])*(x-xrift[2])^2*(y>yrift[2]) - 0.6*(x>xrift[1])*(x-xrift[1])^2*(y<yrift[1])
\ee
where the two-dimensional $\vx = (x,y)^T$ and $\mathbbm{1}_A$ is the indicator function that takes value $1$ when statement $A$ is true and 0 otherwise, and ``$\sign$" just returns the sign of its argument. 
The form of this function is shown in figure~\ref{fig_toymod1_a} for the region $\mathcal{X} = \{0<x<2, 0<y<2\}$. We see that it has a discontinuity across the line $y=1$, for $x>0.75$, shown as the black horizontal line, and that the discontinuity begins in the interior of $\mathcal{X}$ at the point $(x=0.75,y=1)$, and ends on the boundary at $(x=2,y=1)$.
It is also clear that the function is smooth everywhere else apart from the discontinuity, an attribute that we would wish to exploit in the emulation process. 

However, if we naively attempt to apply standard GP or Bayes Linear emulation procedures to $f(x,y)$ they will fail, as they will attempt to smooth over the discontinuity leading to two problems (i) the
emulator predictions close to the discontinuity will be highly inaccurate resulting in poor emulator diagnostics, and (ii) the estimation of global emulation parameters (e.g. the correlation lengths $\theta$) may produce strange results that are very sensitive to the design, leading to possible global issues with the emulator. 
%An example of this would be if the design did not contain points close to the discontinuity: then the effect of the discontinuity may go unnoticed and moderate correlations lengths
%estimated, exacerbating poor diagnostics in the vicinity of the discontinuity. However, if the design did have points either side and close to the discontinuity, the effects would be noticed and very small correlation lengths might be favoured (depending on the rest of the design), leading to unnecessarily cautious emulator predictions in the rest of the input space. Either scenario would produce flawed emulators.
We see that the main problem here is that a discontinuity of this form severely violates the assumption of stationarity and also the common assumption of some form of smoothness/differentiability/continuity implicit in the standard emulator covariance structures. 
As argued in the Introduction, attempts to alter these assumptions e.g. by breaking stationarity via input dependent correlation lengths or resorting to full deep GPs, do not adequately address this issue as they are still using essentially continuous structures to represent a discontinuity. Our approach in contrast, uses torn embeddings that naturally capture the essence of the discontinuity.

Another approach worth mentioning would be to tinker with the correlation structure of the emulator directly, to reduce the correlation between outputs either side of the discontinuity. For example, one suggestion is to use the geodesic distance between input points in the correlation function, defined such that viable geodesics do not cross the discontinuity (and hence have to go around it). 
However, this fails as it does not provide a valid covariance structure. This is easy to demonstrate e.g. by using equations~(\ref{eq_cor_struc}) and (\ref{eq_prodgausscor}) to construct the $4\times4$ covariance matrix formed from the four outputs $f(x_A), f(x_B), f(x_C), f(x_D)$ corresponding to the four input points $x_A=(0.5,1),x_B=(0.75,1),x_C=(1,1^+),x_D=(1,1^-)$, and noting that it is not positive semi-definite (see appendix~\reff{app_B}{B} for details).
This shows that altering the covariance structure of an emulator to deal with a discontinuity in such ad hoc ways is fraught with danger, even more so for multiple discontinuities of possibly complex, non-linear shape. 

Our proposed approach however, guarantees the validity of the emulator's covariance structure, even in the presence of multiple discontinuities of arbitrary shape, while still providing a flexible choice of emulator form, as we now describe.
\ifthenelse{\value{plotstyle}=1}{
\begin{figure}[t]
\begin{center}
\begin{subfigure}{0.49\columnwidth}
\centering
\includegraphics[page=1,scale=0.35,viewport= 0 0 630 530,clip]{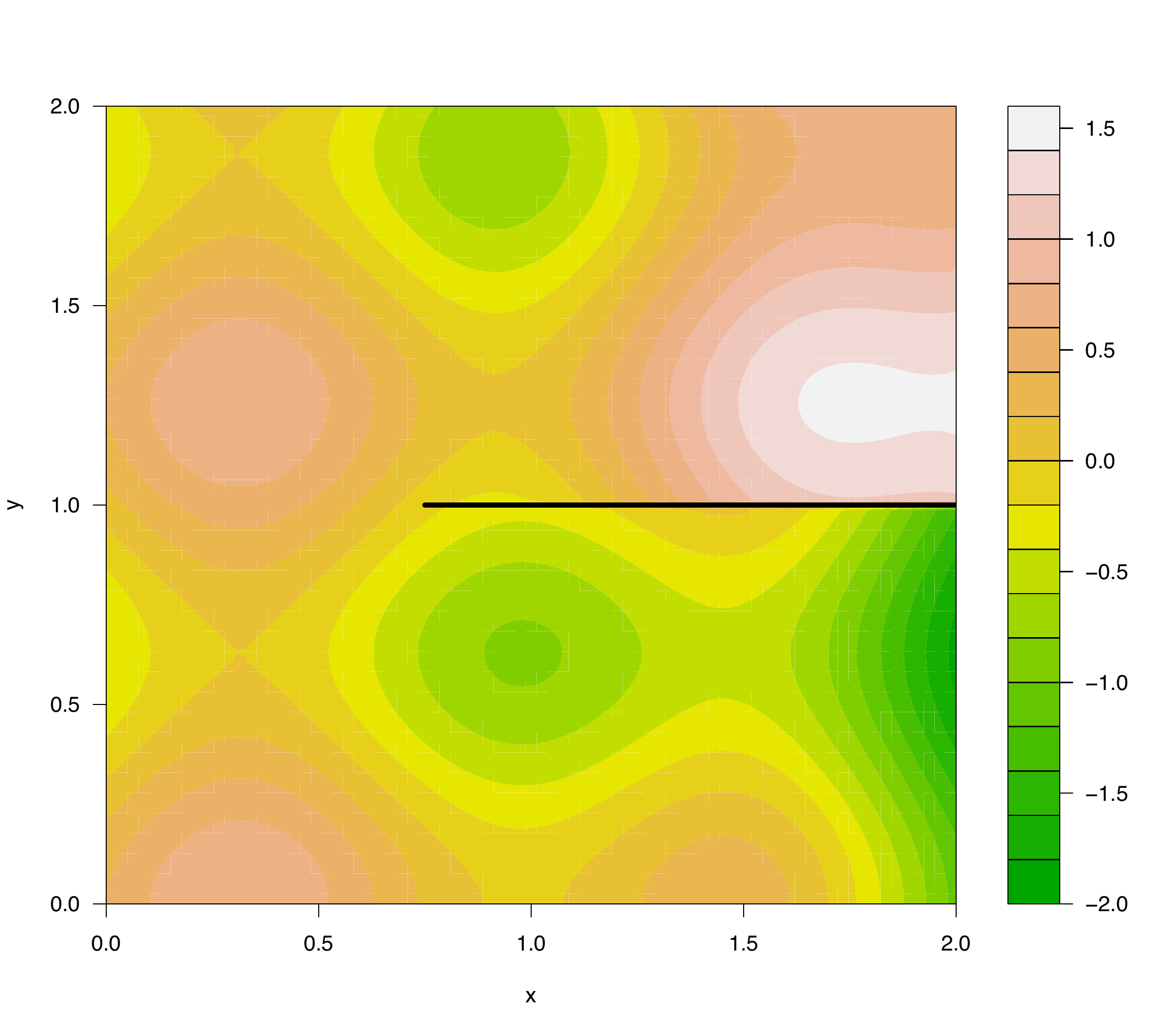} 
\vspace{-0.4cm}
\caption{\footnotesize{The true 2-dimensional function $f(x,y)$.}}
\label{fig_toymod1_a}
\end{subfigure}
\begin{subfigure}{0.49\columnwidth}
\centering
\includegraphics[page=2,scale=0.35,viewport= 0 0 630 530,clip]{plots_paper/plots_examples/embedding2d_simple.pdf}
\vspace{-0.4cm}
\caption{\footnotesize{The embedding surface $v(x,y)$.}}
\label{fig_toymod1_b}
\end{subfigure}
\begin{subfigure}{0.49\columnwidth}
\centering
\vspace{0.4cm}
\includegraphics[page=3,scale=0.35,viewport= 0 0 630 530,clip]{plots_paper/plots_examples/embedding2d_simple.pdf}
\vspace{-0.4cm}
\caption{\footnotesize{The emulator expectation $\ed{D}{f(x,y)}$.}}
\label{fig_toymod1_c}
\end{subfigure}
\begin{subfigure}{0.49\columnwidth}
\centering
\vspace{0.4cm}
\includegraphics[page=4,scale=0.35,viewport= 0 0 630 530,clip]{plots_paper/plots_examples/embedding2d_simple.pdf}
\vspace{-0.4cm}
\caption{\footnotesize{The emulator stan. dev. $\sqrt{\vard{D}{f(x,y)}}$.}}
\label{fig_toymod1_d}
\end{subfigure}
\end{center}
\vspace{-0.4cm}
\caption{\footnotesize{(a) An example toy 2-dimensional function \cm{f(x)} with partial discontinuity located along the black horizontal line. (b) The embedding surface $v(x,y)$, torn along the location of the discontinuity. (c) The emulator expectation \cm{\ed{D}{f(x,y)}} with induced partial discontinuity. (d) The emulator standard deviation $\sqrt{\vard{D}{f(x,y)}}$ with induced partial discontinuity (note the horizontal compression for larger values of $x$).}}
\label{fig_toymod1}
\vspace{-0.cm}
\end{figure}
}{}
\ifthenelse{\value{plotstyle}=2}{
\begin{figure}[t!]
\vspace{-0.2cm}
\begin{center}
\begin{subfigure}{0.49\columnwidth}
%\centering
\hspace{-0.9cm} \includegraphics[page=1,scale=0.335,viewport= 0 0 630 530,clip]{plots_paper/plots_examples/embedding2d_simple.pdf} 
\vspace{-0.7cm}
\caption{\footnotesize{The true 2-dimensional function $f(x,y)$.}}
\label{fig_toymod1_a}
\end{subfigure}
\begin{subfigure}{0.49\columnwidth}
%\centering
\hspace{0.0cm} \includegraphics[page=2,scale=0.335,viewport= 0 0 630 530,clip]{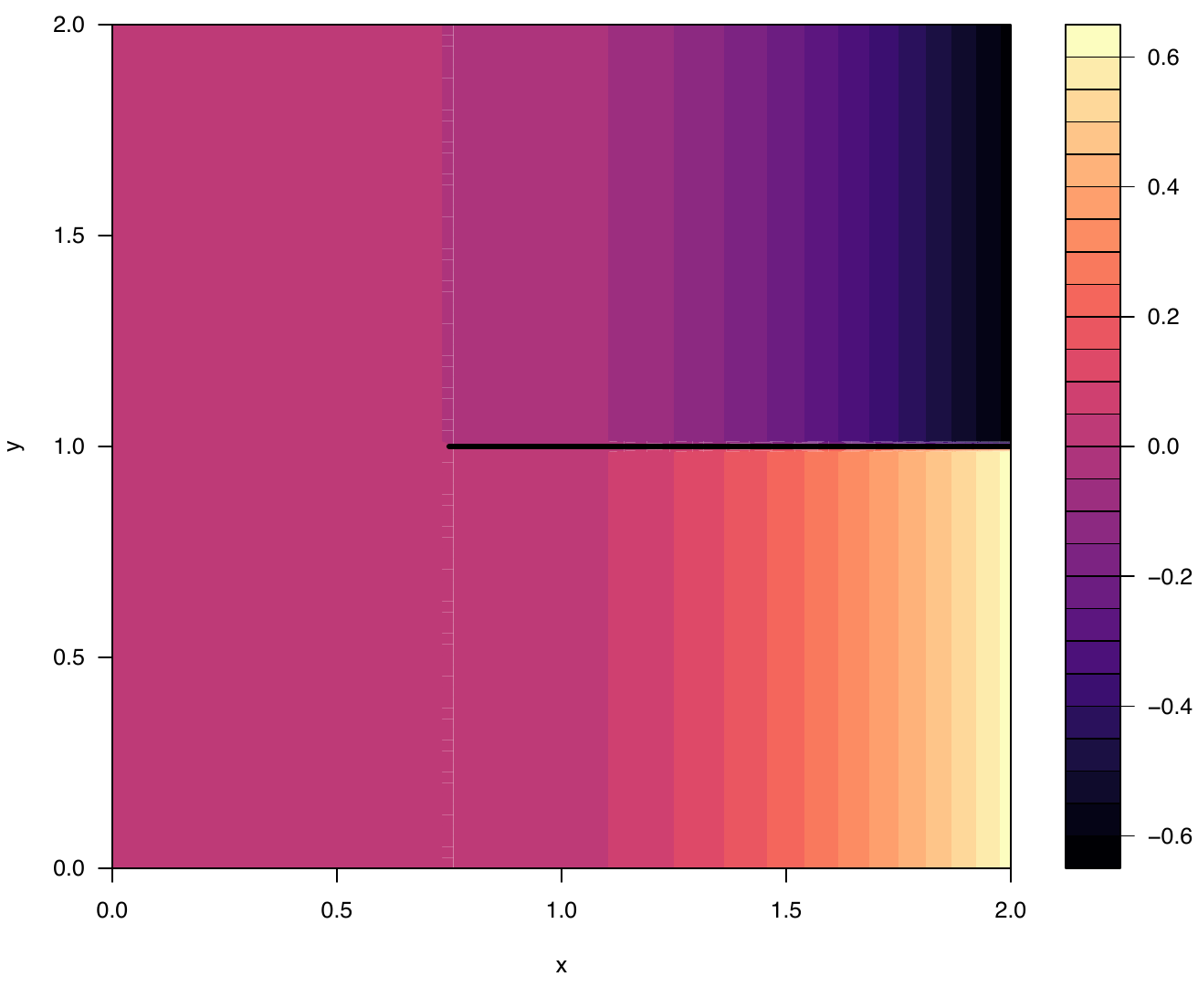}
\vspace{-0.7cm}
\caption{\footnotesize{The embedding surface $v(x,y)$.}}
\label{fig_toymod1_b}
\end{subfigure}
\begin{subfigure}{0.49\columnwidth}
%\centering
\vspace{0.cm}
\hspace{-0.9cm} \includegraphics[page=3,scale=0.335,viewport= 0 0 630 530,clip]{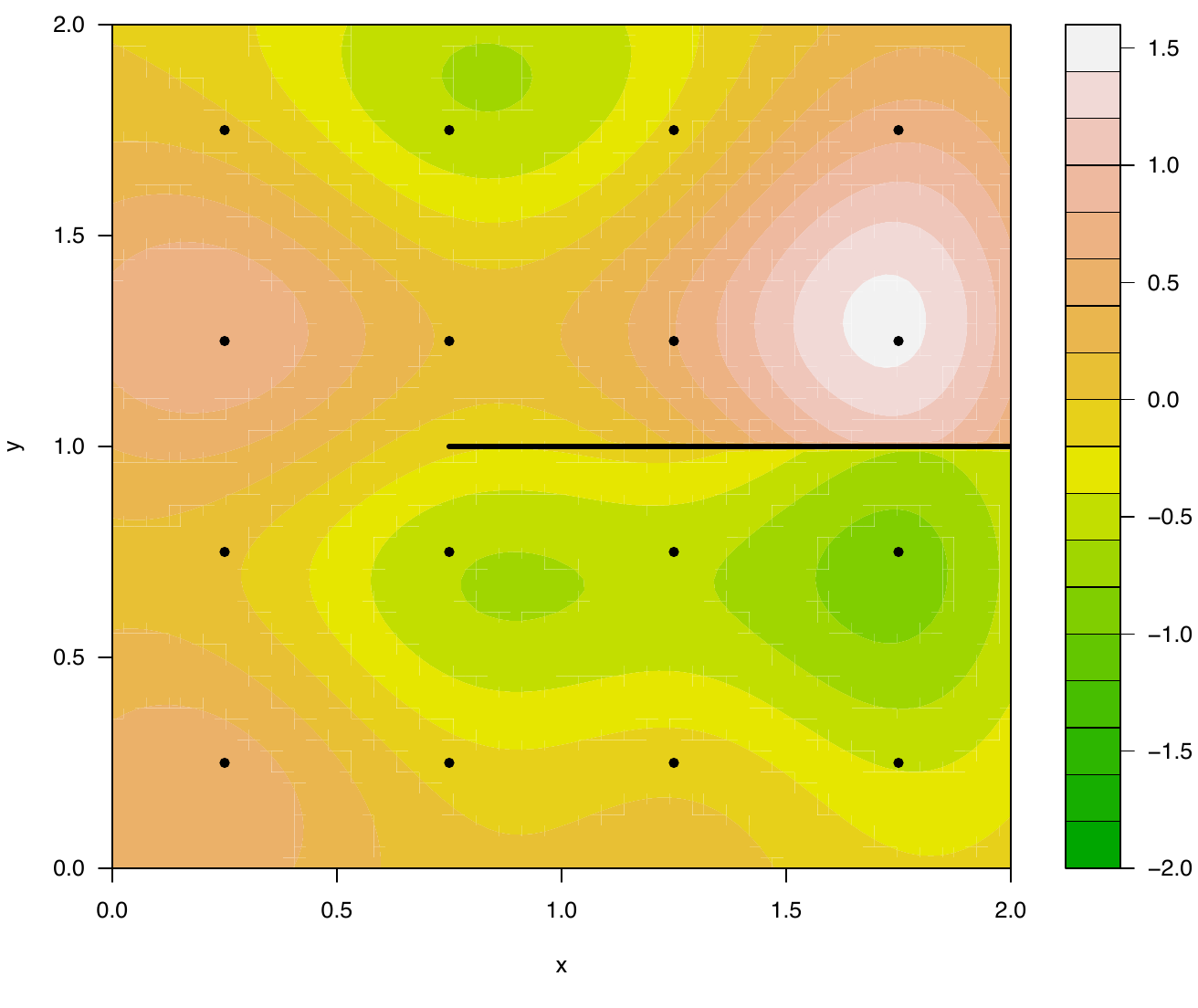}
\vspace{-0.7cm}
\caption{\footnotesize{The emulator expectation $\ed{D}{f(x,y)}$.}}
\label{fig_toymod1_c}
\end{subfigure}
\begin{subfigure}{0.49\columnwidth}
%\centering
\vspace{0.cm}
\hspace{0.0cm} \includegraphics[page=4,scale=0.335,viewport= 0 0 630 530,clip]{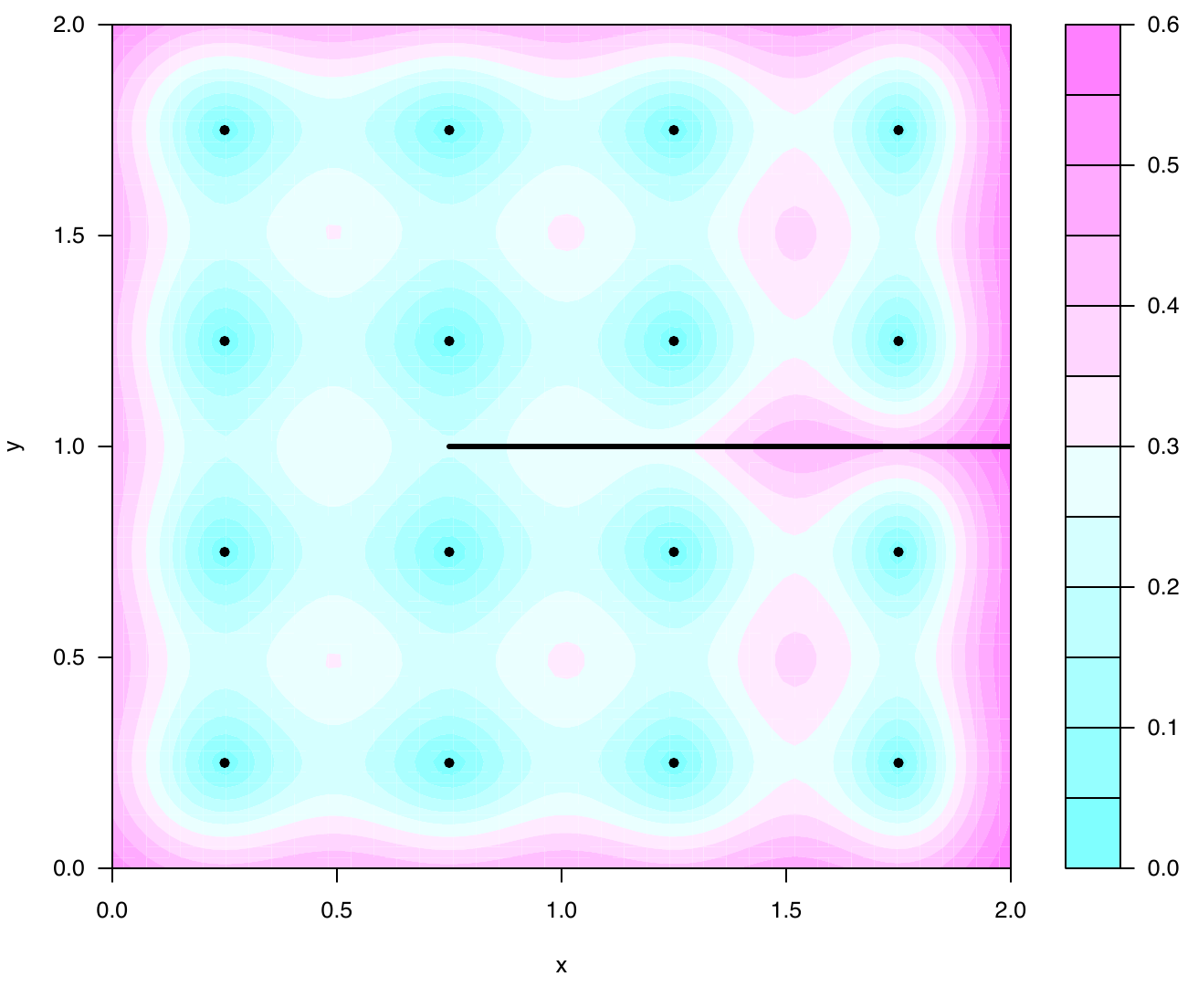}
\vspace{-0.7cm}
\caption{\footnotesize{The emulator stan. dev. $\sqrt{\vard{D}{f(x,y)}}$.}}
\label{fig_toymod1_d}
\end{subfigure}
\end{center}
\vspace{-0.6cm}
\caption{\footnotesize{(a) An example toy 2-dimensional function \cm{f(x)} with partial discontinuity located along the black horizontal line. (b) The embedding surface $v(x,y)$, torn along the location of the discontinuity. (c) The emulator expectation \cm{\ed{D}{f(x,y)}} with induced partial discontinuity. (d) The emulator standard deviation $\sqrt{\vard{D}{f(x,y)}}$ with induced partial discontinuity (note the horizontal compression for larger values of $x$).}}
\label{fig_toymod1}
\vspace{-0.6cm}
\end{figure}
}{}

\subsection{Torn Embedding in a Higher Dimension}\label{ssec_torn_embed}

The challenge is therefore clear: to develop more sophisticated emulators that exploit regions of smoothness/differentiability/continuity while also respecting the effects of multiple partial discontinuities at known, but possibly non-linear, locations, as seen in the TNO Challenge II.
In the interest of clarity, we introduce our approach in terms of a 2-dimensional computer model, but note that the generalisation to higher dimensions is straightforward.
To incorporate discontinuities we employ the following procedure:
%As described above, partial discontinuities present a fundamental challenge to the standard emulation process. To overcome this we employ the following procedure:
\bn
\item We embed the emulator's 2-dimensional input space $\vx \in \mathcal{X} \subset \mathbb{R}^2 $  into a higher 3-dimensional input space 
 $\vv(\vx) \in \mathcal{V} \subset \mathbb{R}^3$ using the embedding surface $v(x,y)$ such that we have
\ifthenelse{\value{plotstyle}=2}{\vspace{-0.5cm}}{}
 \be\label{eq_defn_vec_v}
 \vx =  \begin{pmatrix} x \\ y \end{pmatrix} \quad \quad \text{and} \quad \quad  \vv(\vx) =  \begin{pmatrix} x \\ y \\ v(x,y) \end{pmatrix} 
 \ee
\ifthenelse{\value{plotstyle}=2}{\vspace{-0.6cm}}{}
\item We tear the otherwise smooth 2-dimensional embedding surface \cm{v(x,y)} along the known locations of the discontinuities.
\ifthenelse{\value{plotstyle}=2}{\vspace{-0.1cm}}{}
\item We then set up the emulator as usual using equations~(\ref{eq_cor_struc}), (\ref{eq_prodgausscor}), (\ref{eq_BLm}) and (\ref{eq_BLv}), but now in the full 3-dimensional space, using the 3-dimensional $\vv(\vx)$ as its input. 
Specifically, we can design a space-filling collection of runs at locations $\vx^{(1)},\dots,\vx^{(n)}$ that are embedded in 3-dimensional space as $\vv(\vx^{(1)}),\dots,\vv(\vx^{(n)})$, where the design process can now respect the presence of the discontinuities.
\ifthenelse{\value{plotstyle}=2}{\vspace{-0.1cm}}{}
\item To evaluate the emulator's expectation and variance at a new point $\vx$ we simply evaluate the emulator on the projection of $\vx$ onto the embedding surface,
that is evaluate $\ed{D}{f(\vv(\vx))}$ and $\vard{D}{f(\vv(\vx))}$ using equations (\ref{eq_BLm}) and (\ref{eq_BLv}).
\en
The tears in the embedding surface $v(x,y)$ will induce a discontinuity, of as yet uncertain size, in the unknown output $f(\vx)$, and also in our uncertainty statements for $f(\vx)$, just as we require.

So for example, the covariance structure of the original non-embedded 2-dimensional emulator using the squared exponential covariance function of equation~(\ref{eq_prodgausscor}) was:
\be \label{eq_sigma_2d}
\cm{\! \!{\rm Cov}[f(\vx),f(\vx')] \;=\; \sigma^2 {\rm exp}\left\{- (\vx-\vx')^T \Sigma^{-1}_{2D}(\vx-\vx') \right\}  \!}
\ee
%where $\Sigma^{-1}_{2D}$ governs a more general Mahalanobis distance between input points $\vx$ and $\vx'$.
After the embedding into 3-dimensions, the covariance becomes simply:
%The corresponding 3-dimensional version, after the embedding, would therefore simply be 
\be\label{eq_emul_3D_cov_struct}
\cm{\! \!{\rm Cov}[f(\vv(\vx)),f(\vv(\vx'))] \;=\; \sigma^2 {\rm exp}\left\{- (\vv(\vx)-\vv(\vx'))^T \Sigma^{-1}_{3D}(\vv(\vx)-\vv(\vx')) \right\}  \!}
\ee
i.e. it depends on distances in the new 3-dimensional space via $\vv(\vx) \in \mathcal{V}$,
where $\Sigma_{3D}$ governs the general 3D Mahalanobis distances. The freedom to choose from various allowable forms for $\Sigma_{3D}$ will be an important 
part in the full embedded emulator development as we shall discuss in section~\ref{sec_contr_warping}. In practical terms, when constructing the emulator using the embedding surface, we simply replace equations~(\ref{eq_cor_struc}) and (\ref{eq_prodgausscor}) by  equation~(\ref{eq_emul_3D_cov_struct}).

\underline{Toy Example:} Returning to the toy model of figure~\ref{fig_toymod1_a} and equation~(\ref{eq_toy1}), we specify an embedding surface as
\be\label{eq_v_sim_toy1}
v(x,y) \;\;=\;\;  - 0.4(x-0.75)^2 \sign(y-1)\mathbbm{1}_{\{x>0.75\}}
\ee
which is shown in figure~\ref{fig_toymod1_b}.
The main requirement of the embedding surface at this stage is that it is locally smooth, whilst also being torn along the discontinuity such that the regions above and below the 
discontinuity are sufficiently different in height in the third dimension in order to decorrelate outputs either side of the discontinuity. Note that $v(x,y)$ does not have to 
track the form of the 
actual computer model function $f(x,y)$ at all: in this example $v(x,y)$ above/below the discontinuity goes low/high while the function $f(x,y)$ does the opposite. 

To demonstrate, we design a simple grid of 16 runs $\vx^{(i)},i=1,\dots,16$ in the 2-dimensional region $\mathcal{X}$, shown as the black points in figures~\ref{fig_toymod1_c} and \ref{fig_toymod1_d}, and raise them into 3 dimensions using $\vv^{(i)}= \vv(\vx^{(i)})$. Note that we choose a grid here as its symmetries help to illustrate the emulator's behaviour. We then emulate in the 3-dimensional space as usual, using equations~(\ref{eq_emul_3D_cov_struct}), (\ref{eq_BLm}) and (\ref{eq_BLv}), with $D = \{f(\vx^{(1)}),\dots,f(\vx^{(16)} ) \} $, and using isotropic $\Sigma_{3D} = {\rm diag}\{\theta,\theta,\theta\}$, with $\theta=0.5$ and $\sigma= 0.7$.
The emulator expectation $\ed{D}{f(\vx)} \equiv \ed{D}{f(\vv(\vx))} $ evaluated across a dense grid of $80\times 80$ points over $\mathcal{X}$, is shown in figure~\ref{fig_toymod1_c}. We see that the emulator expectation is smooth away from the discontinuity, but displays a suitable jump across the discontinuity, as desired, hence mimicking the discontinuous behaviour of the real function $f(x)$, given in figure~\ref{fig_toymod1_a}, reasonably well. Note that we do not claim that this emulator is particularly accurate (especially given the simple grid design), just that it has the desired capability to represent smooth regions combined with partial discontinuities. 
Individual realisations of $f(x)$ drawn from the emulator, also must have similar smooth/discontinuous behaviour, as shown in appendix~\reff{app_C}{C}.
The emulator standard deviation $\sqrt{\vard{D}{f(\vx)}} \equiv \sqrt{\vard{D}{f(\vv(\vx))}}$ is shown in figure~\ref{fig_toymod1_d}, and shows the desired behaviour, in that the further we go along the discontinuity (in the positive $x$ direction) the more uncorrelated the two regions (above and below the discontinuity) become. For example, the point $(1.75, 1^-)$ just below the discontinuity has a similarly low level of emulator standard deviation as the point $(1.75,0)$ i.e. a point on the lower boundary. This shows that the emulator at the point $(1.75, 1^-)$ is just as uninformed as on the lower boundary, and is therefore hardly learning anything from the runs above the discontinuity: it is almost uncorrelated with them, as desired. There will be a more detailed discussion of this point and an examination of the underlying induced 2D correlation structure in section~\ref{ssec_TNO_emulation}.
\ifthenelse{\value{plotstyle}=2}{\vspace{-0.1cm}}{}

However, there is a problem: the emulator standard deviation (and expectation) seem compressed slightly, in the $x$ direction, for larger values of $x$. 
This issue is more clearly seen in 
figure~\ref{fig_toy2_2disc} which shows a similar toy model example but now with two discontinuities of different length. Here we have (see figure~\ref{fig_toy2_2disc_a}):
\ifthenelse{\value{plotstyle}=2}{\vspace{-0.1cm}}{}
\be
f(x,y) = 0.4\sin(5x) + 0.4\cos(5y) + 1.2\mathbbm{1}_{\{x>1\}} (x-1)^2\mathbbm{1}_{\{y>1.25\}}  - 0.6 (x-0.6)^2 \mathbbm{1}_{\{x>0.6\}} \mathbbm{1}_{\{y<0.75\}}   \nonumber
\ee
Now we have to use a more complex embedding surface to accommodate the discontinuities of differing length:
\ba
v(x,y) &=& 0.6(x-b(y))^2 \mathbbm{1}_{\{x>b(y)\}}  \mathbbm{1}_{\{y<1.25\}}   \mathbbm{1}_{\{y>0.75\}} 	-		% between two rifts
			  0.6(x-0.6)^2\mathbbm{1}_{\{x>0.6\}} \mathbbm{1}_{\{y<0.75\}}  \nonumber \\
%	g  &=& (1.25-0.75) / (1-0.6) \\
%	x_{int} &=& 0.6 + (y-0.75) / g 			% x point on line between rift left endpoints.
{\rm with} \;\;\;	b(y) &=& 0.6 +  (1-0.6) (y-0.75) / (1.25-0.75)		% x point on line between rift left endpoints.
\ea
where $b(y)$ represents the $x$ coordinate of the line that interpolates the two interior end points $(0.6,0.75)$ and $(1,1.25)$ of the discontinuities (see figure~\ref{fig_toy2_2disc_b}).

Now the emulator standard deviation $\sqrt{\vard{D}{f(\vx)}}$, shown in figure~\ref{fig_toy2_2disc_d} displays clear compression/warping effects in the middle and lower regions for larger $x$, which can be seen to be a direct consequence of the chosen form of $v(x,y)$, as shown in figure~\ref{fig_toy2_2disc_b}. 
This compression is a natural consequence of using a stretched embedding surface (that for example does not conserve 2D distances) whilst using a stationary (isotropic) 3-dimensional covariance structure: paths on steep regions of the embedded surface move ``too fast" into the 3rd dimension, and lead to an induced compression in 2-dimensions. Equivalently, pairs of points in 
2-dimensions end up further apart in 3-dimensions for regions of the embedding $v(x,y)$ that possess large partial derivatives. 

However, we really wish to keep the flexibility of stretched embeddings to ensure that we can always create large enough jumps across discontinuities, and to handle more complex cases, for example, discontinuities that begin and end within the space $\mathcal{X}$, or multiple sets of discontinuities of non-linear form that could be closed, or may even intersect, neither of which could be addressed using say a distance conserving embedding (which notably would mitigate such compression effects, but not entirely remove them). Therefore the compression resulting from use of stretched embeddings represents a serious problem that we will address in the next section.

\ifthenelse{\value{plotstyle}=1}{
\begin{figure}[t]
\begin{center}
\vspace{-0.4cm}
\begin{subfigure}{0.49\columnwidth}
\centering
\includegraphics[scale=0.35,viewport= 0 0 630 530,clip]{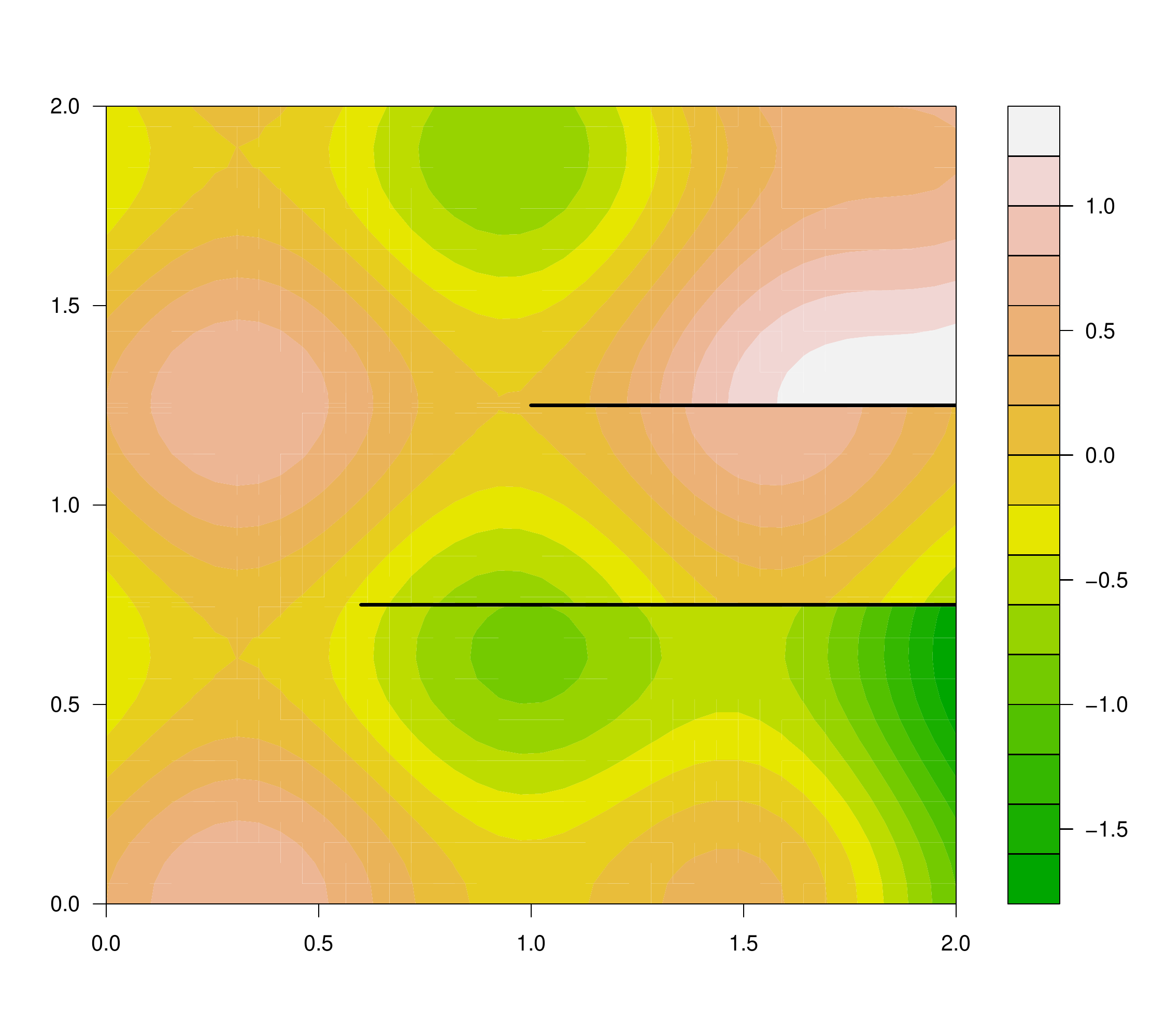} 
\vspace{-0.6cm}
\caption{\footnotesize{The true 2-dimensional function $f(x,y)$.}}
\label{fig_toy2_2disc_a}
\end{subfigure}
\begin{subfigure}{0.49\columnwidth}
\centering
\includegraphics[scale=0.35,viewport= 0 0 630 530,clip]{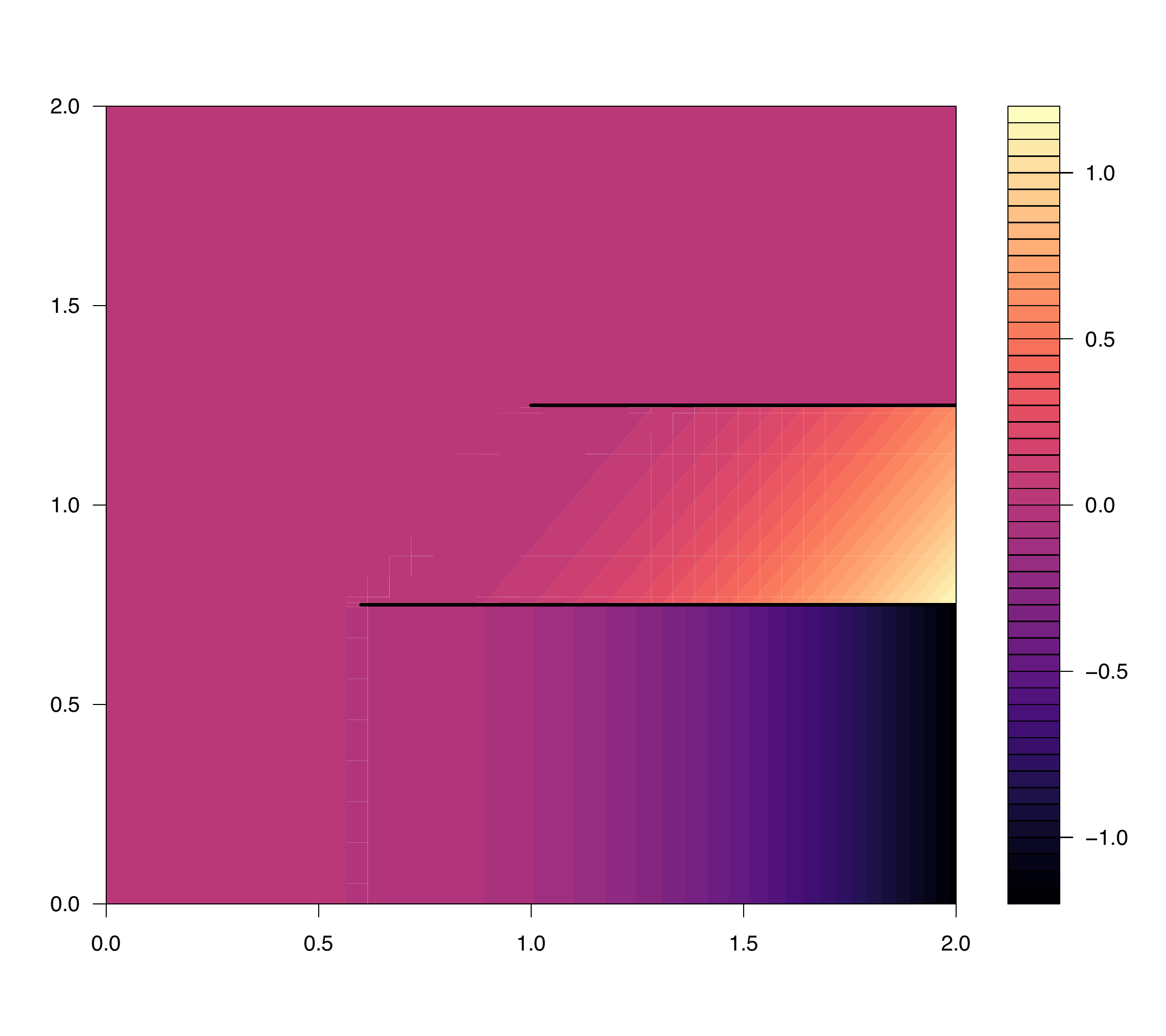}
\vspace{-0.6cm}
\caption{\footnotesize{The embedding surface $v(x,y)$.}}
\label{fig_toy2_2disc_b}
\end{subfigure}
\begin{subfigure}{0.49\columnwidth}
\centering
\vspace{0.3cm}
\includegraphics[scale=0.35,viewport= 0 0 630 530,clip]{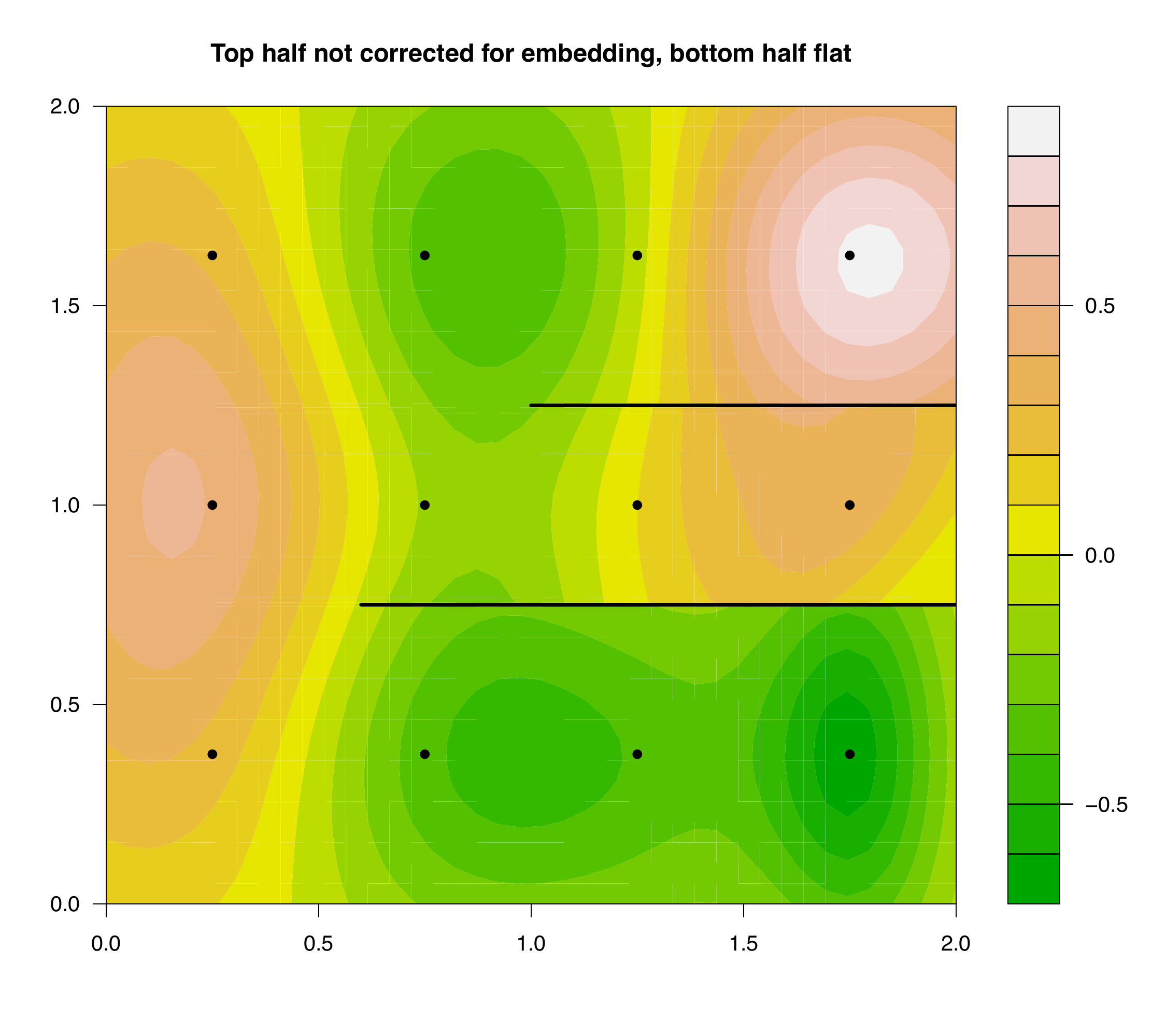}
\vspace{-0.6cm}
\caption{\footnotesize{The emulator expectation $\ed{D}{f(x,y)}$.}}
\label{fig_toy2_2disc_c}
\end{subfigure}
\begin{subfigure}{0.49\columnwidth}
\centering
\vspace{0.3cm}
\includegraphics[scale=0.35,viewport= 0 0 630 530,clip]{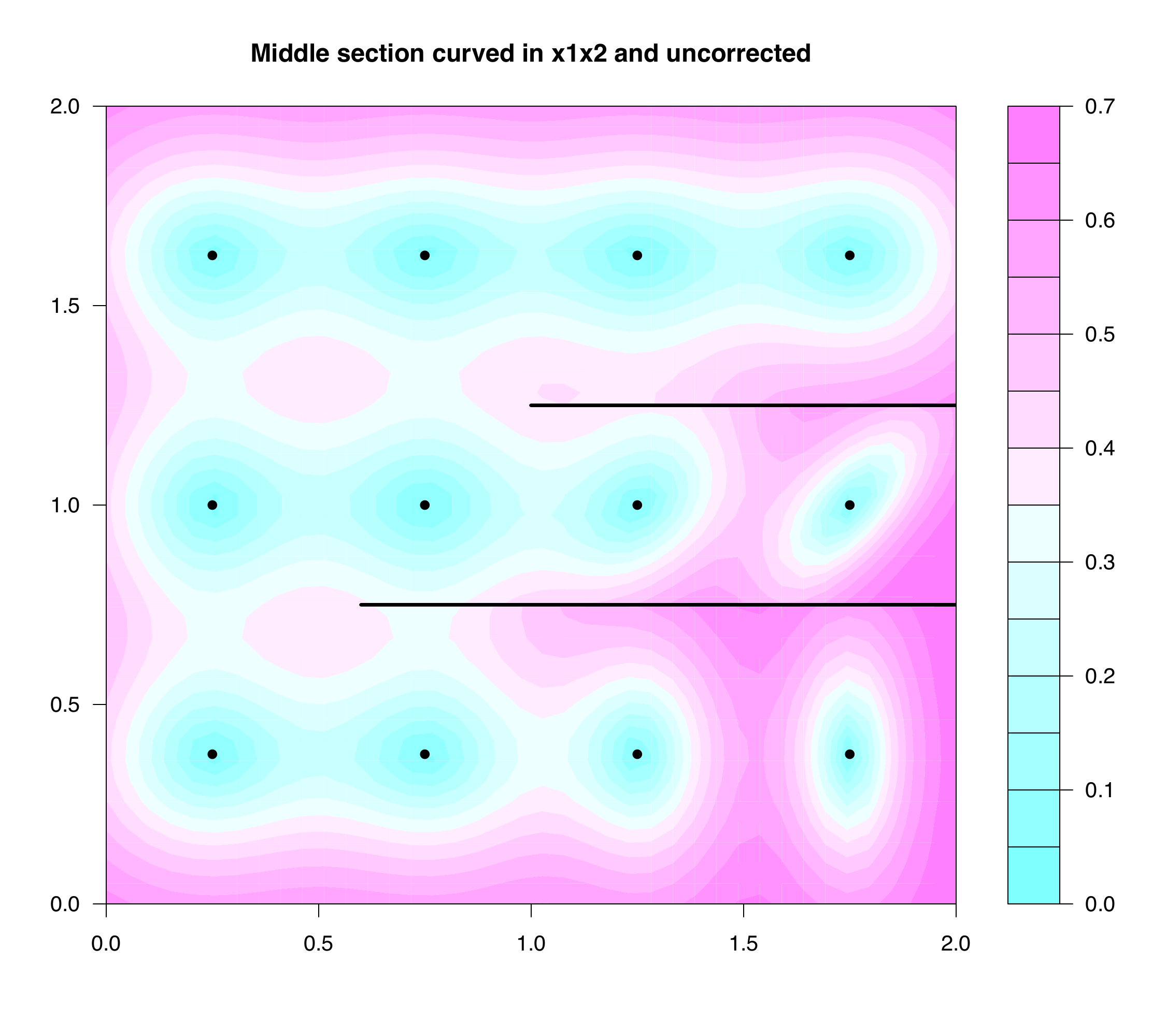}
\vspace{-0.6cm}
\caption{\footnotesize{The emulator stan. dev. $\sqrt{\vard{D}{f(x,y)}}$.}}
\label{fig_toy2_2disc_d}
\end{subfigure}
\begin{subfigure}{0.49\columnwidth}
\centering
\vspace{0.3cm}
\includegraphics[scale=0.35,viewport= 0 0 630 530,clip]{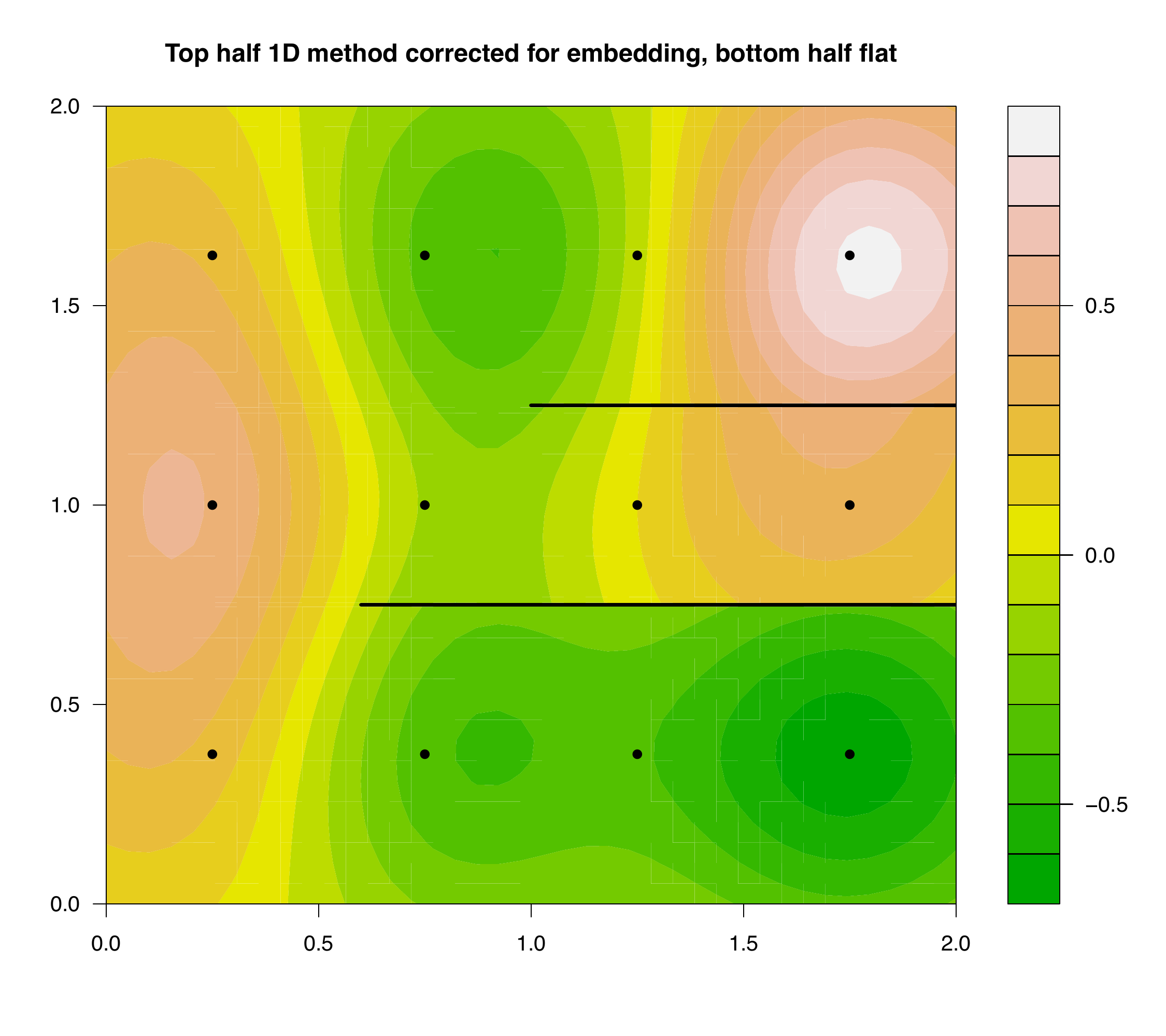} 
\vspace{-0.6cm}
\caption{\footnotesize{The TENSE emulator expectation $\ed{D}{f(x,y)}$.}}
\label{fig_toy2_2disc_e}
\end{subfigure}
\begin{subfigure}{0.49\columnwidth}
\centering
\vspace{0.3cm}
\includegraphics[scale=0.35,viewport= 0 0 630 530,clip]{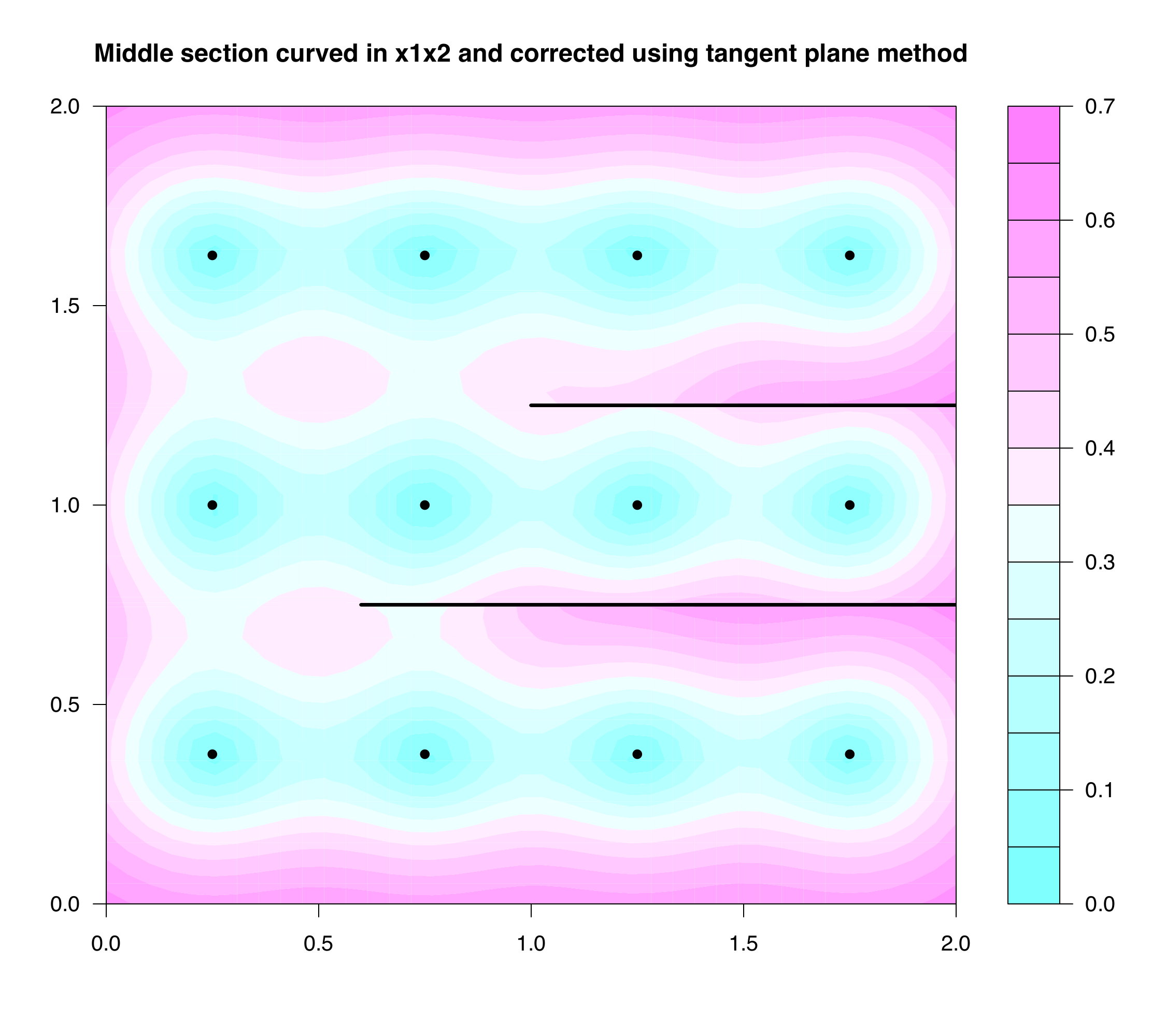} 
\vspace{-0.6cm}
\caption{\footnotesize{The TENSE emulator stan. dev. $\sqrt{\vard{D}{f(x,y)}}$.}}
\label{fig_toy2_2disc_f}
\end{subfigure}
\end{center}
\vspace{-0.4cm}
\caption{\footnotesize{
(a) a 2-dimensional function \cm{f(x,y)} with two partial discontinuities of differing length. (b) The embedding surface \cm{v(x,y)}. (c) The naive emulator expectation \cm{\ed{D}{f(x,y)}} (note warping). (d) Naive emulator standard deviation $\sqrt{\vard{D}{f(x,y)}}$ (note warping due to the embedded surface $v(x,y)$). (e) TENSE emulator expectation \cm{\ed{D}{f(x,y)}} with the warping induced by the use of the embedding surface $\cm{v(x,y)}$ shown in (b), corrected using Non-Stationary Covariance Structures (compare with the uncorrected version given in (c)). (f) TENSE emulator standard deviation $\sqrt{\vard{D}{f(x,y)}}$, again with the warping corrected using Non-Stationary-Covariance Structures (compare with the uncorrected version in (d)).
}}
\label{fig_toy2_2disc}
\vspace{-0.4cm}
\end{figure}
}{}
\ifthenelse{\value{plotstyle}=2}{
\begin{figure}[t]
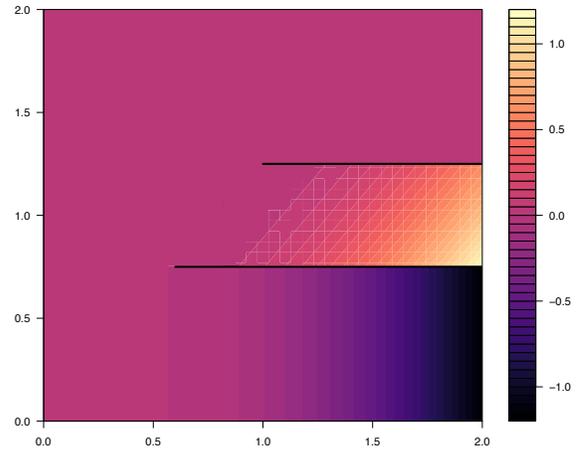
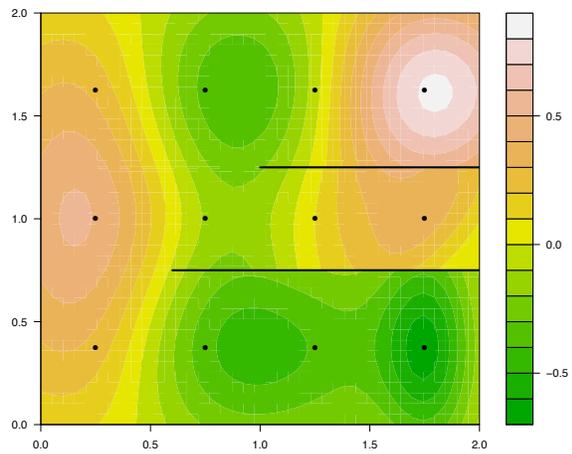
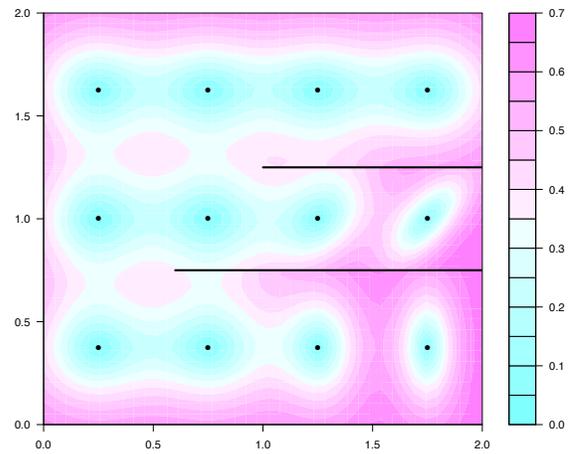
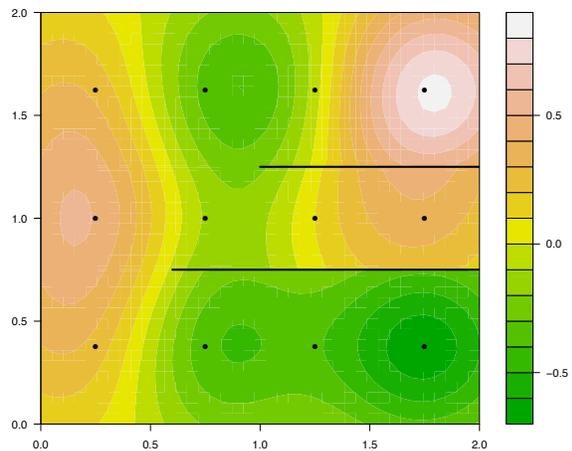
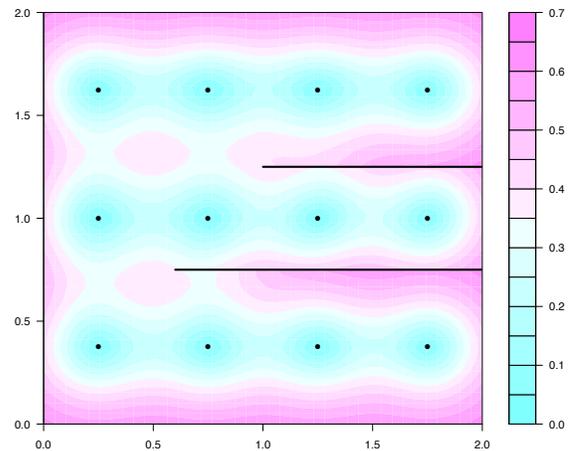

\begin{center}
\vspace{-0.3cm}
\begin{subfigure}{0.49\columnwidth}
%\centering
\hspace{-0.9cm} \includegraphics[scale=0.335,viewport= 0 0 630 530,clip]{plots_paper/plots_examples/2D-Example-2rift-difflength_curved_in_x1x2-true-function.pdf} 
\vspace{-1.0cm}
\caption{\footnotesize{The true 2-dimensional function $f(x,y)$.}}
\label{fig_toy2_2disc_a}
\end{subfigure}
\begin{subfigure}{0.49\columnwidth}
%\centering
\includegraphics[scale=0.335,viewport= 0 0 630 530,clip]{plots_paper/plots_examples/embedding2d_compress_fixed_2rift_diflength_curved_in_x1x2_1.pdf}
\vspace{-1.0cm}
\caption{\footnotesize{The embedding surface $v(x,y)$.}}
\label{fig_toy2_2disc_b}
\end{subfigure}
\begin{subfigure}{0.49\columnwidth}
%\centering
\vspace{0.0cm}
\hspace{-0.9cm} \includegraphics[scale=0.335,viewport= 0 0 630 530,clip]{plots_paper/plots_examples/embedding2d_compress_fixed_2rift_diflength_curved_in_x1x2_2.pdf}
\vspace{-1.0cm}
\caption{\footnotesize{The emulator expectation $\ed{D}{f(x,y)}$.}}
\label{fig_toy2_2disc_c}
\end{subfigure}
\begin{subfigure}{0.49\columnwidth}
%\centering
\vspace{0.0cm}
\includegraphics[scale=0.335,viewport= 0 0 630 530,clip]{plots_paper/plots_examples/embedding2d_compress_fixed_2rift_diflength_curved_in_x1x2_3.pdf}
\vspace{-1.0cm}
\caption{\footnotesize{The emulator stan. dev. $\sqrt{\vard{D}{f(x,y)}}$.}}
\label{fig_toy2_2disc_d}
\end{subfigure}
\begin{subfigure}{0.49\columnwidth}
%\centering
\vspace{0.0cm}
\hspace{-0.9cm} \includegraphics[scale=0.335,viewport= 0 0 630 530,clip]{plots_paper/plots_examples/embedding2d_compress_fixed_2rift_diflength_curved_in_x1x2_4.pdf} 
\vspace{-1.0cm}
\caption{\footnotesize{The TENSE emulator expectation $\ed{D}{f(x,y)}$.}}
\label{fig_toy2_2disc_e}
\end{subfigure}
\begin{subfigure}{0.49\columnwidth}
%\centering
\vspace{0.0cm}
\includegraphics[scale=0.335,viewport= 0 0 630 530,clip]{plots_paper/plots_examples/embedding2d_compress_fixed_2rift_diflength_curved_in_x1x2_5.pdf} 
\vspace{-1.0cm}
\caption{\footnotesize{The TENSE emulator SD $\sqrt{\vard{D}{f(x,y)}}$.}}
\label{fig_toy2_2disc_f}
\end{subfigure}
\end{center}
\vspace{-0.6cm}
\caption{\footnotesize{
(a) a 2-dimensional function \cm{f(x,y)} with two partial discontinuities of differing length. (b) The embedding surface \cm{v(x,y)}. (c) The naive emulator expectation \cm{\ed{D}{f(x,y)}} (note warping). (d) Naive emulator standard deviation $\sqrt{\vard{D}{f(x,y)}}$ (note warping due to the embedded surface $v(x,y)$). (e) TENSE emulator expectation \cm{\ed{D}{f(x,y)}} with the warping induced by the use of the embedding surface $\cm{v(x,y)}$ shown in (b), corrected using Non-Stationary Covariance Structures (compare with the uncorrected version given in (c)). (f) TENSE emulator standard deviation $\sqrt{\vard{D}{f(x,y)}}$, again with the warping corrected using Non-Stationary-Covariance Structures (compare with the uncorrected version in (d)).
}}
\label{fig_toy2_2disc}
\vspace{-0.4cm}
\end{figure}
}{}
%\ifthenelse{\value{plotstyle}=1}{\clearpage}{}
\ifthenelse{\value{plotstyle}=2}{\vspace{-0.3cm}}{}

%%%%%%%%%%%%%%%%%%%%%%%%%%%%%%%%%%%%%%%%%%%%%%%%%%%%%
\section{Controlling the Warping Effect of the Embedding}\label{sec_contr_warping}
%%%%%%%%%%%%%%%%%%%%%%%%%%%%%%%%%%%%%%%%%%%%%%%%%%%%%

\subsection{Reversing the local impact of the embedding}
\ifthenelse{\value{plotstyle}=2}{\vspace{-0.1cm}}{}
\underline{\bf Problem}: The use of the stretched embedding surface \cm{v(x,y)} warps the emulator, compressing the variances and expectations in the examples we have
seen, inducing unwanted \cm{\vx} dependent correlation lengths (and more).
This may lead to inefficient emulators and multiple unintended consequences, and will not reflect our actual prior beliefs about the 2-dimensional computer model. Additionally, we want the freedom to choose a wide variety of embedding surfaces \cm{v(x,y)} without this possibly damaging warping effect occurring.

\vspace{0.0cm}
\noindent \underline{\bf Solution}: We can control this issue using carefully chosen non-stationary covariance structures (NS-CS) defined over the 3-dimensional space.
\vspace{0.0cm}

We now detail a proposed form of the 3-dimensional correlation matrix $\Sigma_{3D}$, used in equation~(\ref{eq_emul_3D_cov_struct}), that is guaranteed to reverse the {\it local} effect of the embedding, that is for input points close together compared to the curvature of the embedding surface. We discuss how to incorporate this choice across the whole input space using necessarily NS-CS in the next section.
We first focus on a reference input point $\vx_0$, and wish to specify a form for $\Sigma_{3D}$ that induces the desired squared exponential 2D covariance structure locally around this point, that 
is such that $\cov{f(\vx)}{f(\vx_0)}$ approximately has the form given by equation~(\ref{eq_sigma_2d}):
\ifthenelse{\value{plotstyle}=2}{\vspace{-0.1cm}}{}
\be \label{eq_sigma_2dB}
\cm{\! \!{\rm Cov}[f(\vx),f(\vx_0)] \;\; \simeq \;\;  \sigma^2 {\rm exp}\left\{- (\vx-\vx_0)^T \Sigma^{-1}_{2D}(\vx-\vx_0) \right\}  \!}
\ee
\ifthenelse{\value{plotstyle}=2}{\vspace{-0.0cm}}{}
for inputs $\vx$ close to $\vx_0$. For definiteness we choose the standard isotropic form of 
\ifthenelse{\value{plotstyle}=2}{\vspace{-0.1cm}}{}
\be\label{eq_sigma_2D_stat}
\cm{\Sigma_{2D} \;\; = \;\; \begin{pmatrix} \theta^2 & 0 \\ 0 & \theta^2 \end{pmatrix} }
\ee
\ifthenelse{\value{plotstyle}=1}{\clearpage}{}
%\ifthenelse{\value{plotstyle}=2}{\vspace{-0.0cm}}{}
\ifthenelse{\value{plotstyle}=1}{\noindent}{}
although everything that follows can be applied to general $\Sigma_{2D}$ by using a simple pre-transformation. 
As the actual covariance structure will be calculated via the embedding $\vv(\vx)$, using equations~(\ref{eq_emul_3D_cov_struct}) and (\ref{eq_sigma_2dB}) we see that we simply require:
%\ba
%{\rm Cov}[f(\vv(\vx)),f(\vv(\vx_0))] &  \simeq &  {\rm Cov}[f(x),f(x_0)] \nonumber \\
%\Leftrightarrow \;\; \sigma^2 {\rm exp}\left\{- (\vv(\vx)-\vv(\vx_0))^T \Sigma^{-1}_{3D}(\vv(\vx)-\vv(\vx_0)) \right\}&  \simeq & \sigma^2 {\rm exp}\left\{- (\vx-\vx_0)^T \Sigma^{-1}_{2D}(\vx-\vx_0) \right\}   \nonumber \\
%\Leftrightarrow \;\; (\vv(\vx)-\vv(\vx_0))^T \Sigma^{-1}_{3D}(\vv(\vx)-\vv(\vx_0))  & \simeq &  (\vx-\vx_0)^T \Sigma^{-1}_{2D}(\vx-\vx_0)  \label{eq_simp_require}
%%{\rm Cov}[f(\vv(\vx)),f(\vv(\vx'))] &  \simeq &  {\rm Cov}[f(x),f(x')] \\
%%\Longleftrightarrow \;\; \sigma^2 {\rm exp}\left\{- (v(x)-v(x_0))^T \Sigma^{-1}_{3D}(v(x)-v(x_0)) \right\}&  \simeq & \sigma^2 {\rm exp}\left\{- (x-x_0)^T \Sigma^{-1}_{2D}(x-x_0) \right\}  \\
%%\Longleftrightarrow \;\; (v(x)-v(x_0))^T \Sigma^{-1}_{3D}(v(x)-v(x_0))  & \simeq &  (x-x_0)^T \Sigma^{-1}_{2D}(x-x_0) 
%\ea
\begin{align}
&& {\rm Cov}[f(\vx),f(\vx_0)] \;\equiv\; {\rm Cov}[f(\vv(\vx)),f(\vv(\vx_0))] &  \;\simeq\;  \sigma^2 {\rm exp}\left\{- (\vx-\vx_0)^T \Sigma^{-1}_{2D}(\vx-\vx_0) \right\}   \nonumber \\
\Leftrightarrow && \sigma^2 {\rm exp}\left\{- (\vv(\vx)-\vv(\vx_0))^T \Sigma^{-1}_{3D}(\vv(\vx)-\vv(\vx_0)) \right\}&  \;\simeq\;  \sigma^2 {\rm exp}\left\{- (\vx-\vx_0)^T \Sigma^{-1}_{2D}(\vx-\vx_0) \right\}   \nonumber \\
\Leftrightarrow && (\vv(\vx)-\vv(\vx_0))^T \Sigma^{-1}_{3D}(\vv(\vx)-\vv(\vx_0))  & \;\simeq \;  (\vx-\vx_0)^T \Sigma^{-1}_{2D}(\vx-\vx_0)  \label{eq_simp_require}
\end{align}
\ifthenelse{\value{plotstyle}=2}{\clearpage}{}

We now approximate $\vv(\vx)$ by its linear Taylor expansion around the point $\vx_0$. This is equivalent to approximating the embedding surface $v(x,y)$ 
by the tangent plane to $v(x,y)$ at the point $\vx_0$ (we will require the tangent plane below for the construction of  $\Sigma^{-1}_{3D}$). Hence we approximate:
\be\label{eq_taylor_expan_v}
v(x,y)  - v(x_0,y_0)  \;=\;   v_x (x - x_0)  +  v_y (y - y_0)  +  \mathcal{O}(\vx^2)
%v(x,y)  - v_0(x_0,y_0)  \;=\;   v_x (x - x_0)  +  v_y (y - y_0)  +  \mathcal{O}(\vx^2)
\ee 
where $v_x = \partial v(x,y)/\partial x$ and $v_y = \partial v(x,y)/\partial y$ are the partial derivatives of $v(x,y)$ evaluated at $\vx_0$, and  
$\mathcal{O}(\vx^2)$ represents second-order terms and above. Similarly for the vector quantity $\vv(\vx)$, we have that, using equations~(\ref{eq_defn_vec_v}) and (\ref{eq_taylor_expan_v}):
\ba
\vv(\vx)  - \vv(\vx_0)  &=&   \begin{pmatrix} x -x_0\\ y -y_0 \\ v(x,y) - v(x_0,y_0) \end{pmatrix}   \;=\;   \begin{pmatrix} x -x_0\\ y -y_0 \\  v_x (x - x_0)  +  v_y (y - y_0)  +  \mathcal{O}(\vx^2) \end{pmatrix}  \nonumber \\
&=& \begin{pmatrix}   1 & 0 \\  0 & 1 \\ v_x & v_y  \end{pmatrix}   \begin{pmatrix}    x -x_0  \\  y -y_0    \end{pmatrix}  +  \mathcal{O}(\vx^2)  \nonumber \\
&=& A (\vx-\vx_0) +  \mathcal{O}(\vx^2), \quad \quad  \quad \quad \text{where} \quad A = \begin{pmatrix}   1 & 0 \\  0 & 1 \\ v_x & v_y  \end{pmatrix} 
\ea 
Replacing this into equation~(\ref{eq_simp_require}) and dropping second-order terms and above, we get
\begin{align}
\Leftrightarrow && (\vx-\vx_0)^T A^T \Sigma^{-1}_{3D}A (\vx-\vx_0) & \; \simeq \; (\vx-\vx_0)^T \Sigma^{-1}_{2D}(\vx-\vx_0)  \\
\Leftrightarrow  &&  A^T \Sigma^{-1}_{3D}A &\; \simeq \; \Sigma^{-1}_{2D}  \label{eq_3D_2D_require1}
\end{align}
Hence we see the intuitive result that in order to counter the linear effect of the embedding surface in the vicinity of $\vx_0$, we just need to choose a form for $\Sigma_{3D}$ that satisfies equation~(\ref{eq_3D_2D_require1}), where 
$A$ represents the linear embedding operator that raises the 2-dimensional position vector $\vx$ onto its corresponding location on the 3-dimensional tangent plane given by $A(\vx-\vx_0)$.
%
%\item These have correlation lengths \cm{\theta_{3\!\!\:D}(v(x))} and more generally correlation matrices \cm{\Sigma_{3\!\!\:D}(v(x))} that change for different values of \cm{x\in \mathcal{X}}.
%\vspace{0.3cm}
%
%\item \cm{\Sigma_{3\!\!\:D}(v(x))} is then carefully chosen to cancel out the warping effect from the embedding surface \cm{v(x)}, ensuring that the induced 2-dimensional correlation is
%corrected and approximates \cm{\Sigma_{2\!\!\:D}(x)}. This works to first order.
%\vspace{0.3cm}
%
%\item Now we have powerful emulators that can handle any number of partial (or complete) discontinuities of general form.
\ifthenelse{\value{plotstyle}=2}{\vspace{-0.2cm}}{}
\subsubsection*{Constructing $\Sigma_{3D}$}
\ifthenelse{\value{plotstyle}=2}{\vspace{-0.1cm}}{}
There are several forms one could choose for $\Sigma_{3D}$ in order to satisfy equation~(\ref{eq_3D_2D_require1}), however, many of these will not facilitate sufficient decorrelation of the emulator across discontinuities in the embedding surface $v(x,y)$.
We hence choose a form for $\Sigma_{3D}$ that is aligned with the tangent plane to $v(x,y)$ at the point $\vx_0$, a form which is specifically selected to provide substantial (possibly maximal) and controllable decorrelation across the discontinuities. 

We first set up a relevant orthonormal basis $\{\vw_1,\vw_2,\vw_3\}$. Setting $g(x,y,z) = z - v(x,y)$ and noting that $g(x,y,z)=0$ defines the embedding surface $z=v(x,y)$, we see, according to standard vector calculus results, that $\nabla g(x,y,z)$ evaluated at $\vx_0$ gives 
the vector normal to the embedding surface (and normal to the tangent plane), which we set as the unit vector $\vw_3$:
\ba
\vw_3  \;\; \propto \;\; \nabla g(x,y,z) \;\;=\;\; -v_x \ve_x -v_y \ve_y + \ve_z 
\ea
%where $v_x = \partial v(x,y)/\partial x$ and $v_y = \partial v(x,y)/\partial y$ are again the partial derivatives of $v(x,y)$ evaluated at $\vx_0$. 
We choose the unit basis vector $\vw_1$ to lie in the tangent plane, but pointing in the direction of maximally increasing $v(x,y)$. Hence $\vw_1$ has 2-dimensional components parallel to $\nabla v(x,y) =  v_x \ve_x +v_y \ve_y$, and hence has the form
 \ba
 \vw_1 & \propto &  v_x \ve_x +v_y \ve_y + \gamma \ve_z 
 \ea
 where as $\vw_1$ lies on the tangent plane we have that $\vw_1 . \vw_3 \;=\; 0$ which implies that $\gamma =  v_x^2 + v_y^2$. The vector $\vw_2$ will be 
 orthogonal to both $\vw_1$ and $\vw_3$, but as $\vw_1$ was chosen to be in the direction of maximally increasing $v(x,y)$, $\vw_2$ must have zero component in the 3rd dimension and so takes the form:
\be
\vw_2 \;\; \propto \;\; \beta \ve_x + \delta \ve_y 
\ee 
 Applying the orthogonality relation $\vw_2 . \vw_1 = 0$ implies $\beta  v_x + \delta v_y  =0$
% \ba
% \vw_2 . \vw_1 &=& 0 \\
% \Leftrightarrow \quad  \beta  v_x + \delta v_y  &=&0
%  \ea 
 which in turn implies that $ \beta = -v_y, \delta = v_x $ are suitable choices, up to an overall normalising constant. To summarise, we have constructed the orthonormal basis  
 $\{\vw_1,\vw_2,\vw_3\}$ given by
%  \ba
%  \vw_1 &=&  \frac{1}{c_1}\left[ v_x \ve_x +v_y \ve_y + (v_x^2 + v_y^2) \ve_z \right], \text{ with }  c_1 = v_x^2 + v_y^2 + (v_x^2 + v_y^2)^2 \\
%  \vw_2 &=&  \frac{1}{c_2}\left[-v_y \ve_1 + v_x \ve_2  \right], \text{ with }  c_2 = v_x^2 + v_y^2  \\
%  \vw_3 &=&   \frac{1}{c_3}\left[-v_x \ve_x -v_y \ve_y + \ve_z \right], \text{ with }  c_3 = v_x^2 + v_y^2 + 1
%  \ea
 \begin{align}
  \vw_1 &\;=\;  \frac{1}{c_1}\left[ v_x \ve_x +v_y \ve_y + (v_x^2 + v_y^2) \ve_z \right],& \text{ where }&  c_1^2 = v_x^2 + v_y^2 + (v_x^2 + v_y^2)^2    \label{eq_w1} \\
  \vw_2 &\;=\;  \frac{1}{c_2}\left[-v_y \ve_x + v_x \ve_y  \right], &\text{ where }&  c_2^2 = v_x^2 + v_y^2   \label{eq_w2}  \\
  \vw_3 &\;=\;   \frac{1}{c_3}\left[-v_x \ve_x -v_y \ve_y + \ve_z \right], &\text{ where }&  c_3^2 = v_x^2 + v_y^2 + 1  \label{eq_w3}
 \end{align}
% \begin{align}
%  \vw_1 &=  \frac{1}{c_1}\left[ v_x \ve_x +v_y \ve_y + (v_x^2 + v_y^2) \ve_z \right],& \text{ with }&  c_1 = v_x^2 + v_y^2 + (v_x^2 + v_y^2)^2 = r^2(1+r^2) \\
%  \vw_2 &=  \frac{1}{c_2}\left[-v_y \ve_1 + v_x \ve_2  \right], &\text{ with }&  c_2 = v_x^2 + v_y^2 = r^2  \\
%  \vw_3 &=   \frac{1}{c_3}\left[-v_x \ve_x -v_y \ve_y + \ve_z \right], &\text{ with }&  c_3 = v_x^2 + v_y^2 + 1 = r^2+1
% \end{align}
where $\vw_1$ and $\vw_2$ lie on the tangent plane at $\vx_0$, while $\vw_3$ is orthogonal to the tangent plane.

We postulate that if we specify $\Sigma_{3D}$ to be diagonal in the above $\{\vw_1,\vw_2,\vw_3\}$ basis, then it will satisfy the desired projection constraint given by equation~(\ref{eq_3D_2D_require1}). We now show this to be true, subject to some additional conditions. For definiteness, say that $\Sigma_{3D}$ is indeed diagonal with respect to the $\{\vw_1,\vw_2,\vw_3\}$ basis
with corresponding eigenvalues $\{\alpha_1^2, \alpha_2^2, \alpha_3^2\}$, hence we can represent $\Sigma_{3D}$ as:
\ba \label{eq_sig_3d_spec_decomp}
\Sigma_{3D} &=& \alpha_1^2\, \vw_1 \vw_1^T + \alpha_2^2\, \vw_2 \vw_2^T + \alpha_3^2 \, \vw_3 \vw_3^T 
\ea
and similarly the inverse $\Sigma_{3D}^{-1}$ as
\ba
\Sigma_{3D}^{-1} &=& \frac{1}{\alpha_1^2} \, \vw_1 \vw_1^T + \frac{1}{\alpha_2^2} \, \vw_2 \vw_2^T + \frac{1}{\alpha_3^2} \, \vw_3 \vw_3^T 
\ea
To evaluate $A^T \Sigma^{-1}_{3D}A $ as required by equation~(\ref{eq_3D_2D_require1}), we first note that 
\be
\vw_3^T A \;\;=\;\; \frac{1}{c_3}\begin{pmatrix} -v_x & -v_y & 1 \end{pmatrix}  \begin{pmatrix} 1&0 \\ 0&1 \\ v_x & v_y \end{pmatrix} \;\;=\;\;  \begin{pmatrix} 0&0  \end{pmatrix}
\ee
and hence we have that
\ifthenelse{\value{plotstyle}=1}{
\begin{align}
A^T \Sigma^{-1}_{3D}A & \;=\; A^T ( \frac{1}{\alpha_1^2} \, \vw_1 \vw_1^T + \frac{1}{\alpha_2^2} \, \vw_2 \vw_2^T + \frac{1}{\alpha_3^2} \, \vw_3 \vw_3^T ) A \nonumber\\
     & \;=\;  \frac{1}{\alpha_1^2} \, A^T \vw_1 \vw_1^T A + \frac{1}{\alpha_2^2} \, A^T \vw_2 \vw_2^T A.   \label{eq_ASigA_w1_w2_terms}
\end{align}
}{}
\ifthenelse{\value{plotstyle}=2}{
\be
A^T \Sigma^{-1}_{3D}A  \;=\; A^T ( \frac{1}{\alpha_1^2} \, \vw_1 \vw_1^T + \frac{1}{\alpha_2^2} \, \vw_2 \vw_2^T + \frac{1}{\alpha_3^2} \, \vw_3 \vw_3^T ) A 
      \;=\;  \frac{1}{\alpha_1^2} \, A^T \vw_1 \vw_1^T A + \frac{1}{\alpha_2^2} \, A^T \vw_2 \vw_2^T A.   \label{eq_ASigA_w1_w2_terms}
\ee
}{}
We see that $A^T \Sigma^{-1}_{3D}A$ does not depend on $\alpha_3$. As will be discussed further below, $\alpha_3$ is a free parameter, which we can choose to control the extent of the decorrelation of the emulator across the discontinuities, and is one of the motivations for choosing the proposed form of $\Sigma_{3D}$ given by equation~(\ref{eq_sig_3d_spec_decomp}).

To evaluate the remaining terms in equation~(\ref{eq_ASigA_w1_w2_terms}), we have that:
\begin{align}
\vw_2^T A &\;\;=\;\; \frac{1}{c_2} \begin{pmatrix} -v_y & v_x & 0 \end{pmatrix} \begin{pmatrix} 1&0 \\ 0&1 \\ v_x & v_y \end{pmatrix} \;\;=\;\; \frac{1}{c_2} \begin{pmatrix} -v_y & v_x  \end{pmatrix} \\
\Rightarrow \quad \frac{1}{\alpha_2^2} \, A^T \vw_2 \vw_2^T A & \;\;=\;\; \frac{1}{\alpha_2^2 c_2^2}  \begin{pmatrix} -v_y \\ v_x  \end{pmatrix} \begin{pmatrix} -v_y & v_x  \end{pmatrix} \;\;=\;\;
\frac{1}{\alpha_2^2 r^2} \begin{pmatrix} v_y^2 & -v_xv_y \\  -v_x v_y & v_x^2  \end{pmatrix}   \label{eq_Aw2wA}
\end{align}
where we have employed the simplifying notation $r^2 \equiv v_x^2+v_y^2 = c_2^2$. Similarly, and using $c_1^2 = r^2+ r^4= r^2(1+r^2)$, we have
\begin{align}
\vw_1^T A &\;=\; \frac{1}{c_1} \begin{pmatrix} v_x & v_y & v_x^2+v_y^2 \end{pmatrix} \begin{pmatrix} 1&0 \\ 0&1 \\ v_x & v_y \end{pmatrix} \;=\; \frac{1}{c_1} \begin{pmatrix} v_x(1+r^2) & v_y(1+r^2)  \end{pmatrix} \nonumber \\
%% Intermediate step included
%\Rightarrow \quad \frac{1}{\alpha_1^2} \, A^T \vw_1 \vw_1^T A & \;\;=\;\; \frac{1}{\alpha_1^2 c_1^2}  \begin{pmatrix} v_x(1+r^2) \\ v_y(1+r^2)  \end{pmatrix} \begin{pmatrix} v_x(1+r^2) & v_y(1+r^2)  \end{pmatrix} \;\;=\;\;
%\frac{1}{\alpha_1^2 c_1^2} \begin{pmatrix} v_x^2( 1+r^2)^2 & v_x v_y ( 1+r^2)^2 \\  v_x v_y ( 1+r^2)^2 &  v_y^2( 1+r^2)^2  \end{pmatrix}
\Rightarrow \quad \frac{1}{\alpha_1^2} \, A^T \vw_1 \vw_1^T A & \;=\; \frac{(1+r^2)^2 }{\alpha_1^2 c_1^2}  \begin{pmatrix} v_x\\ v_y \end{pmatrix} \begin{pmatrix} v_x & v_y \end{pmatrix} \;=\;
\frac{( 1+r^2)}{\alpha_1^2 r^2} \begin{pmatrix} v_x^2 & v_x v_y  \\  v_x v_y  &  v_y^2  \end{pmatrix} \label{eq_Aw1wA}
\end{align}

Combining equations~(\ref{eq_sigma_2D_stat}), (\ref{eq_ASigA_w1_w2_terms}), (\ref{eq_Aw2wA}) and (\ref{eq_Aw1wA}), we see that the projection constraint given by equation~(\ref{eq_3D_2D_require1}) can now be rewritten as
\begin{align}
\Sigma_{2D} &  \;\;=\;\;  A^T \Sigma^{-1}_{3D}A\\
 \Leftrightarrow \quad \quad   \begin{pmatrix} \frac{1}{\theta^2} & 0 \\ 0 & \frac{1}{\theta^2} \end{pmatrix}  
		& \;\;=\;\;    \frac{( 1+r^2)}{\alpha_1^2 r^2} \begin{pmatrix} v_x^2 & v_x v_y  \\  v_x v_y  &  v_y^2  \end{pmatrix} 
		 \;+\;   \frac{1}{\alpha_2^2 r^2} \begin{pmatrix} v_y^2 & -v_xv_y \\  -v_x v_y & v_x^2  \end{pmatrix}  \\
 \Leftrightarrow \quad \quad   \begin{pmatrix} \frac{1}{\theta^2} & 0 \\ 0 & \frac{1}{\theta^2} \end{pmatrix}  & \;\;=\;\;  \frac{1}{r^2}\begin{pmatrix}
   \frac{v_y^2}{\alpha_2^2}  + \frac{( 1+r^2) v_x^2}{\alpha_1^2}    &    \left(\frac{( 1+r^2) }{\alpha_1^2}  - \frac{1}{\alpha_2^2} \right)v_x v_y \\
    \left(\frac{( 1+r^2) }{\alpha_1^2}  - \frac{1}{\alpha_2^2} \right) v_x v_y   &     \frac{v_x^2}{\alpha_2^2}  + \frac{( 1+r^2) v_y^2}{\alpha_1^2}   \label{eq_full_indu_mat1}
\end{pmatrix} 
\end{align}
Equating the off-diagonal terms gives:
\begin{align}
\left(\frac{( 1+r^2) }{\alpha_1^2}  - \frac{1}{\alpha_2^2} \right)v_x v_y &\;=\; 0 \\
\Leftrightarrow \quad \quad  \text{\bf{Case 1:} } \; \alpha_1^2 \;=\; \alpha_2^2 (1+r^2) &  \;\;\text{ or }\;\;   \text{\bf{Case 2:} } \;  v_x =0  \;\;\text{ or }\;\;  
\text{\bf{Case 3:} } \;  v_y =0    \nonumber
%\Leftrightarrow  \quad \quad \text{\bf{Case 1:} } \; \alpha_1^2 \;=\; \alpha_2^2 (1+r^2) &
\end{align}
For {\bf Case 1} we replace $\alpha_1^2 =\alpha_2^2 (1+r^2) $ into equation~(\ref{eq_full_indu_mat1}) giving
\be
 \begin{pmatrix} \frac{1}{\theta^2} & 0 \\ 0 &\frac{1}{ \theta^2} \end{pmatrix}  \;\;=\;\;   \frac{1}{r^2}\begin{pmatrix}
   \frac{v_y^2}{\alpha_2^2}  + \frac{( 1+r^2) v_x^2}{\alpha_2^2(1+r^2)}    &   0 \\
   0 &     \frac{v_x^2}{\alpha_2^2}  + \frac{( 1+r^2) v_y^2}{\alpha_2^2(1+r^2)}   \end{pmatrix}   \;\;=\;\;  \begin{pmatrix} \frac{1} {\alpha_2^2} & 0 \\ 0 & \frac{1}{\alpha_2^2} \end{pmatrix}
 \quad  \Leftrightarrow \quad \alpha_2^2\;=\;\theta^2
\ee
For  {\bf Case 2} we replace $v_x =0$ (which implies $r^2=v_y^2$) into equation~(\ref{eq_full_indu_mat1}) giving
\be
 \begin{pmatrix} \frac{1}{\theta^2} & 0 \\ 0 &\frac{1}{ \theta^2} \end{pmatrix}  \;\;=\;\;  \begin{pmatrix}
   \frac{1}{\alpha_2^2}      &   0 \\
   0 &      \frac{( 1+r^2) }{\alpha_1^2}    \end{pmatrix}  
 \quad  \Leftrightarrow \quad \alpha_2^2\;=\;\theta^2 \text{ and } \alpha_1^2 =\alpha_2^2 (1+r^2)
\ee
which is exactly the same result as {\bf Case 1}. {\bf Case 3} gives the same answer also, due to the symmetry between $x$ and $y$. 
Therefore, we finally see that the projection requirement given by equation~(\ref{eq_3D_2D_require1}) is satisfied by specifying the first two eigenvalues $\alpha_1^2$ and $\alpha_2^2$ of $\Sigma_{3D}$ to be
\be \label{eq_alph1_alph2}
\alpha_1^2 \;=\;  \theta^2(1+r^2) \quad \quad \text{and} \quad \quad \alpha_2^2 \;=\;  \theta^2
\ee
with $r^2 = v_x^2 + v_y^2$. This demonstrates that the choice of form of $\Sigma_{3D}$ as proposed in equation~(\ref{eq_sig_3d_spec_decomp}) is indeed valid. 
The constraint on the eigenvalues is intuitive from a geometric perspective especially when considering the choice of the basis $\{\vw_1,\vw_2,\vw_3\}$: as $\vw_2$ points along a direction 
in which the embedding surface $v(x,y)$ is not (locally) increasing, there will be no warping/compression of the emulator along this direction, in which case $\alpha_2^2$ must equal the desired 2D correlation length of $\theta^2$. Conversely, $\vw_1$ was defined to point in the direction of maximally increasing $v(x,y)$, and the gradient of $v(x,y)$ in this direction is 
$|\nabla v(x,y)| = r$ hence $\alpha_1^2$ must be increased to counteract the warping/compression along this direction that would otherwise be induced by the use of such a stretched embedding surface. Finally, as $\vw_3$ by construction is orthogonal to $v(x,y)$ at $\vx_0$, and as we are only interested in points that lie on $v(x,y)$, there must be no constraint imposed at this stage on $\alpha_3^2$, and hence it will be a free parameter that we can choose or indeed infer.

We need an explicit representation for $\Sigma_{3D}$ (in the standard Cartesian basis) for use in the non-stationary emulators employed in the next section, and we now have all the pieces required to build this representation, using equations~(\ref{eq_sig_3d_spec_decomp}), (\ref{eq_alph1_alph2}) and the definition of the basis vectors (equations~(\ref{eq_w1}), (\ref{eq_w2}) and (\ref{eq_w3})), as follows. We have 
\begin{align}
\alpha_1^2 \vw_1\vw_1^T &\;=\; \frac{\alpha_1^2}{c_1^2} \begin{pmatrix} v_x \\ v_y \\ v_x^2+v_y^2 \end{pmatrix}   \begin{pmatrix} v_x & v_y & v_x^2+v_y^2 \end{pmatrix} &\;=\; & &
\frac{\theta^2}{r^2}  \begin{pmatrix}  
						v_x^2 & v_xv_y & v_x r^2 \\
						v_x v_y & v_y^2 & v_y r^2 \\
						v_x r^2 & v_y r^2 & r^4
 				\end{pmatrix}& \nonumber \\
\alpha_2^2 \vw_2\vw_2^T &\;=\; \frac{\alpha_2^2}{c_2^2} \begin{pmatrix} -v_y \\ v_x \\ 0 \end{pmatrix}   \begin{pmatrix} -v_y & v_x & 0 \end{pmatrix}  &\;=\; & &\frac{\theta^2}{r^2}  
				\begin{pmatrix}  
						v_y^2 & -v_xv_y & 0 \\
						-v_x v_y & v_x^2 & 0 \\
						0 & 0 & 0
 				\end{pmatrix} & \nonumber \\
\alpha_3^2 \vw_3\vw_3^T &\;=\; \frac{\alpha_3^2}{c_3^2} \begin{pmatrix} -v_x \\ -v_y \\ 1 \end{pmatrix}   \begin{pmatrix} -v_x & v_y & 1 \end{pmatrix}  &\;=\; & &\frac{\alpha_3^2}{r^2+1}
				\begin{pmatrix}  
						v_x^2 & v_xv_y & -v_x \\
						v_x v_y & v_y^2 & -v_y \\
						-v_x & -v_y & 1
 				\end{pmatrix} 	  &  \nonumber
\end{align}
We can hence explicitly construct $\Sigma_{3D} = \alpha_1^2\, \vw_1 \vw_1^T + \alpha_2^2\, \vw_2 \vw_2^T + \alpha_3^2 \, \vw_3 \vw_3^T$ giving
\be \label{eq_sigma3D_full_exp1}
 \Sigma_{3D}(\vx_0)  \;\;=\;\; \begin{pmatrix}
		\theta^2 + \dfrac{\alpha_3^2 v_x^2}{r^2+1}   &   \dfrac{\alpha_3^2v_x v_y}{r^2+1}      &     v_x\left(\theta^2 - \dfrac{\alpha_3^2}{r^2+1} \right)  \\
		\dfrac{\alpha_3^2v_x v_y}{r^2+1}                 &     \theta^2 + \dfrac{\alpha_3^2 v_y^2}{r^2+1}    &    v_y\left(\theta^2 - \dfrac{\alpha_3^2}{r^2+1} \right)  \\
		v_x\left(\theta^2 - \dfrac{\alpha_3^2}{r^2+1} \right)   &    v_y\left(\theta^2 - \dfrac{\alpha_3^2}{r^2+1} \right)   & \theta^2 r^2  + \dfrac{\alpha_3^2}{r^2+1} 
		\end{pmatrix} 
\ee
where we make the dependence on $\vx_0$ explicit. Using this expression for $\Sigma_{3D}(\vx_0)$ in the covariance structure of the embedded emulator as given in equation~(\ref{eq_emul_3D_cov_struct}), will yield for points 
close to $\vx_0$, the desired induced covariance structure as represented by $\Sigma_{2D}$ in equation~(\ref{eq_sigma_2D_stat}). For embeddings with zero curvature, this correction is exact.

\subsection{Controlling the Global Impact of the Embedding Using Non-Stationary Emulation}

The above form of $\Sigma_{3D}(\vx_0)$ as given by equation~(\ref{eq_sigma3D_full_exp1}), will correct for the impact of the embedding surface on the emulator's covariance structure, but only locally around the point $\vx_0$, as $v_x$, $v_y$ and $r^2$ are all evaluated at $\vx_0$. This is not enough for our needs, as we wish to correct the whole emulator globally 
over all of $\mathcal{X}$.
Therefore we employ a non-stationary covariance structure as follows. We define an $\vx$ dependent covariance matrix $\Sigma_{3D}(\vx)$ exactly of the form given by equation~(\ref{eq_sigma3D_full_exp1}), but now evaluated at general point $\vx$. 
As this covariance matrix $\Sigma_{3D}(\vx)$  will vary over the input space for general embeddings $v(x,y)$ (except in the trivial case of a linear embedding), 
we employ the non-stationary apparatus recently used by \cite{Dunlop:2018uz}, first derived by \cite{Paciorek:2003tg}, in order to define a valid covariance structure. 

In the standard non-stationary scenario (i.e. without any embedding surface) \cite{Dunlop:2018uz} 
use the generalised non-stationary squared exponential covariance function which essentially averages an $\vx$ dependent covariance matrix $\Sigma(\vx)$ as follows, while guaranteeing a valid covariance structure over the whole input space. 
They define the quadratic form $Q(\vx, \vx^\prime)$ for an $\vx$ dependent covariance matrix $\Sigma(\vx)$ as
\be \label{eq:Quadratic-Form-for-non-stationary-cov-func}
Q(\vx, \vx^\prime) \;=\; \left( \vx- \vx^\prime \right)^T \left( \frac{\Sigma(\vx) + \Sigma(\vx^\prime)}{2} \right)^{-1} \left( \vx - \vx^\prime \right)
\ee
and then the corresponding non-stationary squared exponential covariance function for use in the emulator is given, for $d$-dimensional $\vx$, as 
\be \label{eq:Squared-exponential-covariance-structure-higher-dimensional-long}
\Cov[f(\vx), f(\vx^\prime)] \;\;=\;\; \sigma^2 \frac{2^{\frac{d}{2}} \left| \Sigma(\vx) \right|^{\frac{1}{4}} \left| \Sigma(\vx^\prime) \right|^{\frac{1}{4}} }{\left| \Sigma(\vx) + \Sigma(\vx^\prime) \right|^{\frac{1}{2}}} \exp\left \{ - Q(\vx, \vx^\prime) \right \}
\ee

For our use we simply elevate this non-stationary structure to lie on the embedding surface $v(x,y)$ in the 3D space, hence we instead define the quadratic form via the position vector on the embedding surface $\vv(\vx)$ as
\be \label{eq:Quadratic-Form-for-non-stationary-cov-func_embedded}
Q(\vv(\vx), \vv(\vx^\prime)) \;=\; \left( \vv(\vx)- \vv(\vx^\prime) \right)^T \left( \frac{\Sigma_{3D}(\vv(\vx)) + \Sigma_{3D}(\vv(\vx^\prime))}{2} \right)^{-1} \left( \vv(\vx) - \vv(\vx^\prime) \right)
\ee
and similarly the corresponding non-stationary squared exponential covariance function in the embedded 3D input space is given as 
\be \label{eq_NS_squared_cov_struc_3D_full}
\Cov[f(\vv(\vx)), f(\vv(\vx^\prime))] \;\;=\;\; \sigma^2 \frac{2^{\frac{3}{2}} \left| \Sigma_{3D}(\vv(\vx)) \right|^{\frac{1}{4}} \left| \Sigma_{3D}(\vv(\vx^\prime)) \right|^{\frac{1}{4}} }{\left| \Sigma_{3D}(\vv(\vx)) + \Sigma_{3D}(\vv(\vx^\prime)) \right|^{\frac{1}{2}}} \exp\left \{ - Q(\vv(\vx), \vv(\vx^\prime)) \right \}
\ee
where $\Sigma_{3D}(\vv(\vx)) \equiv \Sigma_{3D}(\vx)$ is given by equation~(\ref{eq_sigma3D_full_exp1}) with $\vx_0$ replaced by $\vx$.
This again guarantees a valid covariance structure throughout both the 3D space and the induced 2D space. Note that this construction generalises to a wide class of covariance structures~\citep{Dunlop:2018uz}.

We see that now for {\it any} pair of input points $\vx$ and $\vx^\prime$ that are close together relative to the curvature of the embedding surface, the non-stationary 
covariance structure as given by equation~(\ref{eq_NS_squared_cov_struc_3D_full}), which essentially averages the covariance matrices $\Sigma_{3D}(\vx)$ and 
$\Sigma_{3D}(\vx^\prime)$ defined at each of the points, will counteract the local warping effect of the embedding surface, to first order.
For pairs of input points that are further apart, non-linear effects may become noticeable, however, for modest choices of correlation length $\theta$ these effects will typically be suppressed as the covariance rapidly drops to zero for points that are further apart than the correlation length.
%(  mention long range of effects of squared exponential function?).
Therefore, an emulator constructed using the non-stationary 
covariance structure given by equation~(\ref{eq_NS_squared_cov_struc_3D_full}) will a) allow us the freedom to choose from a wide class of torn embedding surfaces $v(x,y)$ to handle unlimited numbers of discontinuities of complex configuration and to ensure that the emulator is decorrelated across them, as discussed in section~\ref{ssec_torn_embed}, and b) will approximately induce the desired stationary 2D covariance structure across local regions that do not contain discontinuities. We refer to this general framework as the Torn Embedding Non-Stationary Emulation (TENSE) approach.

In figures~\ref{fig_toy2_2disc_e} and \ref{fig_toy2_2disc_f} we apply the TENSE approach to the toy model discussed in section~\ref{ssec_torn_embed}. Comparing with the uncorrected version, as seen in figures~\ref{fig_toy2_2disc_c} and \ref{fig_toy2_2disc_d}, we see that the emulator standard deviation $\sqrt{\vard{D}{f(x)}}$ now displays no noticeable warping effects and maintains the symmetry we would expect around each of the run locations (the black points) especially in the top, middle and lower regions for large $x$, while also displaying suitable uncorrelated behaviour across the discontinuities.
The emulator expectation also looks far more reasonable, displaying no noticeable warping. 

Although we demonstrate this framework in 2D/3D and for squared exponential covariances, it is simple to extend in various ways. For example, the above calculations extend to any covariance structure of the form $r(a)$ where $a$ is the general Mahalanobis distance between $\vx$ and $\vx'$ and $r(.)$ is a valid covariance function, e.g. the Matérn~\citep{GPML}, using the general form for equation~(\ref{eq_NS_squared_cov_struc_3D_full}) \citep{Dunlop:2018uz}. If one desires a non-stationary induced 2D covariance structure we can achieve this using a similar strategy by inserting a point-wise 2D pre-transformation. Similarly this torn embedding strategy can be extended to higher dimensional input spaces with more complex discontinuities, e.g. a $d$-dimensional input space containing discontinuities residing on $d-1$ dimensional hypersurfaces would be embedded in a $d+1$ dimensional space. Note that more complex networks of $m$ discontinuities may require embedding in a higher dimensional space, e.g. of dimension $d+m$, to avoid unwanted effects due to neighbouring discontinuities, however as we would still be operating on a $d$-dimensional surface, we may not be penalised too severely by the use of $m$ extra dimensions.
%but as we just use the extra dimensions locally we don't get burnt too much by the extra dimensions.

\subsection{Emulating with Discontinuities on Non-Linear Locations}\label{ssec_nonlinlocations}

An attractive feature of the Torn Embedding Non-Stationary Emulation (TENSE) approach is that it can be applied to a broad class of discontinuities, for example, when the discontinuities are situated on non-linear locations. 
An example of this is provided by the function $f(x,y)$, shown in figure~\ref{fig_toy3_curved_disc_a} (see appendix~\reff{app_D}{D} for the full definition). A suitable embedding surface $v(x,y)$ is shown in figure~\ref{fig_toy3_curved_disc_b}.
Note again the difference in form between $v(x,y)$ and $f(x,y)$: e.g. in the top/bottom regions $v(x,y)$ is flat while $f(x,y)$ tends downwards, and in the right/left regions $v(x,y)$ tends downwards/upwards respectively whilst $f(x,y)$ tends upwards in each case.

The TENSE emulator expectation $\ed{D}{f(x,y)}$ and standard deviation $\sqrt{\vard{D}{f(x,y)}}$ with $v(x,y)$ induced warping corrected,
are shown in figures~\ref{fig_toy3_curved_disc_c} and \ref{fig_toy3_curved_disc_d} respectively, based on a 16 point grid design given as the black points. 
Comparing figure~\ref{fig_toy3_curved_disc_c} with \ref{fig_toy3_curved_disc_a} we see that the emulator expectation captures the form of $f(x,y)$ well, and handles the curved discontinuities with ease.
We note that one could apply the Treed GP method \citep{doi:10.1198/016214508000000689} here, that divides the input space up by partitioning on individual inputs, effectively creating rectangular subregions in which independent GPs are trained. However, although this method may learn the locations of the discontinuities, it may perform poorly here, as it is very inefficient to represent curved discontinuities using rectangular regions, and many more runs may be required to train
the independent GPs, instead of the single emulator used in the TENSE approach. 
%We note that an alternative approach that could be applied here is the popular Treed GP method\citep{doi:10.1198/016214508000000689}, that divides the input space up by partitioning on individual inputs, effectively creating rectangular subregions in which independent GPs are trained. However, this may perform poorly here, as it is very inefficient to represent curved discontinuities using rectangular regions, and many more runs may be required to train the independent GPs, instead of the single emulator used in the TENSE approach.
\ifthenelse{\value{plotstyle}=2}{\vspace{-0.2cm}}{}

\ifthenelse{\value{plotstyle}=1}{
\begin{figure}[t]
\begin{center}
\begin{subfigure}{0.49\columnwidth}
\centering
\includegraphics[page=2,scale=0.35,viewport= 0 0 630 530,clip]{plots_paper/plots_examples/embedding2d_curved_discontinuites_1A.pdf} 
\vspace{-0.4cm}
\caption{\footnotesize{The true 2-dimensional function $f(x,y)$.}}
\label{fig_toy3_curved_disc_a}
\end{subfigure}
\begin{subfigure}{0.49\columnwidth}
\centering
\includegraphics[page=1,scale=0.35,viewport= 0 0 630 530,clip]{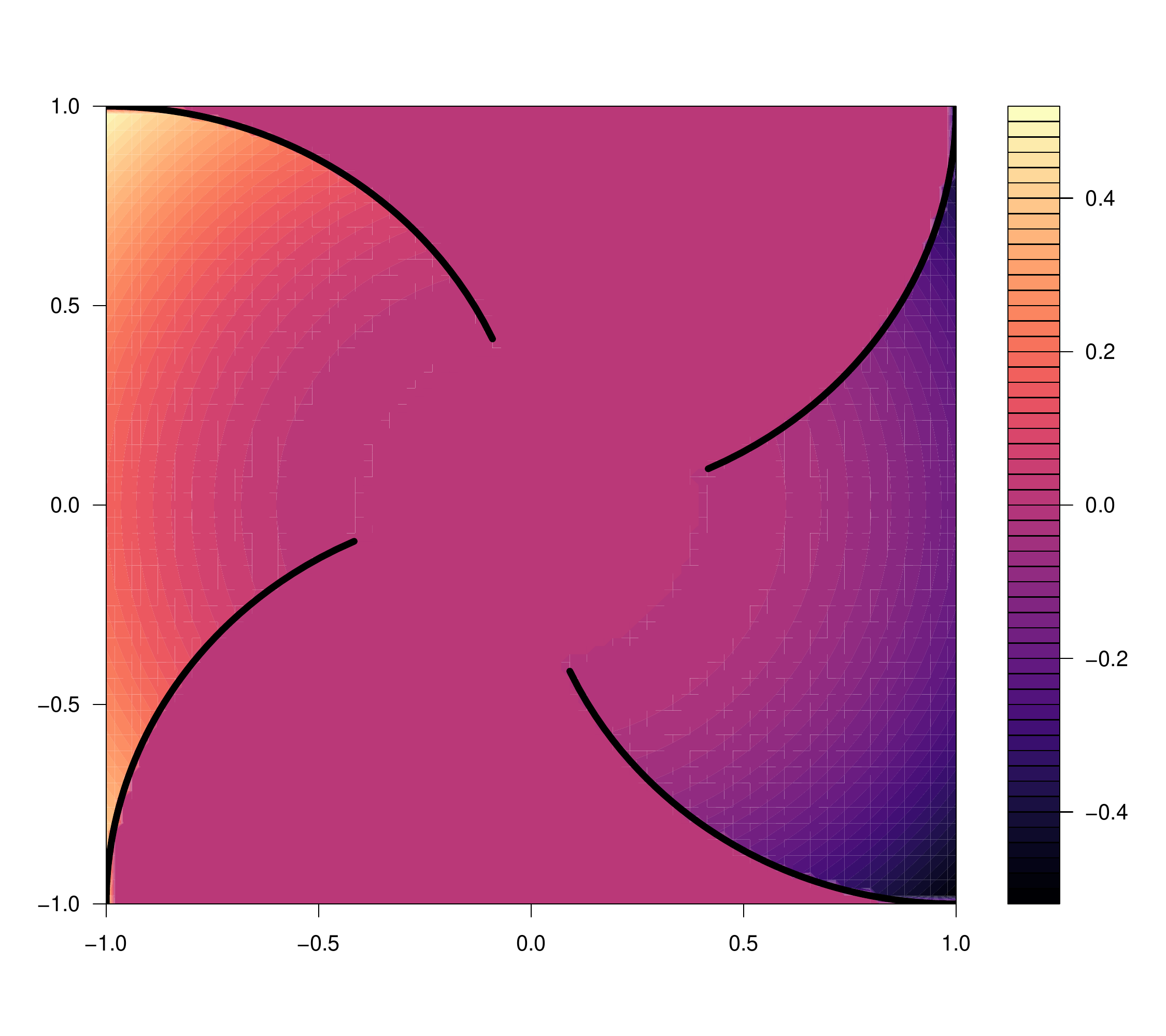} 
\vspace{-0.4cm}
\caption{\footnotesize{The embedding surface $v(x,y)$.}}
\label{fig_toy3_curved_disc_b}
\end{subfigure}
\begin{subfigure}{0.49\columnwidth}
\centering
\vspace{0.4cm}
\includegraphics[page=3,scale=0.35,viewport= 0 0 630 530,clip]{plots_paper/plots_examples/embedding2d_curved_discontinuites_1A.pdf} 
\vspace{-0.4cm}
\caption{\footnotesize{The emulator expectation $\ed{D}{f(x,y)}$.}}
\label{fig_toy3_curved_disc_c}
\end{subfigure}
\begin{subfigure}{0.49\columnwidth}
\centering
\vspace{0.4cm}
\includegraphics[page=4,scale=0.35,viewport= 0 0 630 530,clip]{plots_paper/plots_examples/embedding2d_curved_discontinuites_1A.pdf} 
\vspace{-0.4cm}
\caption{\footnotesize{The emulator stan. dev. $\sqrt{\vard{D}{f(x,y)}}$.}}
\label{fig_toy3_curved_disc_d}
\end{subfigure}
\end{center}
\vspace{-0.4cm}
\caption{\footnotesize{(a) A 2-dimensional function \cm{f(x,y)} with curved discontinuity locations given by the curved black lines. (b) the embedding surface \cm{v(x,y)}. (c) the TENSE emulator expectation \cm{\ed{D}{f(x,y)}} with \cm{v(x,y)} warping corrected using NS-CS. (d) the TENSE emulator variance \cm{\vard{D}{f(x,y)}} with \cm{v(x,y)} warping corrected using NS-CS.
}}
\label{fig_toy3_curved_disc}
\vspace{-0.cm}
\end{figure}
}{}
\ifthenelse{\value{plotstyle}=2}{
\begin{figure}[t!]
\vspace{-0.4cm}
\begin{center}
\begin{subfigure}{0.49\columnwidth}
%\centering
%\includegraphics[scale=0.335,viewport= 0 0 630 530,clip]{plots_paper/plots_examples/2D-Example-2rift-difflength_curved_in_x1x2-true-function.pdf} 
\hspace{-0.9cm} \includegraphics[page=2,scale=0.335,viewport= 0 0 630 530,clip]{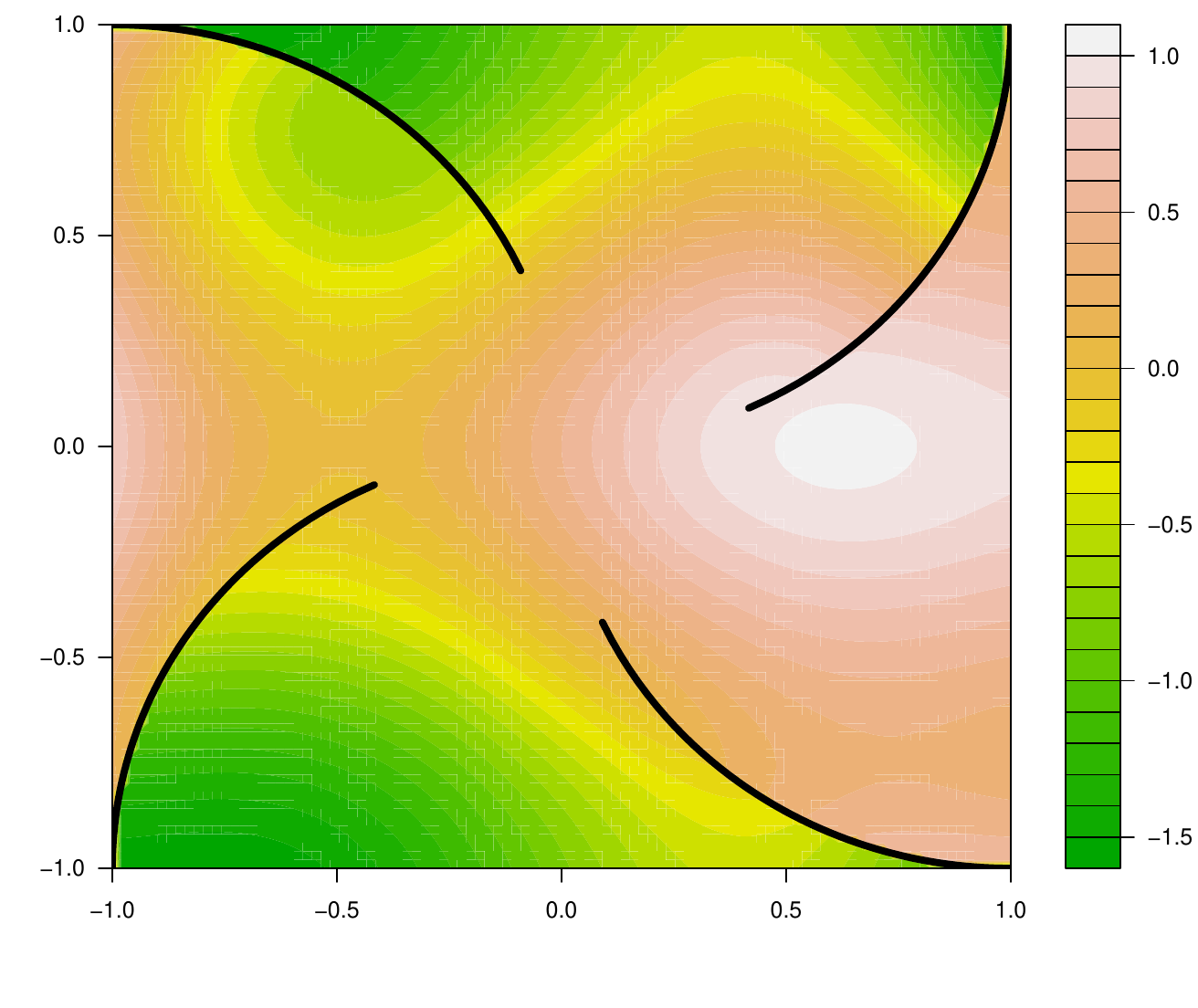} 
\vspace{-1.0cm}
\caption{\footnotesize{The true 2-dimensional function $f(x,y)$.}}
\label{fig_toy3_curved_disc_a}
\end{subfigure}
\begin{subfigure}{0.49\columnwidth}
%\centering
%\includegraphics[scale=0.335,viewport= 0 0 630 530,clip]{plots_paper/plots_examples/embedding2d_compress_fixed_2rift_diflength_curved_in_x1x2_1.pdf}
\hspace{0.0cm} \includegraphics[page=1,scale=0.335,viewport= 0 0 630 530,clip]{plots_paper/plots_examples/embedding2d_curved_discontinuites_1A.pdf} 
\vspace{-1.0cm}
\caption{\footnotesize{The embedding surface $v(x,y)$.}}
\label{fig_toy3_curved_disc_b}
\end{subfigure}
\begin{subfigure}{0.49\columnwidth}
%\centering
\vspace{0.0cm}
\hspace{-0.9cm} \includegraphics[page=3,scale=0.335,viewport= 0 0 630 530,clip]{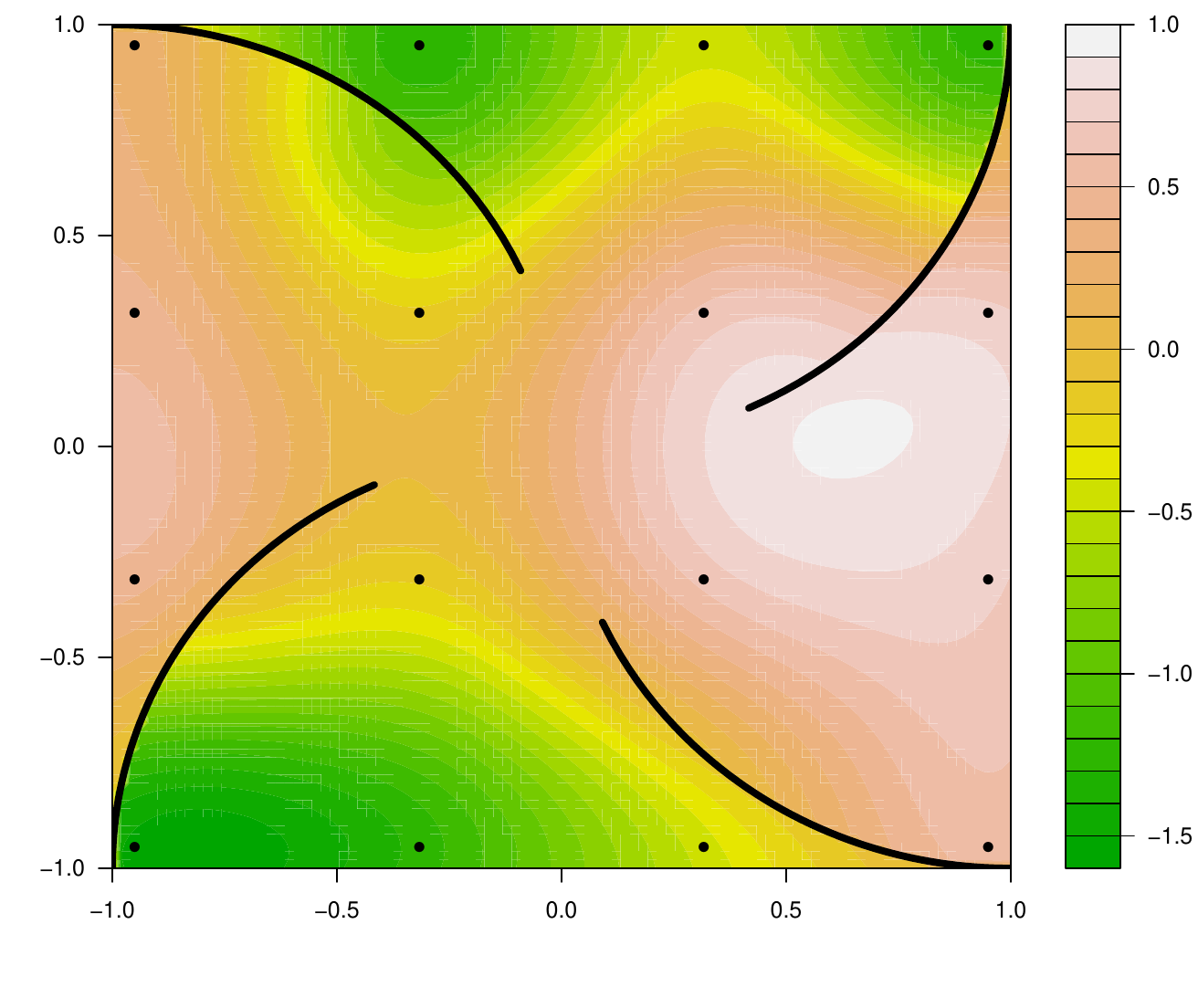} 
\vspace{-1.0cm}
\caption{\footnotesize{The emulator expectation $\ed{D}{f(x,y)}$.}}
\label{fig_toy3_curved_disc_c}
\end{subfigure}
\begin{subfigure}{0.49\columnwidth}
%\centering
\vspace{0.0cm}
\hspace{0.0cm} \includegraphics[page=4,scale=0.335,viewport= 0 0 630 530,clip]{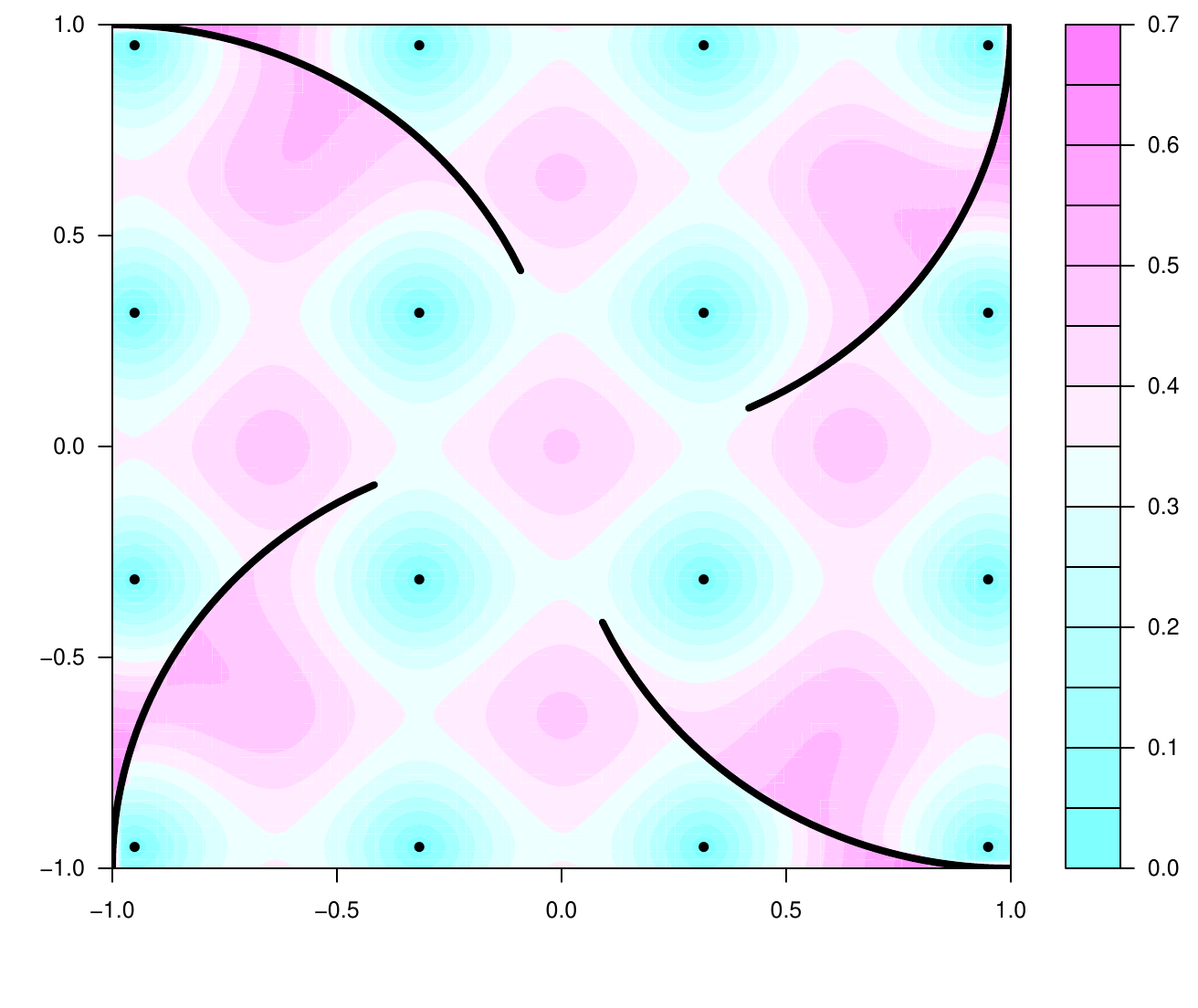} 
\vspace{-1.0cm}
\caption{\footnotesize{The emulator stan. dev. $\sqrt{\vard{D}{f(x,y)}}$.}}
\label{fig_toy3_curved_disc_d}
\end{subfigure}
\end{center}
\vspace{-0.6cm}
\caption{\footnotesize{(a) A 2-dimensional function \cm{f(x,y)} with curved discontinuity locations given by the curved black lines. (b) The embedding surface \cm{v(x,y)}. (c) The TENSE emulator expectation \cm{\ed{D}{f(x,y)}} with \cm{v(x,y)} warping corrected using NS-CS. (d) The TENSE emulator variance \cm{\vard{D}{f(x,y)}} with \cm{v(x,y)} warping corrected using NS-CS.
}}
\label{fig_toy3_curved_disc}
\vspace{-0.6cm}
\end{figure}
}{}

%%%%%%%%%%%%%%%%%%%%%%%%%%%%%%%%%%%%%%%%%%%%%%%%%%%%%
\section{Application: TNO 2 Well Placement Challenge}\label{sec_TNO}
%%%%%%%%%%%%%%%%%%%%%%%%%%%%%%%%%%%%%%%%%%%%%%%%%%%%%
\ifthenelse{\value{plotstyle}=2}{\vspace{-0.0cm}}{}
\subsection{Problem Setup: Multiple Partial Discontinuities}
\ifthenelse{\value{plotstyle}=2}{\vspace{-0.0cm}}{}
The motivation for developing the TENSE framework is in direct response to the following problem posed within the oil industry. 
The TNO OLYMPUS Field Development Optimisation Challenge was devised by the Netherlands Organisation for Applied Scientific Research (TNO) in collaboration with Delft University of Technology (TU Delft), and industrial partners Eni S.p.A, Equinor ASA and Petrobras. 
The TNO challenge is based around the fictitious oil reservoir model named OLYMPUS~\citep{2017:TNO:OLYMPUS-oil-Reservoir-Model}, and was designed to mimic realistic simulation, optimisation and decision problems faced by the oil industry.
It has attracted much attention from industry and academia with results from the active competition period presented and compared at the EAGE/TNO Workshop on OLYMPUS Field Development Optimization~\citep{2018:EAGE-TNO-Workshop-on-OLYMPUS-Field-Development-Optimization}.

The TNO Challenge I concerns well control, however the TNO Challenge II, which we exclusively focus on here, concerns well placement. 
The challenge is to choose a configuration of oil well placement to optimise the Net Present Value (NPV) over a 20 year period for the OLYMPUS reservoir model. NPV essentially represents the discounted profits over the 20 year period.
%(with a substantial discount rate of 8\%)
As the reservoir model, used to calculate the NPV, has complex features including geological uncertainty and is expensive to evaluate, and as multiple wells may be used, this represents a demanding task. Figure~\ref{fig_olympus_model_a} shows an image from above of the Olympus reservoir in physical coordinates, coloured by oil volume per unit area. 
We can choose to locate production wells or injection wells at any location over this 2D map, with each configuration yielding a certain NPV value.
Note however in figure~\ref{fig_olympus_model_a} the black lines extending into the map from the northern edge: these are geological faults in the model, with know location, that will inhibit the flow of oil and water across them. This will induce a sharp discontinuity in the NPV response as the possible well is moved either side of the fault. 
Away from such faults, we anticipate the NPV surface to be far smoother.

%% Two panel version %%%
\ifthenelse{\value{plotstyle}=1}{
\begin{figure}[t]
\begin{center}
\begin{subfigure}{0.49\columnwidth}
\centering
\vspace{0.5cm}
%\scalebox{1}[-1]{\includegraphics[width=8cm,height=6cm,angle=90]{plots_paper/plots_olympus/TNO_physical_map.png}}
\scalebox{1}[-1]{\includegraphics[width=8cm,height=6.5cm,angle=0]{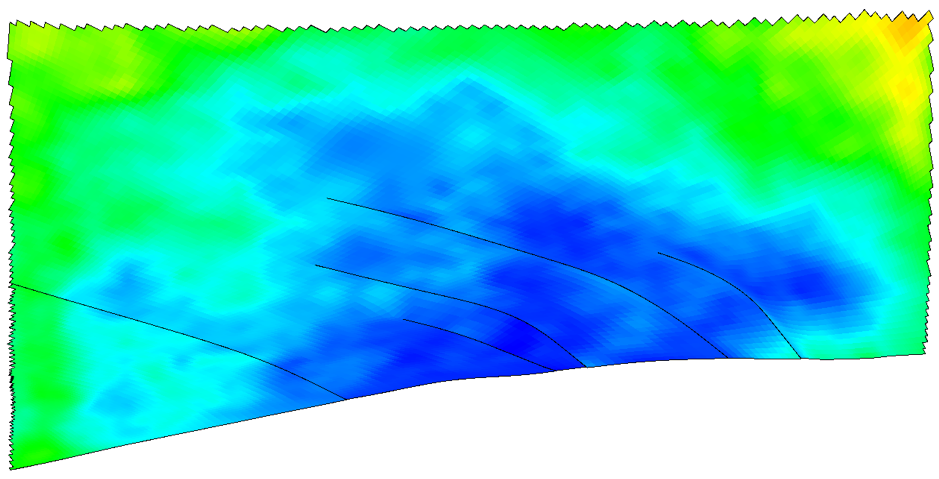}}
\vspace{1.4cm}
\caption{\footnotesize{The Olympus oil reservoir in physical coordinates.}}
\label{fig_olympus_model_a}
\end{subfigure}
\begin{subfigure}{0.49\columnwidth}
\centering
\includegraphics[page=1,scale=0.44,viewport= 5 15 620 550, clip]{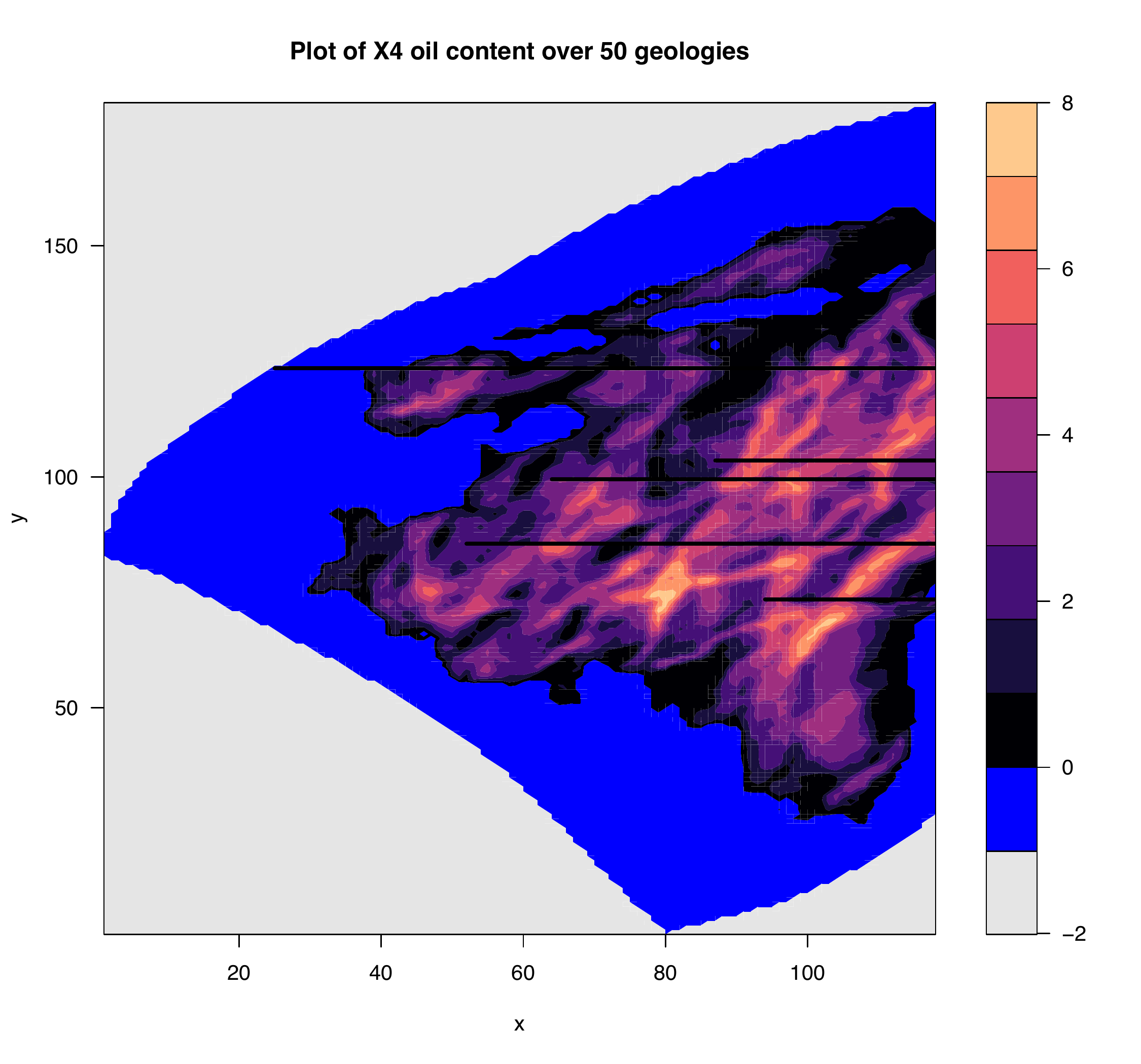}
\vspace{-0.4cm}
\caption{\footnotesize{The Olympus oil reservoir in transformed coordinates.}}
\label{fig_olympus_model_b}
\end{subfigure}
\end{center}
\vspace{-0.4cm}
\caption{\footnotesize{The TNO II Challenge Olympus oil reservoir model. (a) An image in physical coordinates, with the blue areas representing higher oil volume per unit area. The black lines show the locations of curved geological faults that will cause discontinuities in the Net Present Value (NPV) surface defined over the 2D map.
(b) In transformed grid aligned coordinates, showing the oil volume per unit area of one of the 50 geological realisations. Note that the geological faults (black horizontal lines) are now straight: this is not required for the TENSE methodology, but is useful and worth exploiting. The non-oil containing region is coloured blue.}}
\label{fig_olympus_model}
\vspace{-0.cm}
\end{figure}
}{}
\ifthenelse{\value{plotstyle}=2}{
\begin{figure}[t]
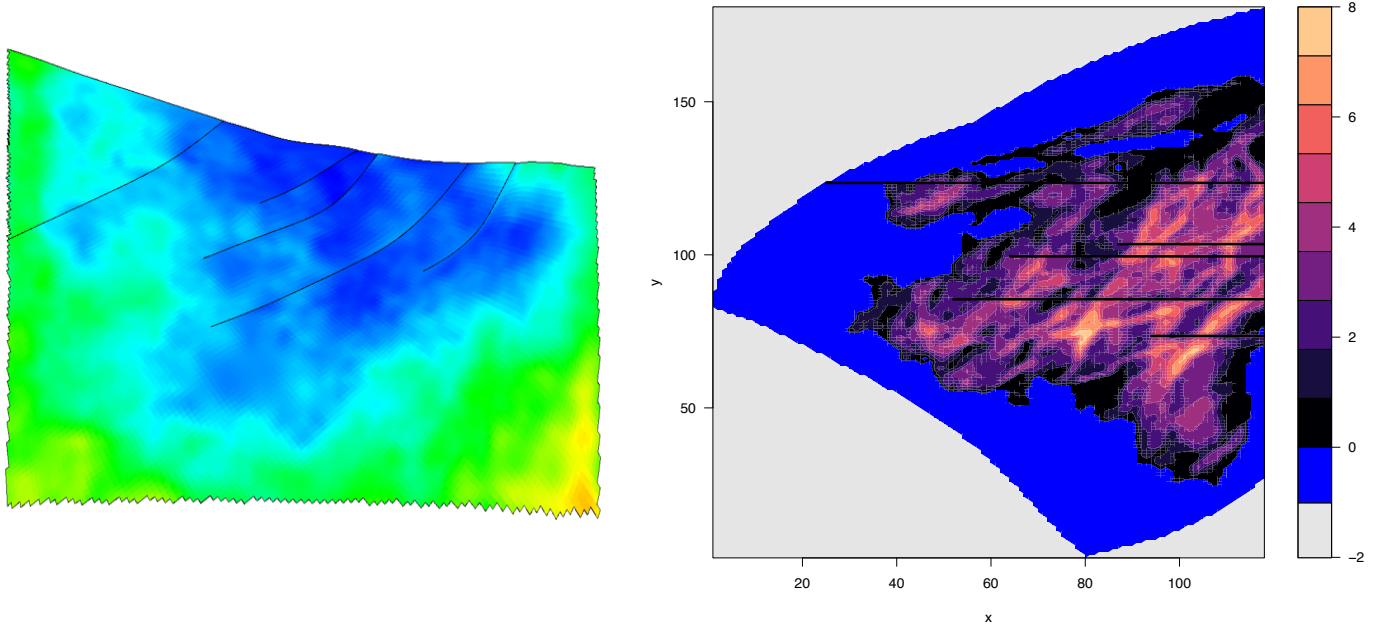

\vspace{-0.2cm}
\begin{center}
\begin{subfigure}{0.49\columnwidth}
%\centering
\vspace{-0.0cm}
%\scalebox{1}[-1]{\includegraphics[width=8cm,height=6cm,angle=90]{plots_paper/plots_olympus/TNO_physical_map.png}}
\hspace{-0.5cm}\scalebox{1}[-1]{\includegraphics[width=6.8cm,height=5.7cm,angle=0]{plots_paper/plots_olympus/TNO_physical_map.png}}
\vspace{0.45cm}
\caption{\footnotesize{Olympus reservoir in physical coordinates.~~~~~}}
\label{fig_olympus_model_a}
\end{subfigure}
\begin{subfigure}{0.49\columnwidth}
%\centering
\hspace{-0.2cm}\includegraphics[page=12,scale=0.34,viewport= 5 15 620 550, clip]{plots_paper/plots_olympus/TNO_map_mean_sd_real_of_5_geologies_with_faults.pdf}
\vspace{-0.6cm}
\caption{\footnotesize{Olympus reservoir in transformed coordinates.}}
\label{fig_olympus_model_b}
\end{subfigure}
\end{center}
\vspace{-0.6cm}
\caption{\footnotesize{The TNO II Challenge Olympus oil reservoir model. (a) An image in physical coordinates, with the blue areas representing higher oil volume per unit area. The black lines show the locations of curved geological faults that will cause discontinuities in the Net Present Value (NPV) surface defined over the 2D map.
(b) In transformed grid aligned coordinates, showing the oil volume per unit area of one of the 50 geological realisations. Note that the geological faults (black horizontal lines) are now straight: this is not required for the TENSE methodology, but is useful and worth exploiting. The non-oil containing region is coloured blue.}}
\label{fig_olympus_model}
\vspace{-0.6cm}
\end{figure}
}{}

In the Olympus model the location of the faults is fixed and known, however many other geological aspects (e.g. the permeability/porosity fields) are treated as uncertain and represented via 50 geological realisations provided by the TNO consortium, derived from an underlying geology model which was not made freely available. An example of one of the geological realisations is given in figure~\ref{fig_olympus_model_b}, coloured by the oil volume per unit area, and more realisations are given in figure~\reff{fig_Olymp_info}{9}, appendix~\reff{app_F}{F}, along with additional plots of the mean and SD of the oil volume per unit area of the 50 realisations. 
In these plots the physical 2D coordinates have been transformed into grid aligned 2D coordinates, which has the added effect of transforming the faults so that they lie along constant horizontal (black) lines. As demonstrated in section~\ref{ssec_nonlinlocations}, the TENSE approach does not require linear discontinuity locations, but this transformation, available due to the way the OLYMPUS model was constructed, simplifies subsequent specifications e.g. for the embedding surface, hence it would be remiss of us not to exploit it here.

The precise remit of the TNO Olympus Challenge II is to choose well locations to optimise the mean NPV over the 50 geological realisations (each geological realisation will generate its own NPV). For example, for a single vertical producer well located at position $\vx = (x,y)$, we could evaluate the NPV for any $\vx$ and for any of the $i = 1,\dots,50$ geological realisations, giving output $NPV^{(i)}(\vx)$. We hence define our primary computer model of interest $f(\vx)$ to be the mean over 50 realisations for a single producer well located at $\vx$, in accordance with the challenge:
\ifthenelse{\value{plotstyle}=2}{\vspace{-0.0cm}}{}
\be\label{eq_NPVmean1}
f(\vx) \;\; \equiv \;\; \overline{NPV}(\vx) \;\;=\;\; \frac{1}{50}\sum_{i=1}^{50} NPV^{(i)}(\vx)
\ee
(see appendix~\reff{app_E}{E} for details).
%where the definition naturally generalises to $f(X)$ for configurations $X$ of $m$ wells at locations $X= \{\vx^{k_1}_1,\dots,\vx^{k_m}_m\}$, where $k_i$ labels the $i$th well as a ``producer" or ``injector". 
Obviously there are several uncertainties and features that one might want to include in a more detailed analysis, that are missing from the original TNO Challenge. These include the effects of the finite sample size of geological realisations, uncertainties due to oil price and water cost, model discrepancy due to the imperfection of the reservoir (and geology) model itself, the benefits of sequential decision making, and indeed whether the NPV should even be identified with the utility of the 
decision makers. See \cite{eage:/content/papers/10.3997/2214-4609.202035109} for discussion of several of these issues, and also \cite{House09_ExchMod2} for a relevant treatment of exchangeable computer models.
However, here we are primarily interested in the following emulation problem. 

Concern has been expressed in the oil industry over the transparency of various black-box optimisers that can produce counterintuitive well configurations of unfamiliar form (and of unknown optimality), that made some engineers nervous.
%of unknown optimality and robustness. For example, many of the submitted entries to the challenge reported different values for the optimal NPV, hence all but (at most) one had not actually found the optimum. In addition they often suggested counterintuitive well configurations of unfamiliar form, that made some engineers nervous.
We were hence approached and asked if we could help visualise the NPV surface, to aid interpretation and insight in various situations that may occur within a more human informed optimisation process. 
Specifically a key request was to visualise the mean NPV surface for a single producer well, as represented by $f(\vx)$, over the full reservoir map $\mathcal{X}$ in the presence of multiple discontinuities, using only a limited set of evaluations of the expensive OLYMPUS model. This was the original motivation for developing TENSE. 
\ifthenelse{\value{plotstyle}=2}{\vspace{-0.3cm}}{}

\subsection{Constructing the Embedding Surface $v(x,y)$}\label{ssec_TNO_emulation}
\ifthenelse{\value{plotstyle}=2}{\vspace{-0.1cm}}{}

We proceed to apply the TENSE framework to the function $f(\vx)$ representing the mean NPV of a single producer well as follows.
We specify an embedding surface $v(\vx)=v(x,y)$ by tearing along the five discontinuities shown in figure~\ref{fig_olympus_model_b} and bending alternate regions higher and lower into the 3D space using quadratic forms, exploiting a similar strategy to that employed in section~\ref{ssec_torn_embed}.
The embedding surface is shown in figure~\ref{fig_embed_olympus}, with the full definition given in appendix~\reff{app_G}{G}. 
\ifthenelse{\value{plotstyle}=2}{\vspace{-0.1cm}}{}

\ifthenelse{\value{plotstyle}=1}{
\begin{figure}[t]
\begin{center}
\includegraphics[scale=0.25]{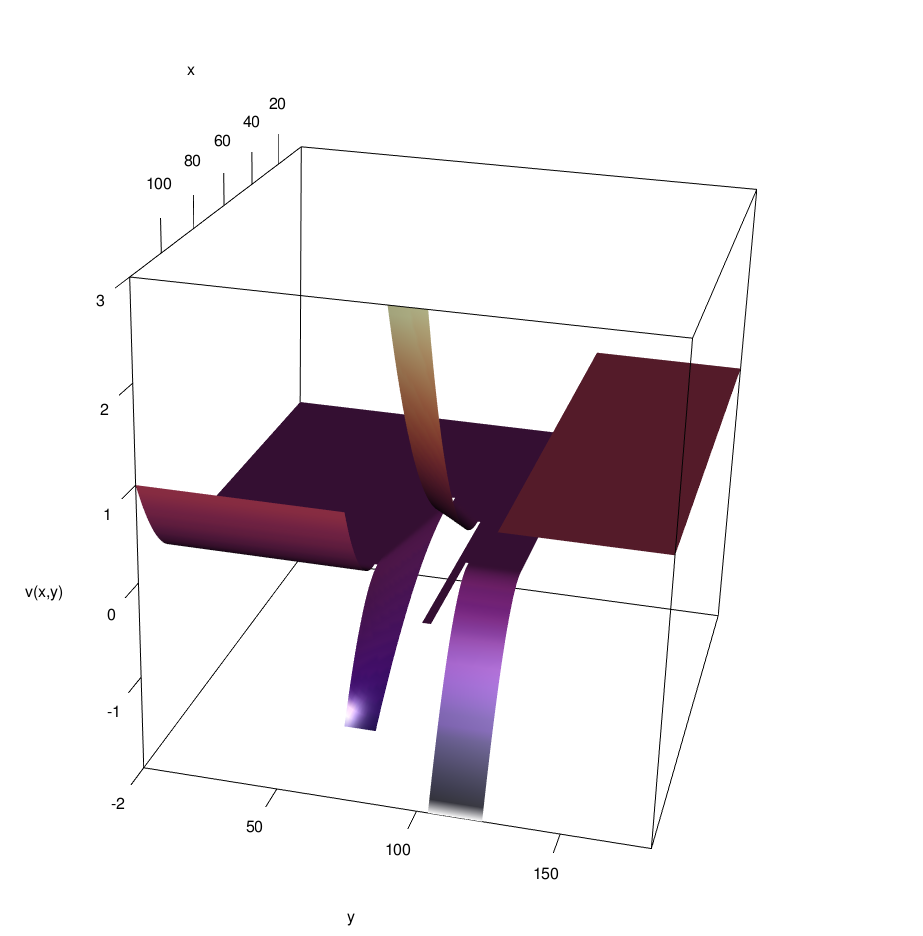} 
\hspace{0.5cm} \includegraphics[scale=0.25]{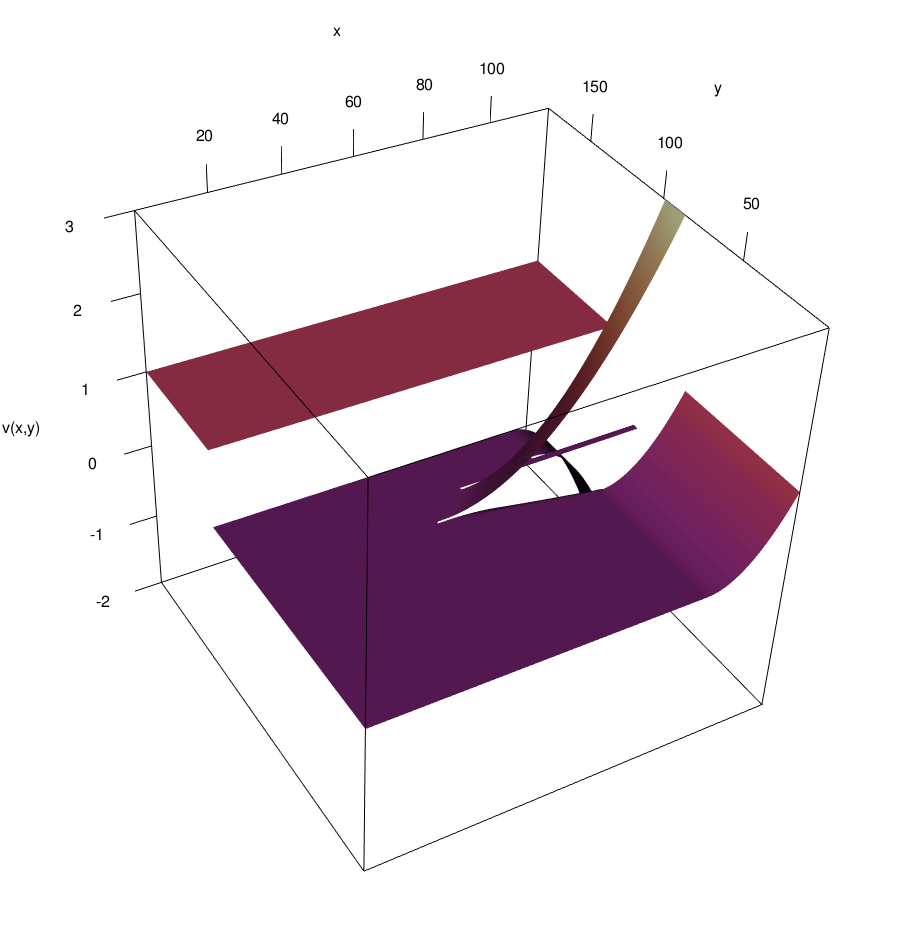}
\end{center}
\caption{\footnotesize{The embedding surface $v(x,y)$ used for the TNO Challenge II Olympus model example. The surface is cut along each of the locations of the geological faults (shown as black lines in figure~\ref{fig_olympus_model}) and various regions of the surface are bent up and down to induce the discontinuities. A 2D (zoomed) view of this embedding surface can also be seen in figure~\ref{fig_2dzoomed_ind_cor}.}}\label{fig_embed_olympus}
\end{figure}
}{}
\ifthenelse{\value{plotstyle}=2}{
\begin{figure}[t]
\vspace{-0.4cm}
\begin{center}
\begin{tabular}{cc}
\hspace{-0.8cm} \includegraphics[scale=0.218]{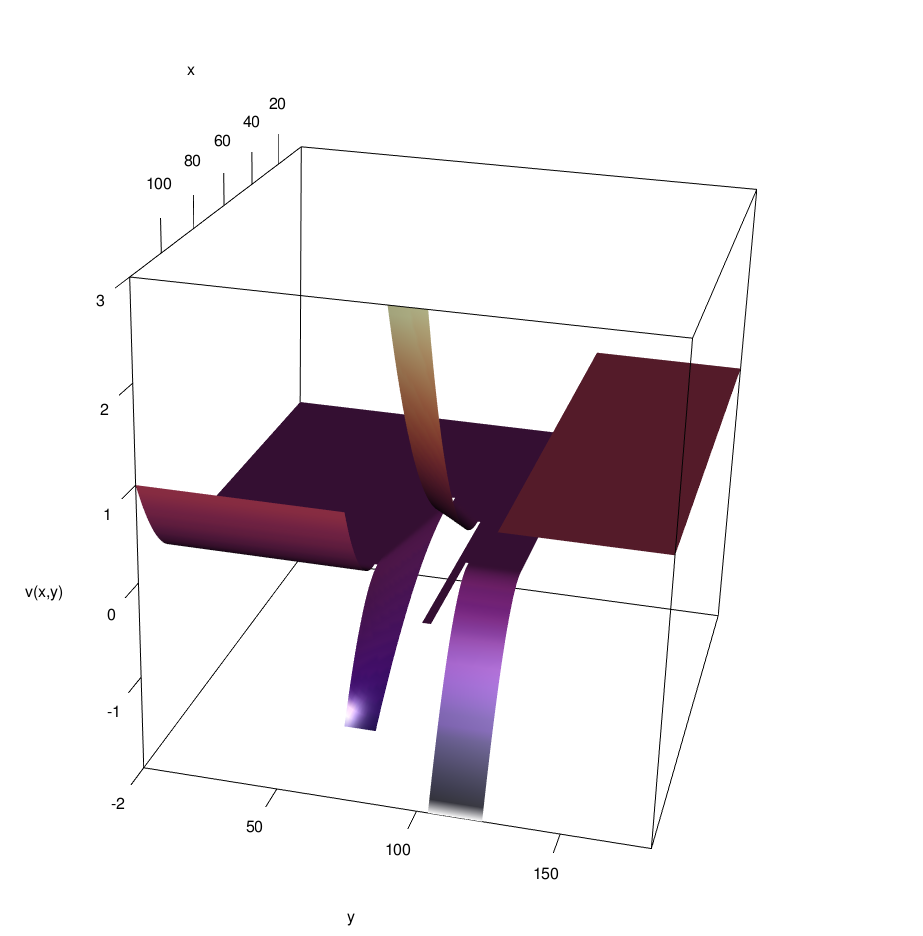} &
\hspace{-0.6cm} \includegraphics[scale=0.218]{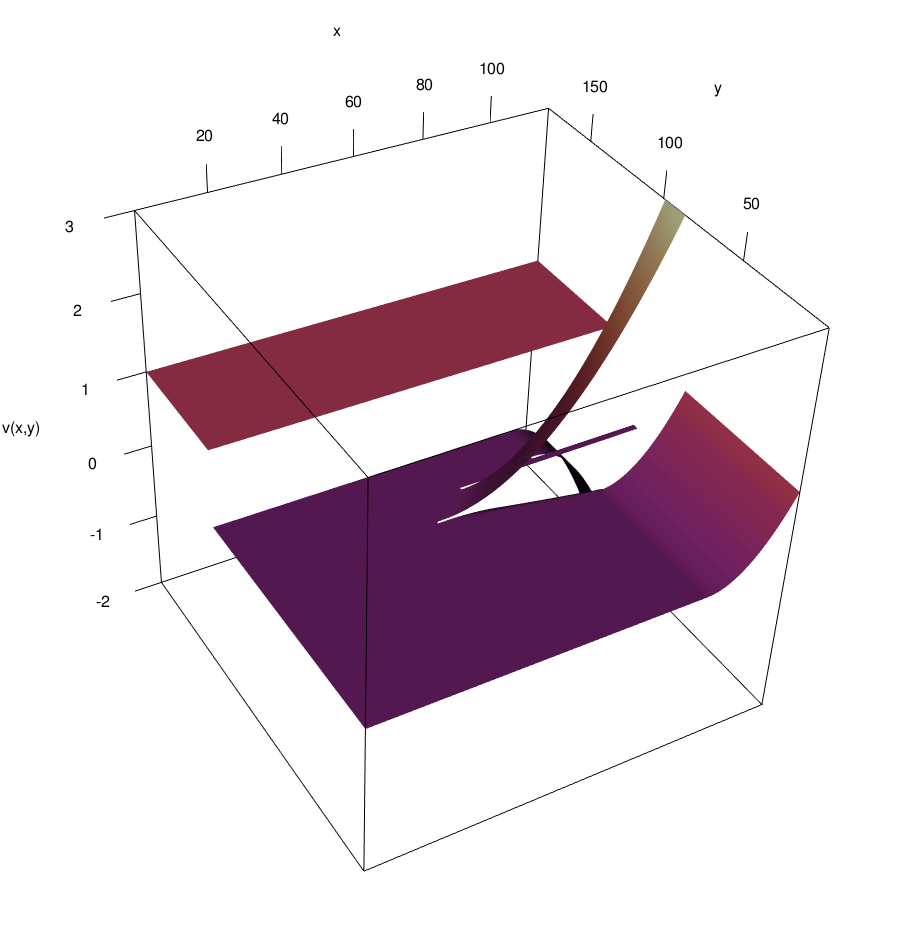}
\end{tabular}
\end{center}
\vspace{-0.8cm}
\caption{\footnotesize{The embedding surface $v(x,y)$ used for the TNO Challenge II Olympus model example. The surface is cut along each of the locations of the geological faults (shown as black lines in figure~\ref{fig_olympus_model}) and various regions of the surface are bent up and down to induce the discontinuities. A 2D (zoomed) view of this embedding surface can also be seen in figure~\ref{fig_2dzoomed_ind_cor}.}}\label{fig_embed_olympus}
\vspace{-0.6cm}
\end{figure}
}{}

To check that this choice of embedding will produce the desired behaviour of allowing the emulator to exhibit discontinuous jumps over the discontinuities with minimal warping, we examine the induced covariance structure of vertical lines of points that cross all the discontinuities, as shown in figure~\ref{fig_2dzoomed_ind_cor}. For example, figure~\ref{fig_2dzoomed_ind_cor_a} shows a zoomed in section of the embedding surface $v(x,y)$ with the discontinuities as horizontal black lines (as in figure~\ref{fig_olympus_model_b}), but also highlights a green vertical line of points at $x=42$, while figure~\ref{fig_2dzoomed_ind_cor_b} shows the induced 2D emulator correlation matrix corresponding to this green line of points. The correlation matrix is formed from ${\rm Cov}[f(\vv(\vx)),f(\vv(\vx'))]$ using equations~(\ref{eq_NS_squared_cov_struc_3D_full}), (\ref{eq:Quadratic-Form-for-non-stationary-cov-func_embedded}) and (\ref{eq_sigma3D_full_exp1}).
\ifthenelse{\value{plotstyle}=2}{\vspace{-0.1cm}}{}

We see that the two regions $y>123.5$ and $y<123.5$ either side of the highest fault are uncorrelated as desired, and that the correlation structure resorts to the usual squared exponential form within each region. Figures~\ref{fig_2dzoomed_ind_cor_c} and \ref{fig_2dzoomed_ind_cor_d} are defined similarly, but for the line $x=78$. Now we see that the regions either side of the fault at 
$y=85.5$ are almost entirely uncorrelated, while either side of the fault at $y=99.5$ the regions have suppressed correlation, as the start of the fault is relatively close to the green line. In figures~\ref{fig_2dzoomed_ind_cor_e} and \ref{fig_2dzoomed_ind_cor_f} the more extreme case of $x=116$ is examined, where we see six uncorrelated regions separated by the five faults, precisely as desired. Due to the TENSE approach of embedding in a higher dimension, all these correlation matrices are guaranteed to be valid. 
Note that we choose to directly specify the form of the embedding surface $v(x,y)$ here, as it is feasible to do this in a controlled way as to ensure each region either side of a discontinuity is well separated in the third dimension. One could of course treat $v(x,y)$ as uncertain, possibly of parameterised form but still torn along the locations of the discontinuities, and then use the run data to learn about $v(x,y)$. However, this may lead to several identifiability issues, and there may not be a strong signal as to the particular form for $v(x,y)$, so we leave such considerations to future work.

\ifthenelse{\value{plotstyle}=1}{
\begin{figure}[t!]
\begin{center}
\vspace{-0.4cm}
\begin{subfigure}{0.49\columnwidth}
\centering
\includegraphics[width=0.92\linewidth, page=1, viewport= 0 0 630 530, clip]{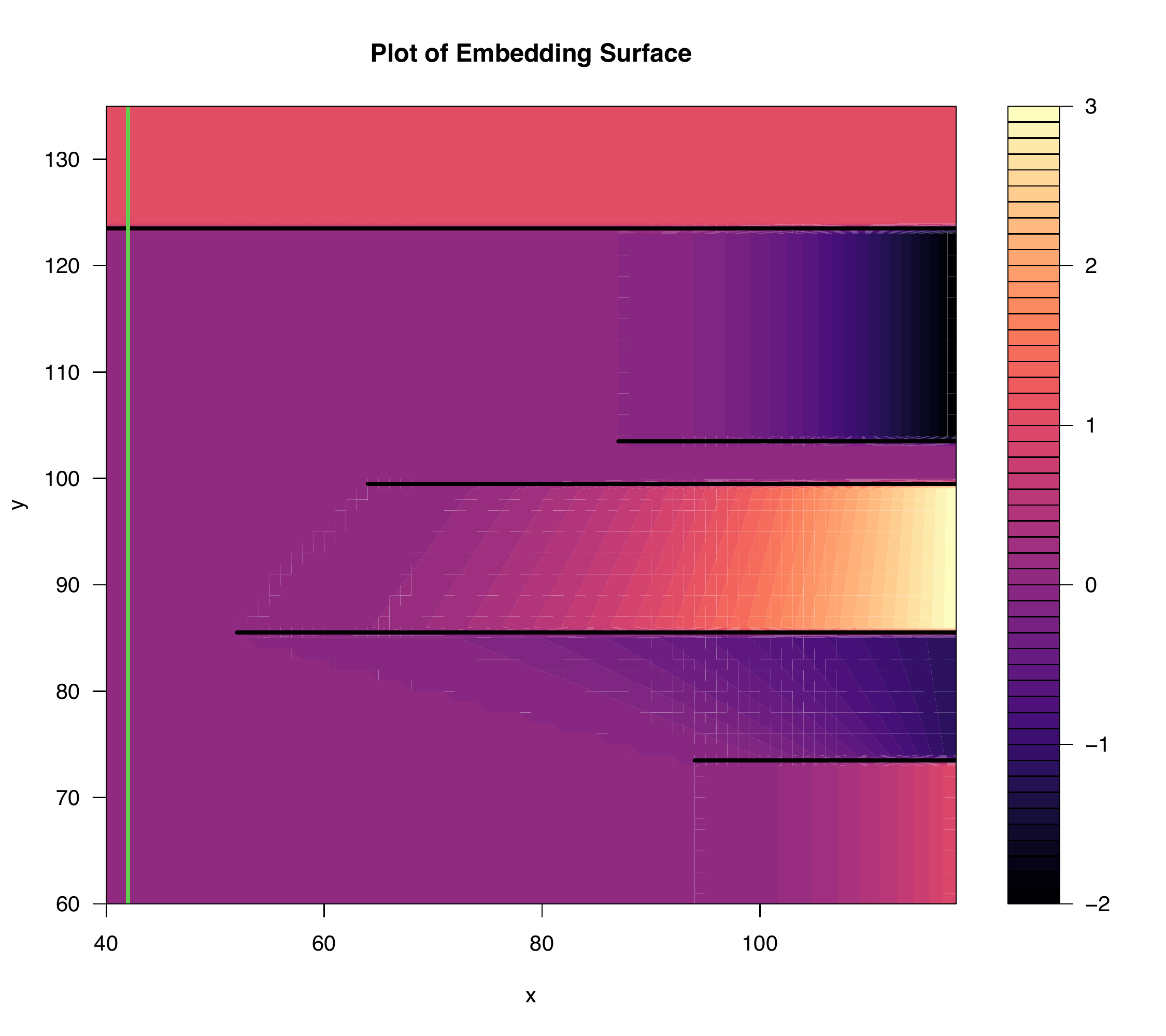}
\vspace{-0.4cm}
\caption{\footnotesize{$v(x,y)$ with vertical line at $x=42$.}}
\label{fig_2dzoomed_ind_cor_a}
\end{subfigure}
\begin{subfigure}{0.49\columnwidth}
\centering
\includegraphics[width=0.92\linewidth, page=1, viewport= 0 0 630 530, clip]{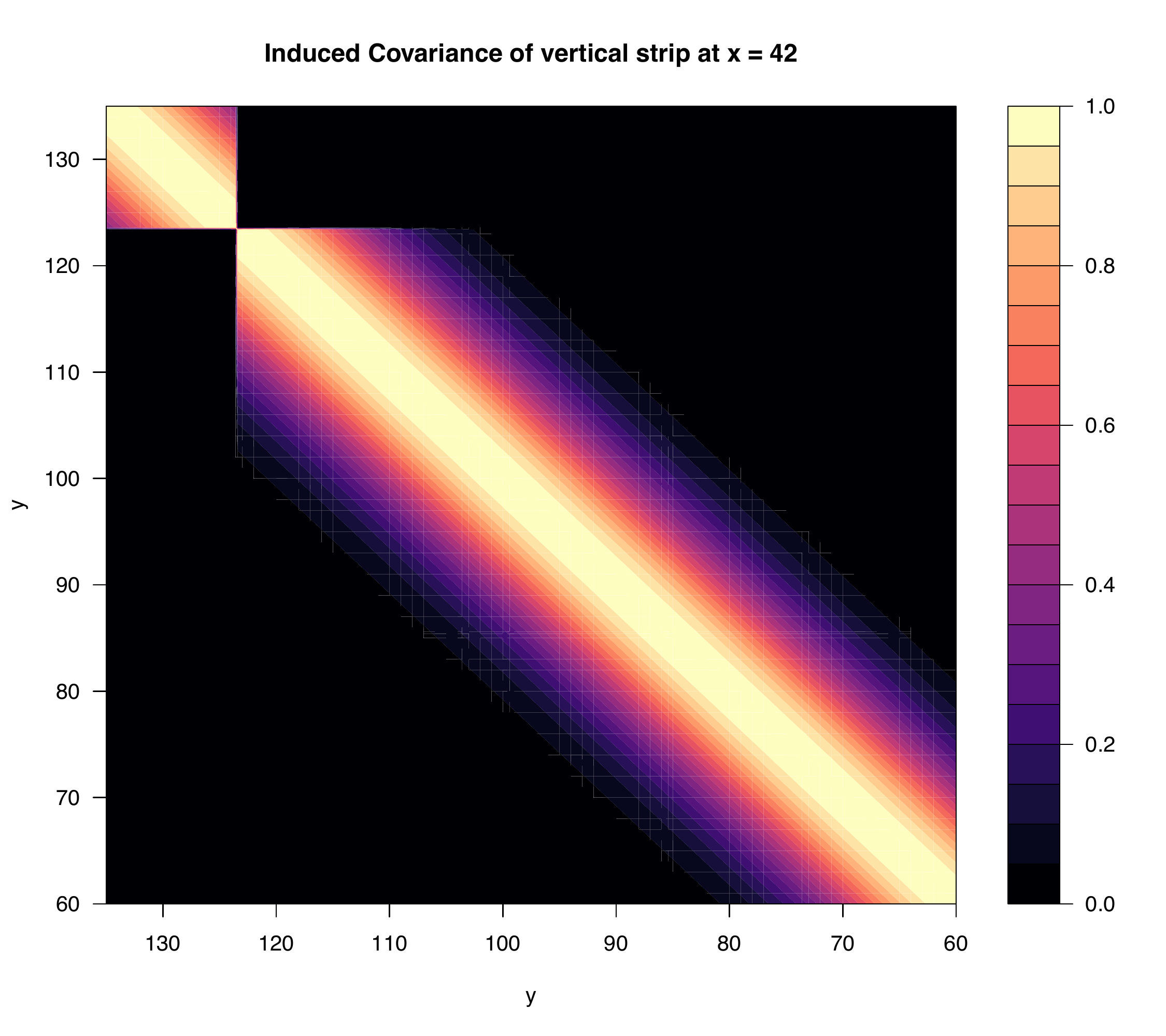} 
\vspace{-0.4cm}
\caption{\footnotesize{Induced emulator correlation matrix at $x=42$.}}
\label{fig_2dzoomed_ind_cor_b}
\end{subfigure}
\begin{subfigure}{0.49\columnwidth}
\centering
\vspace{0.1cm}
\includegraphics[width=0.92\linewidth, page=2, viewport= 0 0 630 530, clip]{plots_paper/plots_olympus/covmat_vertical_strip_locations_page2_20_29.pdf}
\vspace{-0.4cm}
\caption{\footnotesize{$v(x,y)$ with vertical line at $x=78$.}}
\label{fig_2dzoomed_ind_cor_c}
\end{subfigure}
\begin{subfigure}{0.49\columnwidth}
\centering
\vspace{0.1cm}
\includegraphics[width=0.92\linewidth, page=2, viewport= 0 0 630 530, clip]{plots_paper/plots_olympus/covmat_var_B_vertical_strip_zoomedmap_reversed_page2_20_29.pdf} 
\vspace{-0.4cm}
\caption{\footnotesize{Induced emulator correlation matrix at $x=78$.}}
\label{fig_2dzoomed_ind_cor_d}
\end{subfigure}
\begin{subfigure}{0.49\columnwidth}
\centering
\vspace{0.1cm}
\includegraphics[width=0.92\linewidth, page=3, viewport= 0 0 630 530, clip]{plots_paper/plots_olympus/covmat_vertical_strip_locations_page2_20_29.pdf}
\vspace{-0.4cm}
\caption{\footnotesize{$v(x,y)$ with vertical line at $x=116$.}}
\label{fig_2dzoomed_ind_cor_e}
\end{subfigure}
\begin{subfigure}{0.49\columnwidth}
\centering
\vspace{0.1cm}
\includegraphics[width=0.92\linewidth, page=3, viewport= 0 0 630 530, clip]{plots_paper/plots_olympus/covmat_var_B_vertical_strip_zoomedmap_reversed_page2_20_29.pdf}
\vspace{-0.4cm}
\caption{\footnotesize{Induced emulator correlation matrix at $x=116$.}}
\label{fig_2dzoomed_ind_cor_f}
\end{subfigure}
\end{center}
\vspace{-0.4cm}
\caption{\footnotesize{Left panels (a), (c) and (e): the torn surface $v(x,y)$ embedded in 3D used to induce the discontinuities along the five geological faults, shown as the horizontal black lines, in the Olympus model. Right panels (b), (d) and (f): the induced emulator correlation matrix of the set of points along the green vertical line highlighted in the corresponding left panel, at locations $x=42, 78$ and $116$ respectively. }}
\label{fig_2dzoomed_ind_cor}
\vspace{-0.cm}
\end{figure}
}{}
\ifthenelse{\value{plotstyle}=2}{
\begin{figure}[t!]
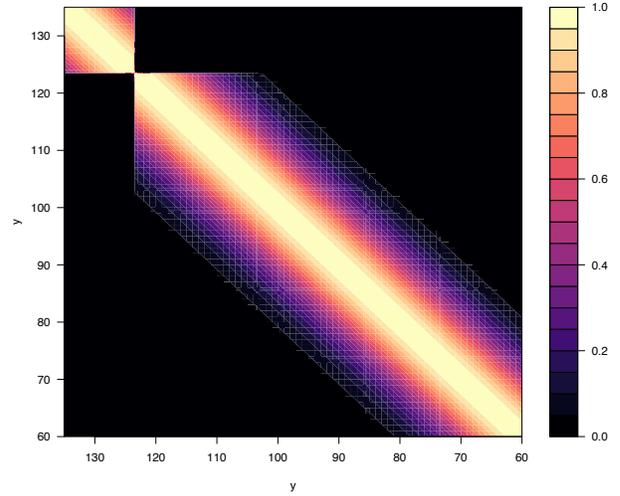
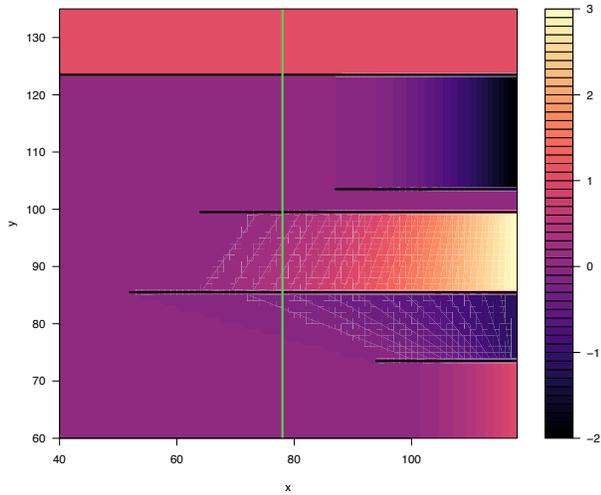
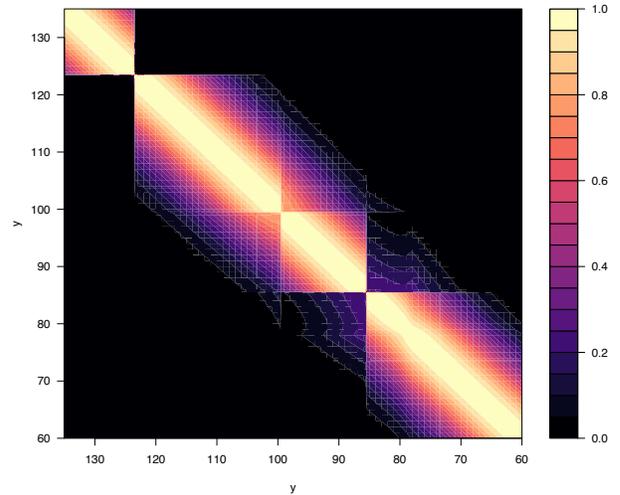
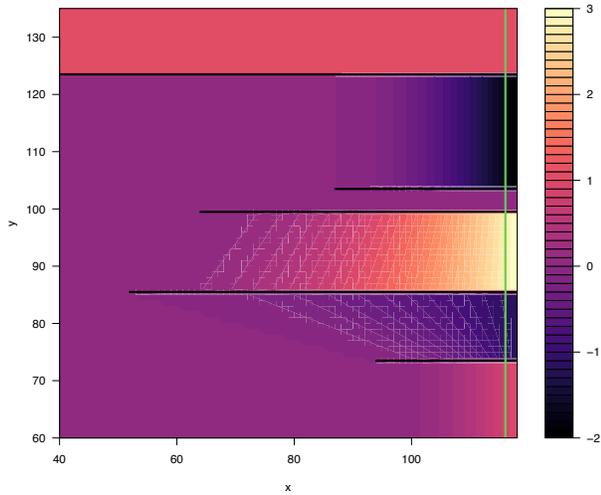
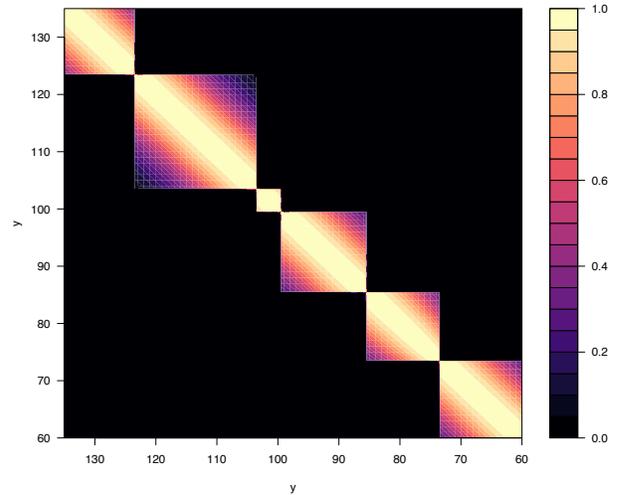

\begin{center}
\vspace{-0.4cm}
\begin{subfigure}{0.49\columnwidth}
%\centering
\hspace{-0.7cm}\includegraphics[scale=0.335, page=2, viewport= 0 0 630 530, clip]{plots_paper/plots_olympus/covmat_vertical_strip_locations.pdf}
\vspace{-0.8cm}
\caption{\footnotesize{$v(x,y)$ with vertical line at $x=42$.~~~}}
\label{fig_2dzoomed_ind_cor_a}
\end{subfigure}
\begin{subfigure}{0.49\columnwidth}
%\centering
\hspace{0.0cm} \includegraphics[scale=0.335, page=2, viewport= 0 0 630 530, clip]{plots_paper/plots_olympus/covmat_var_B_vertical_strip_zoomedmap_reversed.pdf} 
\vspace{-0.8cm}
\caption{\footnotesize{Induced emulator correlation matrix at $x=42$.}}
\label{fig_2dzoomed_ind_cor_b}
\end{subfigure}
\begin{subfigure}{0.49\columnwidth}
%\centering
\vspace{-0.0cm}
\hspace{-0.7cm}\includegraphics[scale=0.335, page=20, viewport= 0 0 630 530, clip]{plots_paper/plots_olympus/covmat_vertical_strip_locations.pdf}
\vspace{-0.8cm}
\caption{\footnotesize{$v(x,y)$ with vertical line at $x=78$.~~~}}
\label{fig_2dzoomed_ind_cor_c}
\end{subfigure}
\begin{subfigure}{0.49\columnwidth}
%\centering
\vspace{-0.0cm}
\includegraphics[scale=0.335, page=20, viewport= 0 0 630 530, clip]{plots_paper/plots_olympus/covmat_var_B_vertical_strip_zoomedmap_reversed.pdf} 
\vspace{-0.8cm}
\caption{\footnotesize{Induced emulator correlation matrix at $x=78$.}}
\label{fig_2dzoomed_ind_cor_d}
\end{subfigure}
\begin{subfigure}{0.49\columnwidth}
%\centering
\vspace{-0.0cm}
\hspace{-0.7cm}\includegraphics[scale=0.335, page=39, viewport= 0 0 630 530, clip]{plots_paper/plots_olympus/covmat_vertical_strip_locations.pdf}
\vspace{-0.8cm}
\caption{\footnotesize{$v(x,y)$ with vertical line at $x=116$.~~~}}
\label{fig_2dzoomed_ind_cor_e}
\end{subfigure}
\begin{subfigure}{0.49\columnwidth}
%\centering
\vspace{-0.0cm}
\includegraphics[scale=0.335, page=39, viewport= 0 0 630 530, clip]{plots_paper/plots_olympus/covmat_var_B_vertical_strip_zoomedmap_reversed.pdf}
\vspace{-0.8cm}
\caption{\footnotesize{Induced emulator correlation matrix at~$x=116$}}
\label{fig_2dzoomed_ind_cor_f}
\end{subfigure}
\end{center}
\vspace{-0.4cm}
\caption{\footnotesize{Left panels (a), (c) and (e): the torn surface $v(x,y)$ embedded in 3D used to induce the discontinuities along the five geological faults, shown as the horizontal black lines, in the Olympus model. Right panels (b), (d) and (f): the induced emulator correlation matrix of the set of points along the green vertical line highlighted in the corresponding left panel, at locations $x=42, 78$ and $116$ respectively. }}
\label{fig_2dzoomed_ind_cor}
\vspace{-0.cm}
\end{figure}
}{}

%\ifthenelse{\value{plotstyle}=1}{\clearpage}{}
%\ifthenelse{\value{plotstyle}=2}{\vspace{-0.4cm}}{}

\ifthenelse{\value{plotstyle}=2}{\vspace{-0.3cm}}{}

\subsection{Emulating the Net Present Value Surface}\label{ssec_emul_NPV_surface}
\ifthenelse{\value{plotstyle}=2}{\vspace{-0.1cm}}{}

Having defined the embedding surface $v(x,y)$, we are now able to construct an emulator for the NPV output as represented by $f(\vx)$, corresponding to a single producer well at location  $\vx \in \mathcal{X} \subset \mathbb{R}^2 $, in the presence of the discontinuities caused by the geological faults. 
However, there is additional prior information about the Olympus model that we can include. We know, without performing any model evaluations, that if a well is placed outside of the oil containing region of the reservoir, there will be no oil production and the NPV will be zero (or a small negative value).
For linear boundaries, one can in fact incorporate known model behaviour on the boundary, within the emulator analytically in any dimension (see for example~\cite{Vernon:2019aa} and \cite{Vernon:2022aa}). However, here the boundary around the edge of the oil containing region is complex, and so we simply add a set of 36 ``ghost points" just outside the oil containing region, with the NPV value of each set to $f(\vx)=0$. The effect of this prior information is shown in figure~\ref{fig_Olymp_em1_a} which gives the prior emulator expectation $\e{f(\vx)}$ over $\mathcal{X}$, and shows the ghost points as red points located within the grey non-oil region.

The initial space filling set of (wave 1) runs was designed respecting the following considerations. The Olympus model is computationally intensive and our collaborator was uncertain as to how much (cloud) computational resources would be available, implying early termination of the design was possible. We therefore constructed the design one point at a time, with each 
\ifthenelse{\value{plotstyle}=1}{\clearpage}{}
\ifthenelse{\value{plotstyle}=1}{\noindent}{}
point chosen to minimise the mean emulator variance over $\mathcal{X}$, given the previous design points. As this calculation uses the emulator's correlation structure, it respects the discontinuities and specifically the low correlation between certain regions as highlighted in figure~\ref{fig_2dzoomed_ind_cor}. In addition, due to the sequential nature of the design construction, even early termination would result in a well spaced and informative set of runs. 
Some pragmatic choices were used in the design calculation e.g. within an isotropic $\Sigma_{2D}$ we specified a fixed 2D correlation length of $\theta =12$, a judgement informed by the local correlation seen in the oil volume per unit area of the geological realisations (figure~\ref{fig_olympus_model_b}), and employed a nearest neighbour approximation in the emulator variance calculation, to greatly improve efficiency. Finally, we added three pairs of points to the design, either side of three of the major faults to give more direct information regarding the discontinuities in those regions.
The resulting 47 point wave 1 
design is shown in figure~\ref{fig_Olymp_em1_b} as the green points. It displays good space filling properties, 
\ifthenelse{\value{plotstyle}=2}{\clearpage}{}
\ifthenelse{\value{plotstyle}=2}{\noindent}{} while adequately exploring each of the uncorrelated regions in between the faults. At each of the 47 points $\vx^{(i)}$, all 50 of the geological realisations were evaluated giving $NPV^{(j)}(\vx^{(i)})$, $j=1,\dots,50$, and the mean calculated, giving $f(\vx^{(i)})$ and hence the first batch of runs, denoted $D_1 = \{f(\vx^{(1)}),\dots,f(\vx^{(47)} ) \}$, for use in the emulator equations. 

\ifthenelse{\value{plotstyle}=1}{
\begin{figure}[t!]
\begin{center}
\begin{subfigure}{0.41\columnwidth}
\centering
\hspace{-1.2cm}\includegraphics[page=1,scale=0.45,viewport= 25 30 535 550, clip]{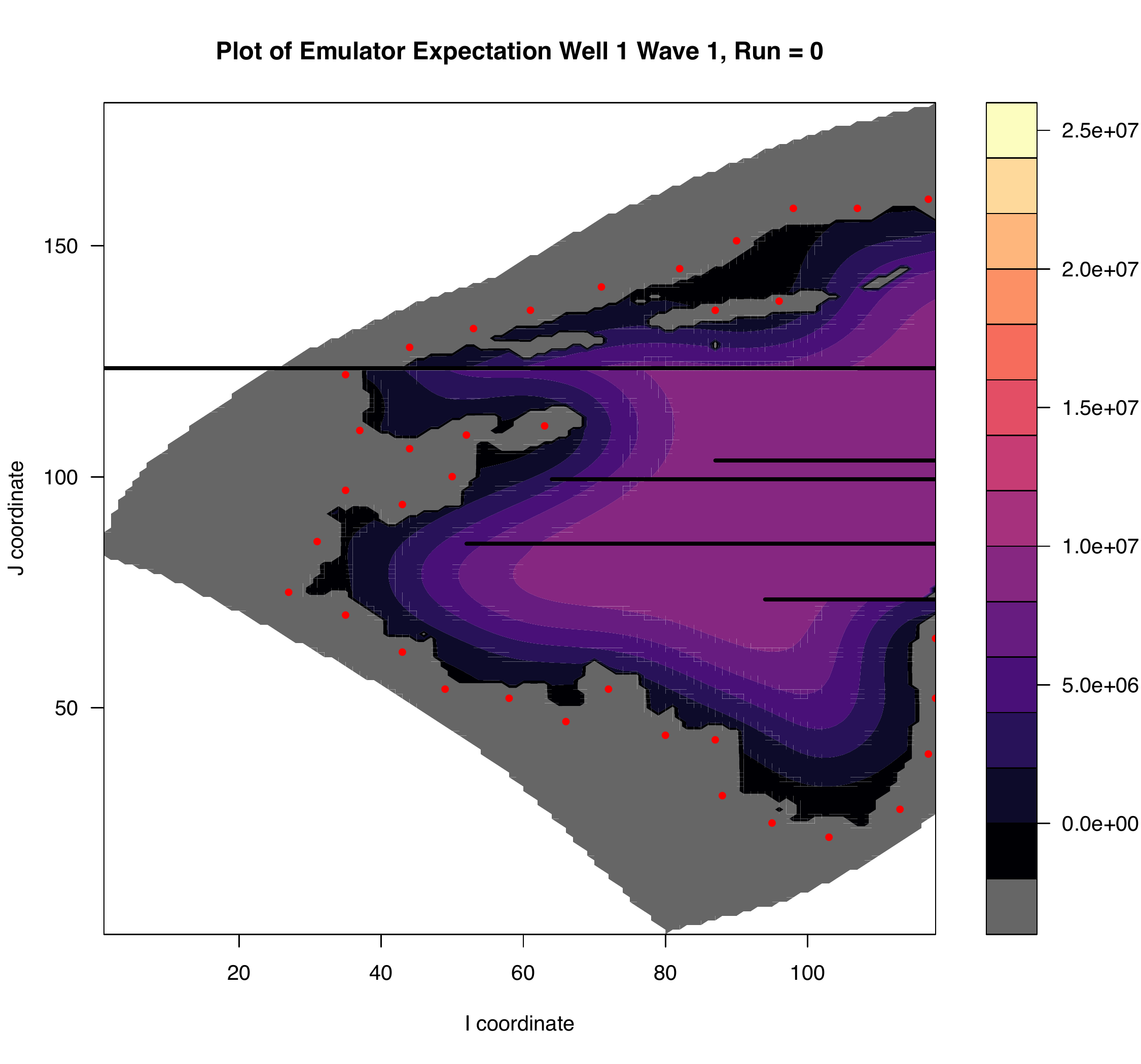} 

\vspace{-0.3cm}
\caption{\footnotesize{The prior emulator expectation $\e{f(\vx)}$.}}
\vspace{0.5cm}
\label{fig_Olymp_em1_a}
\end{subfigure}
\begin{subfigure}{0.49\columnwidth}
\centering
\includegraphics[scale=0.45,viewport= 25 30 650 550, clip]{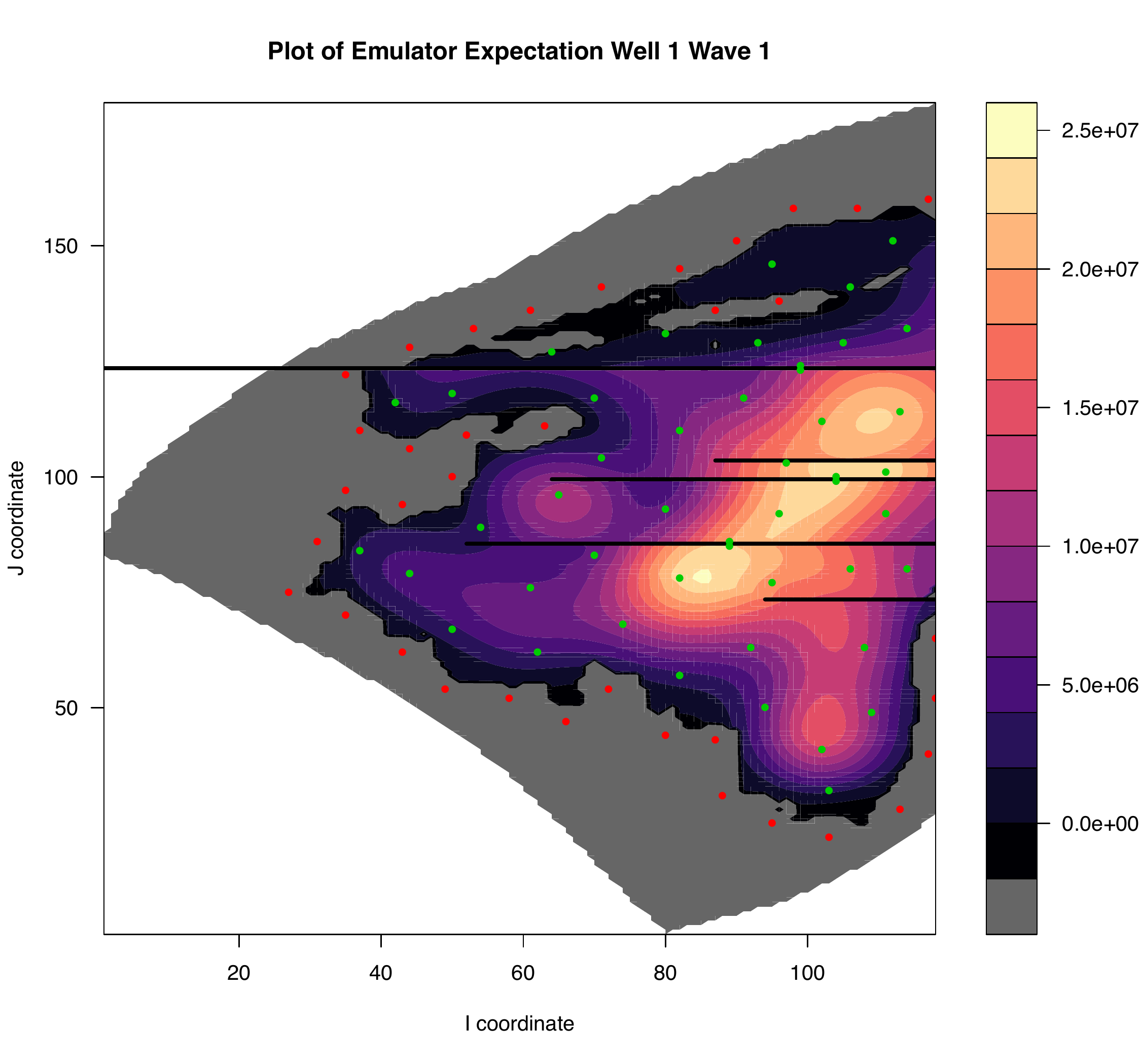} 

\vspace{-0.3cm}
\caption{\footnotesize{The wave 1 emulator expectation $\ed{D_1}{f(\vx)}$.}}
\vspace{0.5cm}
\label{fig_Olymp_em1_b}
\end{subfigure}
\begin{subfigure}{0.41\columnwidth}
\centering
\hspace{-1.2cm}\includegraphics[page=1,scale=0.45,viewport= 25 30 535 550, clip]{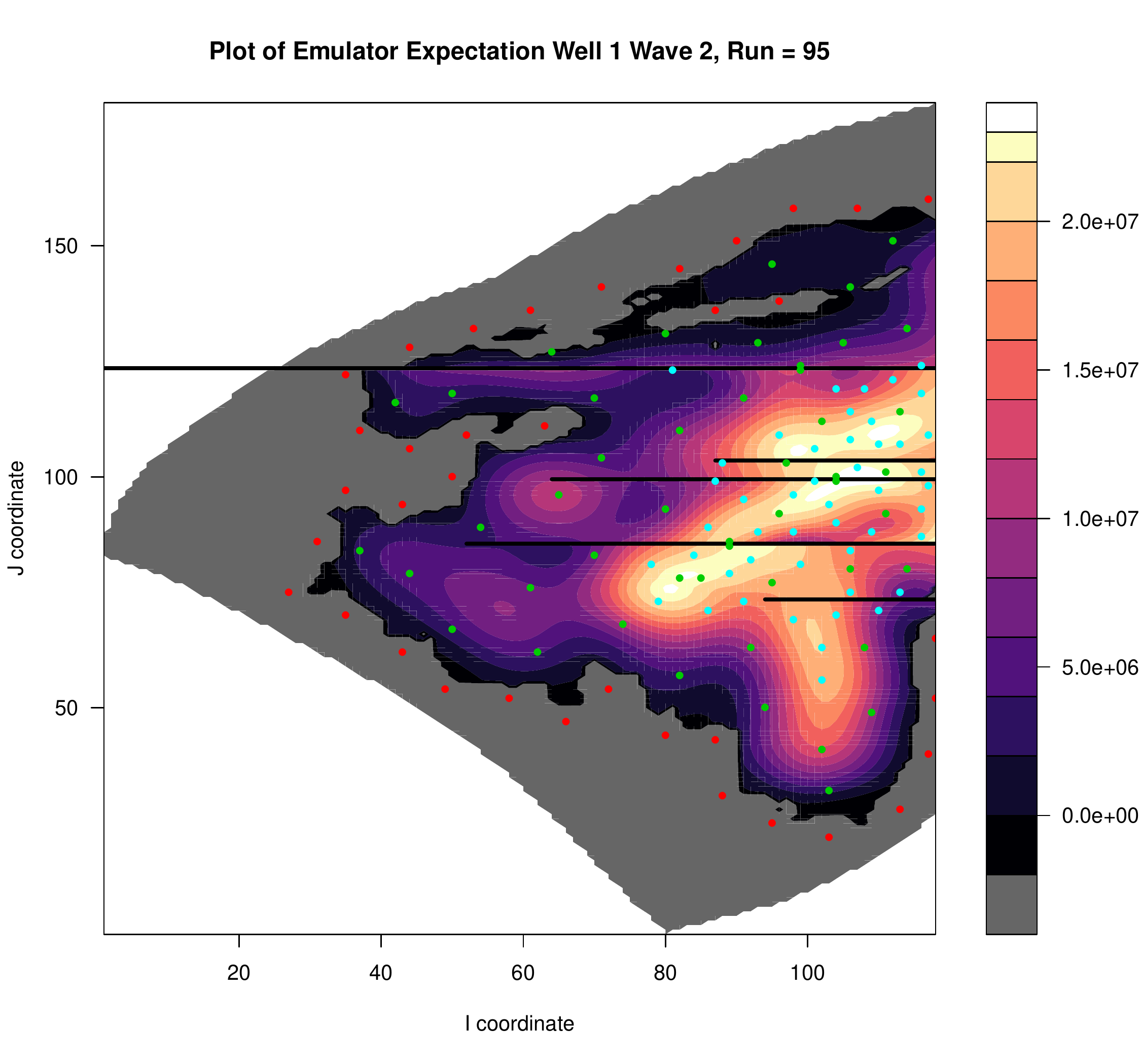} 

\vspace{-0.3cm}
\caption{\footnotesize{Wave 2 emulator expectation $\ed{D_1 \cup D_2}{f(\vx)}$.}}
\label{fig_Olymp_em1_c}
\end{subfigure}
\begin{subfigure}{0.49\columnwidth}
\centering
\includegraphics[page=3,scale=0.45,viewport= 25 30 650 550, clip]{plots_paper/plots_olympus/Emulator_detailed_mean_w2maxlik_GPmeanw2_para_0to95pts0A.pdf} 

\vspace{-0.3cm}
\caption{\footnotesize{Regions of high expected NPV.}}
\label{fig_Olymp_em1_d}
\end{subfigure}
\end{center}
\caption{\footnotesize{The output of the TENSE emulator as applied to the TNO Challenge II Olympus reservoir model. (a) The prior emulator expectation $\e{f(\vx)}$ trained only on the  ghost runs (red points) located in the non-oil producing (grey) regions. (b) Wave 1 TENSE emulator expectation $\ed{D_1}{f(\vx)}$ trained on 47 wave 1 runs denoted $D_1$ (green points) in addition to the ghost points. Plots (a) and (b) share the same key. (c) The wave 2  TENSE emulator expectation $\ed{D_1 \cup D_2}{f(\vx)}$ trained on an additional set of 48 wave 2 runs denoted $D_2$. (d) The wave 2  TENSE emulator expectation of panel (c) now with the high oil production regions highlighted. Plots (c) and (d) share the same key. In all panels the horizontal black lines show the location of the geological faults which induce the discontinuities.}}
\label{fig_Olymp_em1}
\end{figure}
}{}
\ifthenelse{\value{plotstyle}=2}{
\begin{figure}[t!]
\begin{center}
\begin{subfigure}{0.49\columnwidth}
%\centering
\hspace{-1.2cm}\includegraphics[page=1,scale=0.4,viewport= 25 30 535 550, clip]{plots_paper/plots_olympus/Emulator_sequential_mean_maxlik_para_48pts.pdf} 

\vspace{-0.3cm}
\caption{\footnotesize{The prior emulator expectation $\e{f(\vx)}$.~~~~~~}}
\vspace{0.0cm}
\label{fig_Olymp_em1_a}
\end{subfigure}
\begin{subfigure}{0.49\columnwidth}
%\centering
\hspace{-0.5cm}\includegraphics[scale=0.4,viewport= 25 30 650 550, clip]{plots_paper/plots_olympus/Emulator_sd_and_mean_ghostnoghost_meany1_sdy1_maxlik_para_47pts_2.pdf} 

\vspace{-0.3cm}
\caption{\footnotesize{The wave 1 emulator expectation $\ed{D_1}{f(\vx)}$.}}
\vspace{0.0cm}
\label{fig_Olymp_em1_b}
\end{subfigure}
\begin{subfigure}{0.49\columnwidth}
%\centering
\hspace{-1.2cm}\includegraphics[page=1,scale=0.4,viewport= 25 30 535 550, clip]{plots_paper/plots_olympus/Emulator_detailed_mean_w2maxlik_GPmeanw2_para_0to95pts1A.pdf} 

\vspace{-0.3cm}
\caption{\footnotesize{Wave 2 emulator expectation $\ed{D_1 \cup D_2}{f(\vx)}$.~~~~~~}}
\label{fig_Olymp_em1_c}
\end{subfigure}
\begin{subfigure}{0.49\columnwidth}
%\centering
\hspace{-0.5cm}\includegraphics[page=3,scale=0.4,viewport= 25 30 650 550, clip]{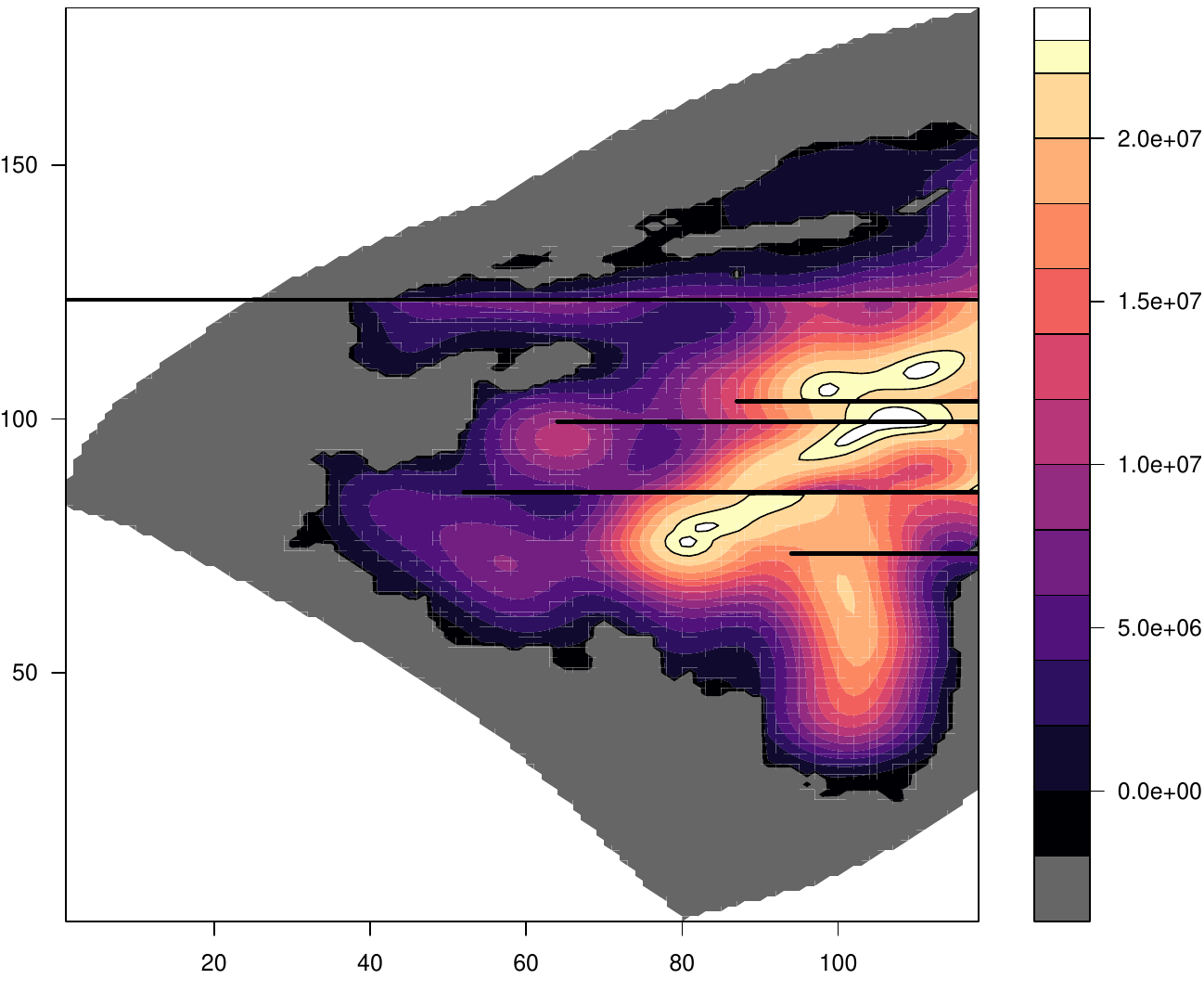} 

\vspace{-0.3cm}
\caption{\footnotesize{Regions of high expected NPV.}}
\label{fig_Olymp_em1_d}
\end{subfigure}
\end{center}
\vspace{-0.4cm}
\caption{\footnotesize{The output of the TENSE emulator as applied to the TNO Challenge II Olympus reservoir model. (a) The prior emulator expectation $\e{f(\vx)}$ trained only on the  ghost runs (red points) located in the non-oil producing (grey) regions. (b) Wave 1 TENSE emulator expectation $\ed{D_1}{f(\vx)}$ trained on 47 wave 1 runs denoted $D_1$ (green points) in addition to the ghost points. Plots (a) and (b) share the same key. (c) The wave 2  TENSE emulator expectation $\ed{D_1 \cup D_2}{f(\vx)}$ trained on an additional set of 48 wave 2 runs denoted $D_2$. (d) The wave 2  TENSE emulator expectation of panel (c) now with the high oil production regions highlighted. Plots (c) and (d) share the same key. In all panels the horizontal black lines show the location of the geological faults which induce the discontinuities.}}
\label{fig_Olymp_em1}
\end{figure}
}{}

The TENSE framework was then applied to $D_1$, employing the embedding surface $v(x,y)$, using equations (\ref{eq_NS_squared_cov_struc_3D_full}), (\ref{eq_sigma3D_full_exp1}), (\ref{eq_app_def_vxy}), (\ref{eq_BLm}) and (\ref{eq_BLv}), with details given in appendix~\reff{app_H}{H}. The resulting emulator expectation $\ed{D_1}{f(\vx)}$ adjusted by the model evaluations $D_1$ is shown in figure~\ref{fig_Olymp_em1_b} as the coloured contours. 
We see that the emulator incorporates jumps in $f(\vx)$ due to the discontinuities caused by the faults, while remaining smooth in all other parts of the space $\mathcal{X}$, as desired. In addition, a clear visualisation of the (expected) NPV surface across the oil reservoir is obtained, and the regions of suspected high NPV identified for further investigation. 

Our primary goal is to visualise this surface, and to identify and examine in more detail regions of higher NPV for consideration by the relevant expert/decision maker.
%APPENDIX? Therefore we employ a simple strategy similar to history matching (  cite, a well used method for identifying regions of input space consistent with observed data), to identify candidate regions that possibly contain high values of NPV. 
%We define an implausibility measure as :
%\be
%I(x) \;\;=\;\; \frac{ (f^+ - \delta) - \ed{D}{f(\vx)} }{ \sqrt{\vard{D}{f(\vx)} + \sigma^2_{\epsilon}} }
%\ee
%where $ \ed{D}{f(\vx)}$ and $\vard{D}{f(\vx)}$ are the usual TENSE emulator expectation and variance trained on data $D$, $f^+$ the NPV value of the highest run found so far,  $\sigma^2_{\epsilon}$ the 
%variance of the structural model discrepancy that accounts for the difference between model and reality, and $ \delta$ a tolerance based on uncertainties in the decision process itself, that allows us to specify how far from the maximum we wish to explore, instead of chasing a single, possibly non-robust maximum   (mention separated decision makers, and cite JO paper 1). We use $I(x)$ to rule out regions of $\mathcal{X}$ highly unlikely to yield values of $f(\vx)$ greater than $f^+$ minus a tolerance $\delta$, by imposing the constraint 
%\be
%I(x) < c
%\ee
%where typically $c$ is chosen to be 3 based on Pukelsheim's 95\% 3-sigma rule for arbitrary unimodal distributions (cite). As 
%the Olympus model is a synthetic example, we can safely set the model discrepancy $\sigma^2_{\epsilon}$ to zero. END APPEND|IX?
We hence use an upper credible interval (UCI) approach to define a region $\mathcal{X}_1$ of possibly high NPV, worthy of further investigation, as
\be\label{eq_non_imp_contraint}
\mathcal{X}_1 \;\;\equiv\;\; \{ \; x\in \mathcal{X} \;\; |\;\;  \ed{D}{f(\vx)} + c\sqrt{\vard{D}{f(\vx)}} \; >\; f^+ - \delta \; \} 
%\mathcal{X}_1 \;\;\equiv\;\; \{ \; x\in \mathcal{X} \;\; |\;\;  \ed{D}{f(\vx)} + c\sqrt{\vard{D}{f(\vx)} +  \sigma^2_{\epsilon}} \; >\; f^+ - \delta \; \} 
\ee 
where $f^+$ is the highest NPV seen so far, $\delta$ is a tolerance based on uncertainties in the decision process itself \citep{eage:/content/papers/10.3997/2214-4609.202035109} and on our desire to explore the region of high NPV and not just to identify a single possibly non-robust maxima, 
%$\sigma^2_{\epsilon}$ is the 
%variance of the structural model discrepancy that accounts for the difference between model and reality (which, as Olympus is synthetic, we subsequently set to zero) 
and $c$ is typically chosen to be 3 based on Pukelsheim's 95\% 3-sigma rule for arbitrary unimodal distributions~\citep{threesigma_short}.
Figures~\ref{fig_Olymp_em2_a} and \ref{fig_Olymp_em2_b} (in appendix~\reff{app_I}{I}) show respectively the prior UCI: $ \e{f(\vx)} + c\sqrt{\var{f(\vx)}}$, and also the wave 1 UCI adjusted by $D_1$: 
$ \ed{D_1}{f(\vx)} + c\sqrt{\vard{D_1}{f(\vx)}}$. 

Following a general history matching strategy~\citep{Vernon10_CS}, we proceeded by designing a second wave of space filling runs over the $\mathcal{X}_1$ region, now chosen to minimise the mean emulator variance over $\mathcal{X}_1$ only. This design is shown as the light blue points in figure~\ref{fig_Olymp_em1_c}. Evaluation of this design using the Olympus model creates a second vector of model outputs denoted $D_2$. We can then adjust the TENSE emulators by $D_1 \cup D_2$, giving
the emulator expectation $\ed{D_1 \cup D_2}{f(\vx)}$ for the mean NPV, were a producer well to be placed at location $\vx$, which is shown in figure~\ref{fig_Olymp_em1_c} as the coloured contours. 
We now have a detailed representation of the high NPV areas of the $\mathcal{X}$ map, naturally incorporating the fault discontinuities. 
Examination of the UCI after the wave 2 runs (appendix~\reff{app_I}{I}, figure~\reff{fig_Olymp_em2_c}{10c}) shows that there is little to be learned about this region by performing further runs. 
Figure~\ref{fig_Olymp_em1_d} highlights the high NPV region in question, with the solid contours corresponding to thresholds of $f^+-\delta = 2.2\times 10^7$ and $2.3\times 10^7$ respectively. This achieves our objective of locating and visualising the high NPV areas for the position of a single producer well.

The TENSE framework can be employed for several further types of analysis e.g. for quantile emulation to examine the uncertainties in the NPV induced by the unknown geology, which we demonstrate in appendix~\reff{app_I}{I}.
%There was also much interest in using TENSE within a greedy algorithm, selecting well types and locations sequentially which (due to the heavy discounting factor) can provide adequate solutions. 
%We report on quantile emulation and the first few steps of the greedy algorithm in appendix~\ref{app_I}.
\ifthenelse{\value{plotstyle}=2}{\clearpage}{}

\subsection{Extensions and Generalisations}\label{sec_future_full_opt}
\ifthenelse{\value{plotstyle}=2}{\vspace{-0.2cm}}{}
These initial investigations of the TNO challenge using TENSE can be extended in multiple ways.
The full problem of optimising the joint location of multiple producer and injector wells is of course the long-term goal. This is a very challenging problem especially when combined with an appropriate level of uncertainty quantification~\citep{eage:/content/papers/10.3997/2214-4609.202035109}. However, due to the localised structure of oil reservoirs, for early/medium times, often small groups of wells (e.g. one injector combined with two producers) are optimised on particular sub-regions of the map, to break the full problem into smaller, tractable pieces. The TENSE framework can be directly extended to such cases. For the example of three wells, a 6-dimensional problem, we would need to employ a torn embedding in a 9-dimensional space to account for the discontinuity effects on each of the three wells. This 9-dimensional space would look like the direct product of three versions of the 3-dimensional space used here in the single well example. 
\ifthenelse{\value{plotstyle}=2}{\vspace{-0.0cm}}{}

In principle the TENSE approach can be generalised to far more wells that just three, however, constructing an accurate emulator over the full input space for larger numbers may require infeasible numbers of runs (and we would waste a lot of runs exploring the
low NPV parts of the space). So a more targeted approach, optimising sets of three wells, combining them and then employing a final wave or two of optimisation on the full set of wells, may be a sensible strategy. 
We leave this, and the various associated design strategies, to future work.

\ifthenelse{\value{plotstyle}=2}{\vspace{-0.4cm}}{}

%%%%%%%%%%%%%%%%%%%%%%%%%%%%%%%%%%%%%%%%%%%%%%%%%%%%%
\section{Conclusion and Future Plans}
%%%%%%%%%%%%%%%%%%%%%%%%%%%%%%%%%%%%%%%%%%%%%%%%%%%%%
\ifthenelse{\value{plotstyle}=2}{\vspace{-0.1cm}}{}
We have introduced the Torn Embedding Non-Stationary Emulation (TENSE) approach for emulating 
expensive functions that possess partial discontinuities of known location and general non-linear form, which possibly begin and/or end within the input space of interest.
This method utilises a torn embedding surface to induce the required discontinuities, combined with a carefully chosen 
non-stationary covariance structure over the embedding space, to correct for the local impact of the use of the non-linear embedding. While we have introduced this in the context of a squared exponential covariance structure in 2D/3D, it can be applied to a wide class of covariance structures and emulator forms, and in principle, extended into higher dimensions.
We demonstrated this approach on various example functions, and then applied it to the realistic OLYMPUS reservoir model, 
showing how it facilitated the design of model evaluations and the construction of appropriate emulators to visualise the NPV surface, both of which respected the presence of the discontinuities. It was also applied to quantile emulation, and the extension to multiple wells and higher dimensions was discussed. 
\ifthenelse{\value{plotstyle}=2}{\vspace{-0.0cm}}{}

There are many possible extensions of this methodology. While we have employed fixed embedding surfaces $v(x,y)$ chosen to suitably decorrelate outputs either side of the discontinuities, one could instead use the TENSE framework to learn about such surfaces to find more accurate embeddings. This could be combined with methods to learn about the location of the discontinuities themselves to provide a more complete analysis, in a fully Bayesian framework. The extension to more complex networks of discontinuities is also very interesting, as it may require embedding in higher dimensional spaces to provide the necessary freedom to ensure sufficient decorrolation across all discontinuities, especially those that intersect, resulting in a challenging embedding problem.

\ifthenelse{\value{plotstyle}=2}{
\begin{supplement}
\stitle{Appendices}
\sdescription{The appendices A-I referred to in the main article.}
\end{supplement}
\vspace{-0.1cm}
\begin{acks}[Acknowledgments]
I.V. gratefully acknowledges UKRI (EP/W011956/1) and Wellcome (218261/Z/19/Z) funding.
J.O. gratefully acknowledges EPSRC iCase Studentship (Smith Institute) funding.
We thank Rock Flow Dynamics for use of the tNavigator simulator.
%And this is an acknowledgements section with a heading that was produced by the
%$\backslash$section* command. Thank you all for helping me writing this
%\LaTeX\ sample file.
\end{acks}
}{}
\vspace{-0.3cm}
\ifthenelse{\value{plotstyle}=2}{
\bibliographystyle{ba}
\bibliography{/Users/ianvernon/Work/Master_Bibliography_and_Related_Files/master_bib.bib}
}{}

%% file: Emulating_over_Discontinuites_using_Torn_Embeddings_appendices.tex
\clearpage

\appendix

\section{More Advanced Emulator Forms}\label{app_A}
The main article focuses on the simple emulator specification as given by equations~(\ref{eq_cor_struc}) and (\ref{eq_prodgausscor}), however, a more advanced and well-used emulator specification is given by \citep{Craig97_Pressure,Vernon10_CS}:
\be\label{eq_fullem}
f(\vx) \;=\; \sum_{j} \beta_j g_j(\vx_A) + u(\vx_A) + w(\vx)
\ee
where the active inputs $\vx_A$ are a subset of $\vx$ that are strongly influential for $f(\vx)$,
the first term on the right hand side is a regression term containing known functions $g_j(\vx_A)$ and possibly unknown $\beta_j$,
$u(\vx_A)$ is a weakly stationary process over the active inputs only, with stationary covariance structure as in equation~(\ref{eq_cor_struc}), and 
$w(\vx)$ is an uncorrelated nugget term, representing the inactive variables and facilitating an effective dimensional reduction.  See \cite{JAC_Handbook} and \cite{Vernon10_CS,Vernon10_CS_rej} for 
discussions of the benefits of using an emulator structure of this kind, and see~\cite{Kennedy01_Calibration,Higdon08a_calibration} for discussions of alternative structures.
The generalisation of our TENSE methodology to more advanced emulator forms, such as given by equation~(\ref{eq_fullem}), is relatively straightforward, in principle.

\section{Non-validity of Geodesic Distance Approach}\label{app_B}

Continuing the discussion in section~\ref{ssec_EPPD} of why the suggestion to use the geodesic distance between input points in the correlation function, defined such that viable geodesics do not cross the discontinuity (and hence have to go around it), does not lead to valid covariance structures.
Using equations~(\ref{eq_cor_struc}) and (\ref{eq_prodgausscor}) we can construct the $4\times4$ covariance matrix formed from the 4 outputs $f(x_A), f(x_B), f(x_C), f(x_D)$ corresponding to the four input points $x_A=(0.5,1),x_B=(0.75,1),x_C=(1,1^+),x_D=(1,1^-)$ located in figure~\ref{fig_toymod1}. If we use geodesics that go around the discontinuity in figure~\ref{fig_toymod1}, then we have that the geodesic distance between points $x_A$ and $x_B$ is 0.25 as usual, however the geodesic distance between points $x_C$ and $x_D$ is $0.25+0.25 = 0.5$. 
By setting $\Sigma_{2D} = \rm{diag} \{\theta,\theta\} $ in equation~(\ref{eq_prodgausscor}) and combining with equation~(\ref{eq_cor_struc}) we obtain the isotropic squared exponential correlation structure as
\be\label{eq_appB_cor}
\cov{f(\vx)}{f(\vx')} \;=\; \sigma^2 r(\vx-\vx') \;=\; \sigma^2 \exp\{-\|\vx-\vx'\|^2/\theta^2\}   % \;=\;   \prod_{i=1}^d \exp\{-|x_i-x_i'|^2/\theta^2\}
\ee
Setting $\theta=1$ and $\sigma=1$, we can now construct the covariance matrix for the random vector $G=\{f(x_A), f(x_B), f(x_C), f(x_D) \}$ as 
\ba
{\rm Var} [G] &=&  
{\rm Cov} \begin{bmatrix}
\begin{pmatrix}
f(x_A) \\ 
f(x_B) \\ 
f(x_C) \\
f(x_D)
\end{pmatrix},
\begin{pmatrix}
f(x_A) \\ 
f(x_B) \\ 
f(x_C) \\
f(x_D)
\end{pmatrix}
\end{bmatrix}  \\
&=&
\begin{pmatrix}
\exp\{ -0^2 \}  &  \exp\{ -0.25^2 \}  &  \exp\{ -0.5^2 \} &  \exp\{ -0.5^2 \}  \\
\exp\{ -0.25^2 \}  &  \exp\{ -0^2 \}  &  \exp\{ -0.25^2 \} &  \exp\{ -0.25^2 \}  \\
\exp\{ -0.5^2 \}  &  \exp\{ -0.25^2 \}  &  \exp\{ -0^2 \} &  \exp\{ -0.5^2 \}  \\
\exp\{ -0.5^2 \}  &  \exp\{ -0.25^2 \}  &  \exp\{ -0.5^2 \} &  \exp\{ -0^2 \}  \\
\end{pmatrix}
\ea
where this matrix is populated by repetitive use of equation~(\ref{eq_appB_cor}) combined with the geodesic distances.
Examination of the eigenstructure of ${\rm Var} [G]$ shows that the smallest eigenvalue is $-0.0251$, hence it is not positive semi-definite, hence not a valid covariance matrix and hence the geodesic distance approach is fundamentally flawed. This problem will be exacerbated if we examine more than just 4 points. 
We note that this problem occurs for any value of the correlation length such that $\theta> 0.25/\sqrt{(0.5\log(2))}$.

\section{Emulator Realisations with a Discontinuity}\label{app_C}

Figure~\ref{fig_toymod1r} shows individual realisations from the induced 2D process as represented by the emulator for $f(x)$ discussed in section~\ref{ssec_torn_embed} and shown in figure~\ref{fig_toymod1}.

\ifthenelse{\value{plotstyle}=1}{
\begin{figure}[t]
\begin{center}
\begin{subfigure}{0.49\columnwidth}
\centering
\includegraphics[page=2,scale=0.34,viewport= 0 0 630 530,clip]{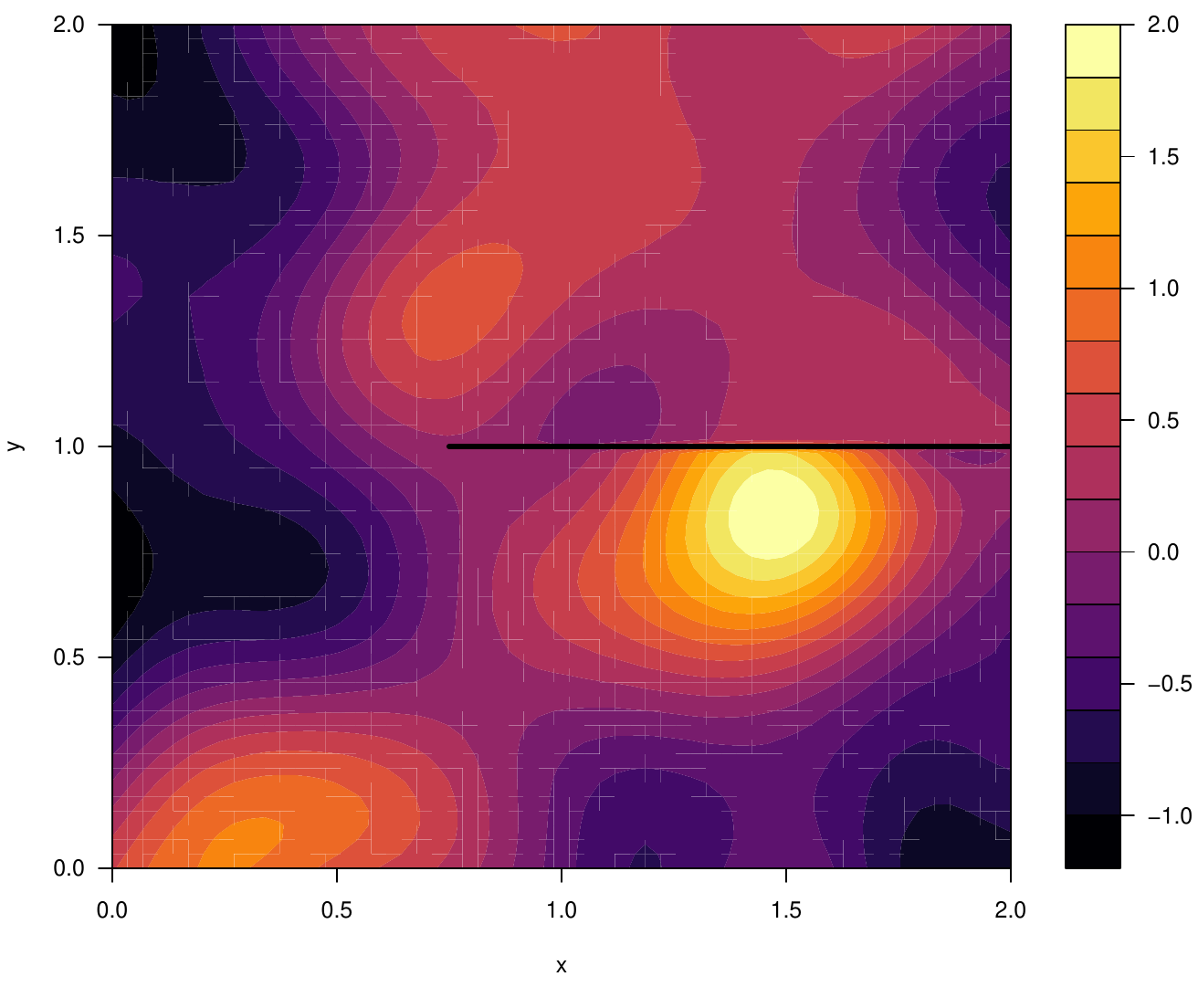} 
\vspace{-0.4cm}
\caption{\footnotesize{A single realisation from the prior $f(x)$.}}
\label{fig_toymod1r_a}
\end{subfigure}
\begin{subfigure}{0.49\columnwidth}
\centering
\includegraphics[page=3,scale=0.34,viewport= 0 0 630 530,clip]{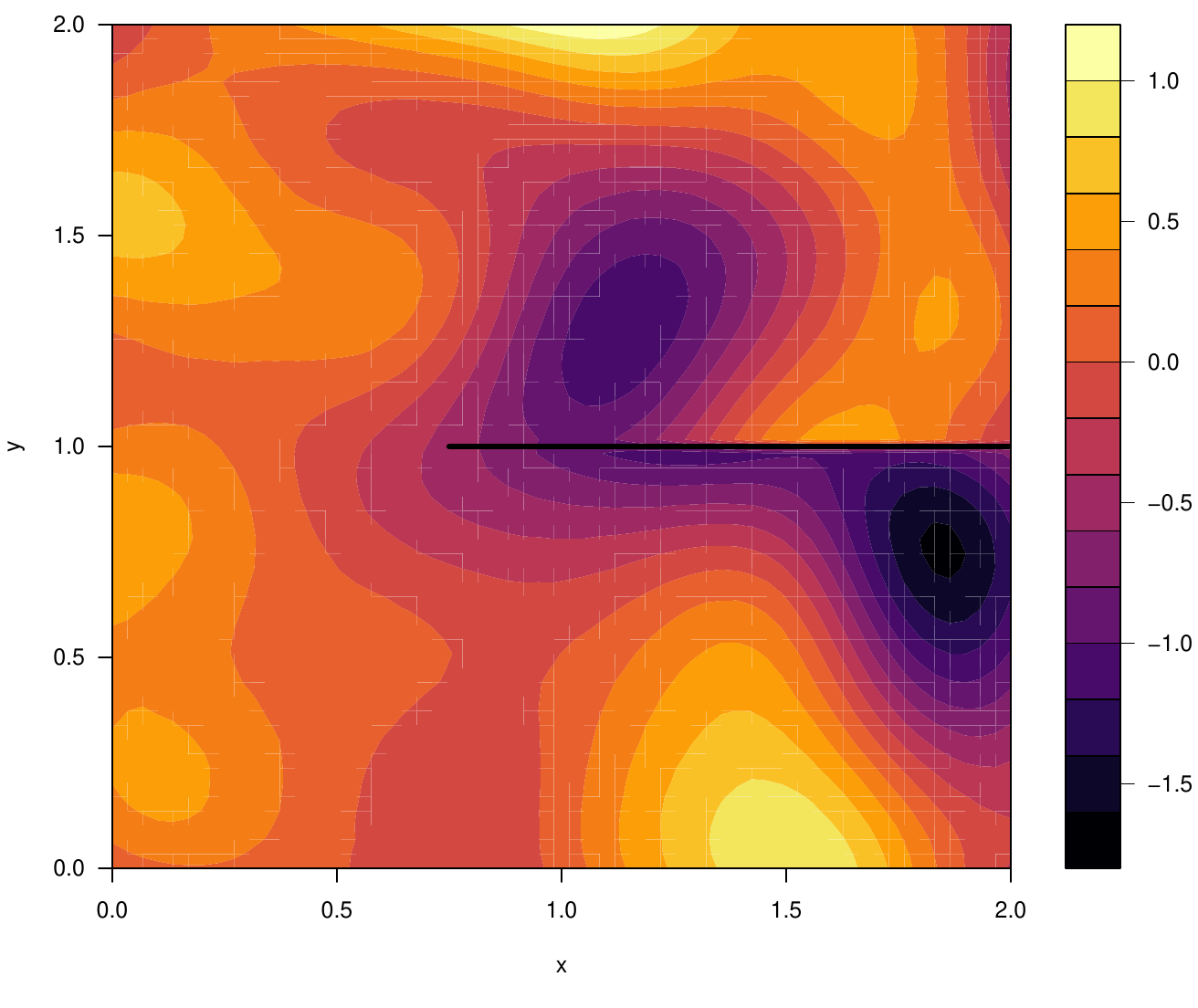}
\vspace{-0.4cm}
\caption{\footnotesize{A single realisation from the prior $f(x)$.}}
\label{fig_toymod1r_b}
\end{subfigure}
\begin{subfigure}{0.49\columnwidth}
\centering
\vspace{0.4cm}
\includegraphics[page=6,scale=0.34,viewport= 0 0 630 530,clip]{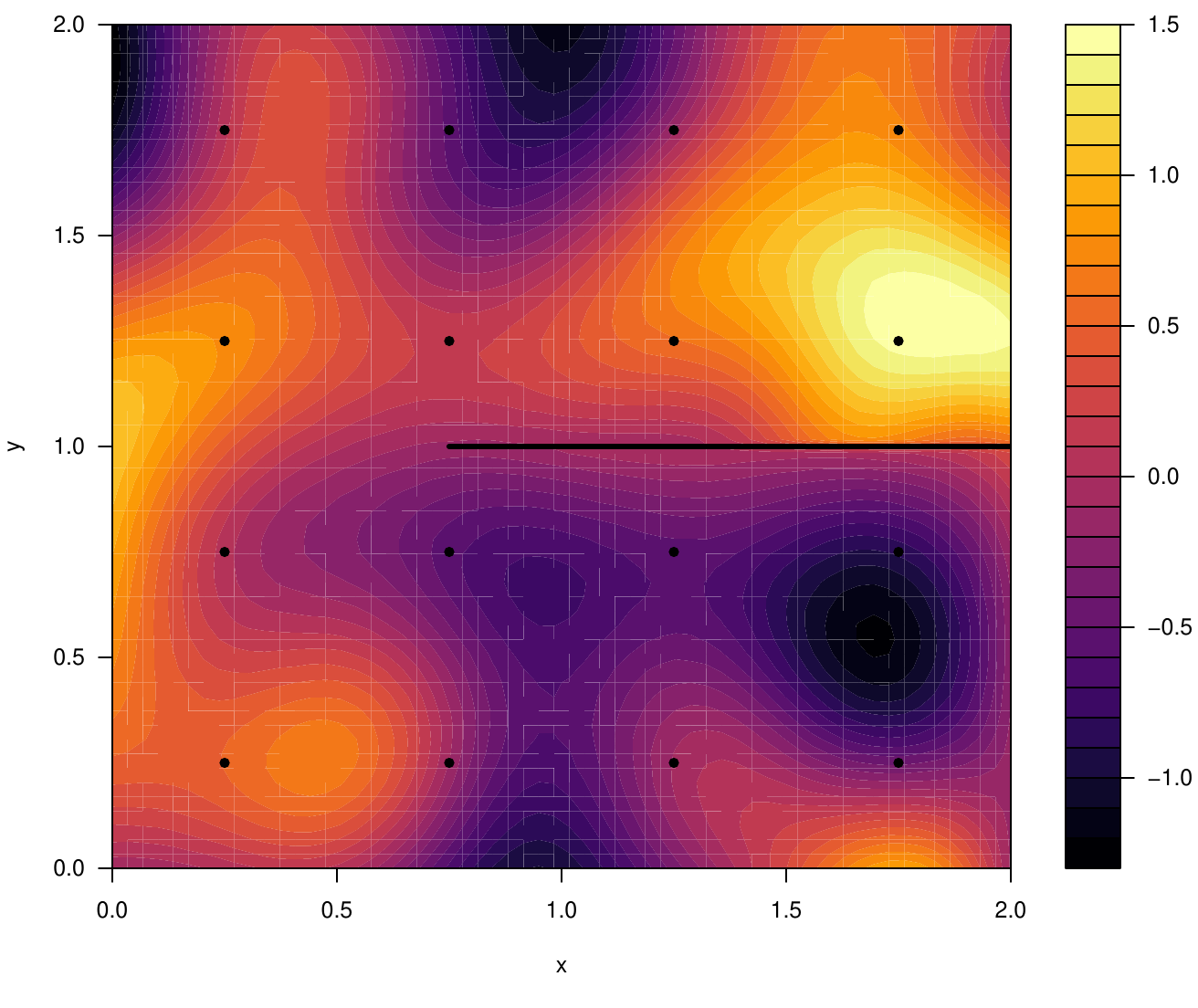}
\vspace{-0.4cm}
\caption{\footnotesize{A single realisation from $f(x)$ adjusted by $D$.}}
\label{fig_toymod1r_c}
\end{subfigure}
\begin{subfigure}{0.49\columnwidth}
\centering
\vspace{0.4cm}
\includegraphics[page=8,scale=0.34,viewport= 0 0 630 530,clip]{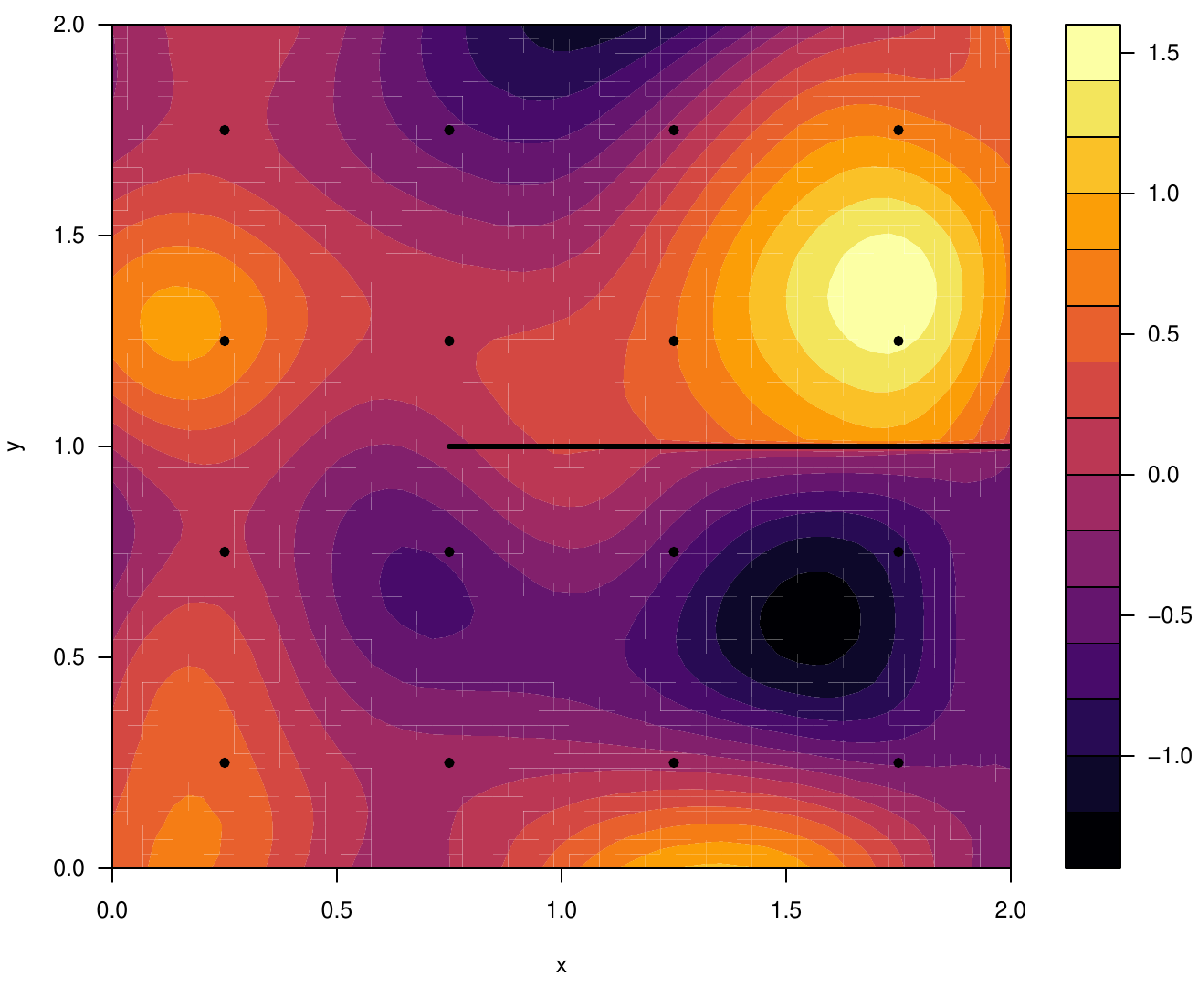}
\vspace{-0.4cm}
\caption{\footnotesize{A single realisation from $f(x)$ adjusted by $D$..}}
\label{fig_toymod1r_d}
\end{subfigure}
\end{center}
\vspace{-0.4cm}
\caption{\footnotesize{Individual realisations from the emulator for $f(x)$ before updating by the runs $D$ (top row) and after 
updating by $D$ (bottom row). The realisations clearly respect the existence of the discontinuity, shown as the horizontal black line. Compare with figure~\ref{fig_toymod1}.}}
\label{fig_toymod1r}
\vspace{-0.cm}
\end{figure}
}{}
\ifthenelse{\value{plotstyle}=2}{
\begin{figure}[t]
\vspace{-0.2cm}
\begin{center}
\begin{subfigure}{0.49\columnwidth}
\centering
\hspace{-0.9cm} \includegraphics[page=2,scale=0.33,viewport= 0 0 630 530,clip]{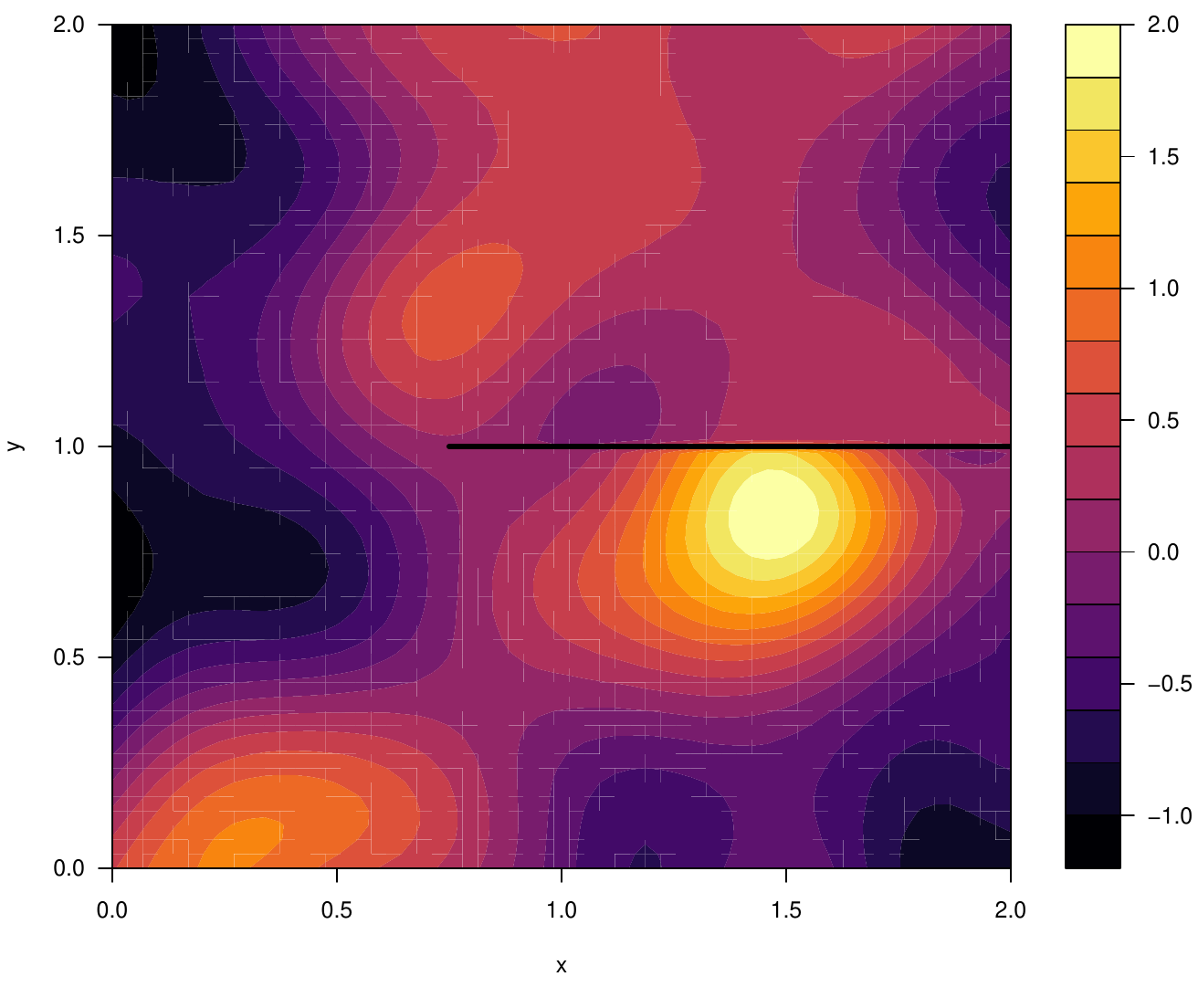} 
\vspace{-0.7cm}
\caption{\footnotesize{A single realisation from the prior $f(x)$.}}
\label{fig_toymod1r_a}
\end{subfigure}
\begin{subfigure}{0.49\columnwidth}
\centering
\includegraphics[page=3,scale=0.33,viewport= 0 0 630 530,clip]{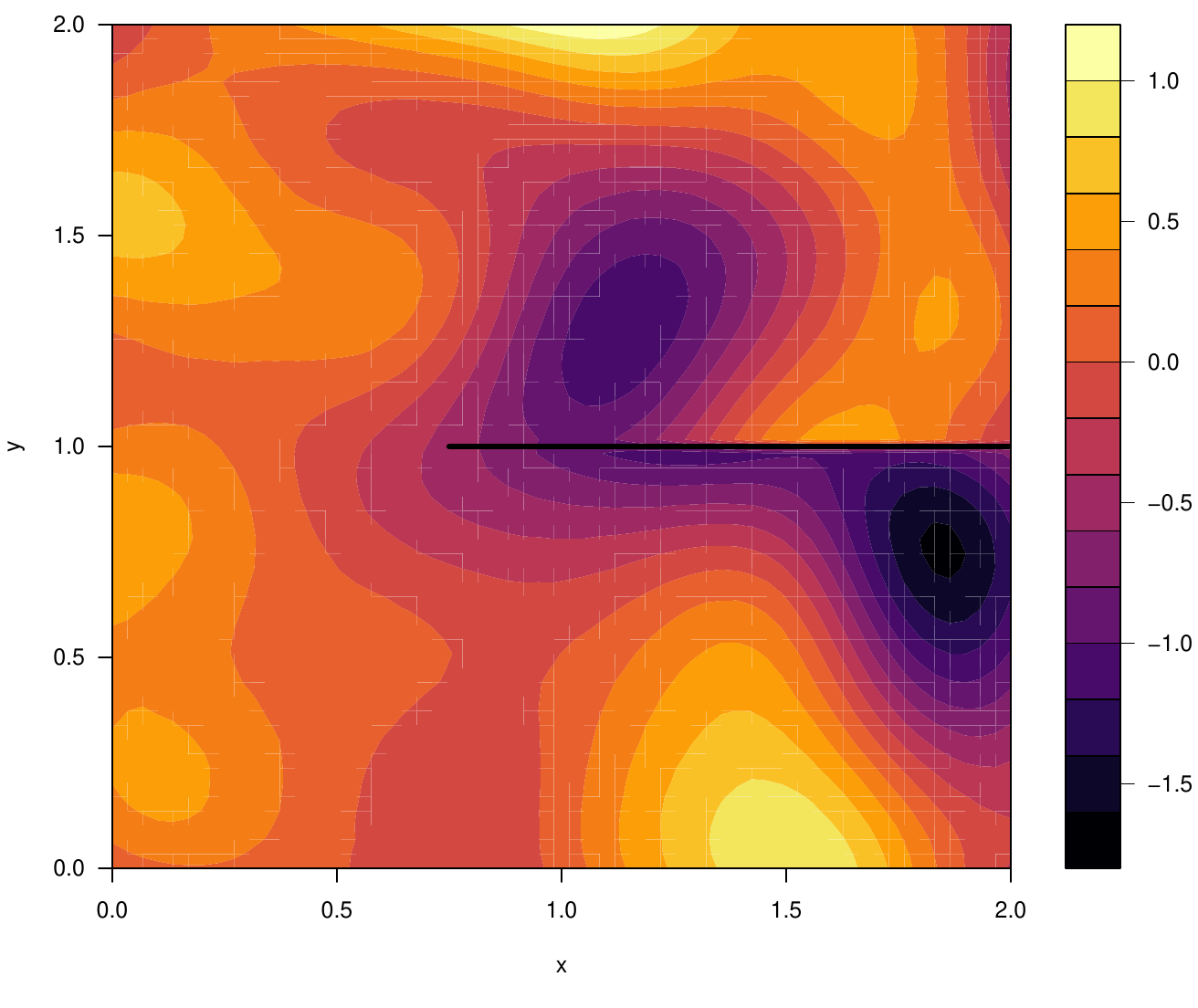}
\vspace{-0.7cm}
\caption{\footnotesize{A single realisation from the prior $f(x)$.}}
\label{fig_toymod1r_b}
\end{subfigure}
\begin{subfigure}{0.49\columnwidth}
\centering
\vspace{0.0cm}
\hspace{-0.9cm}\includegraphics[page=6,scale=0.33,viewport= 0 0 630 530,clip]{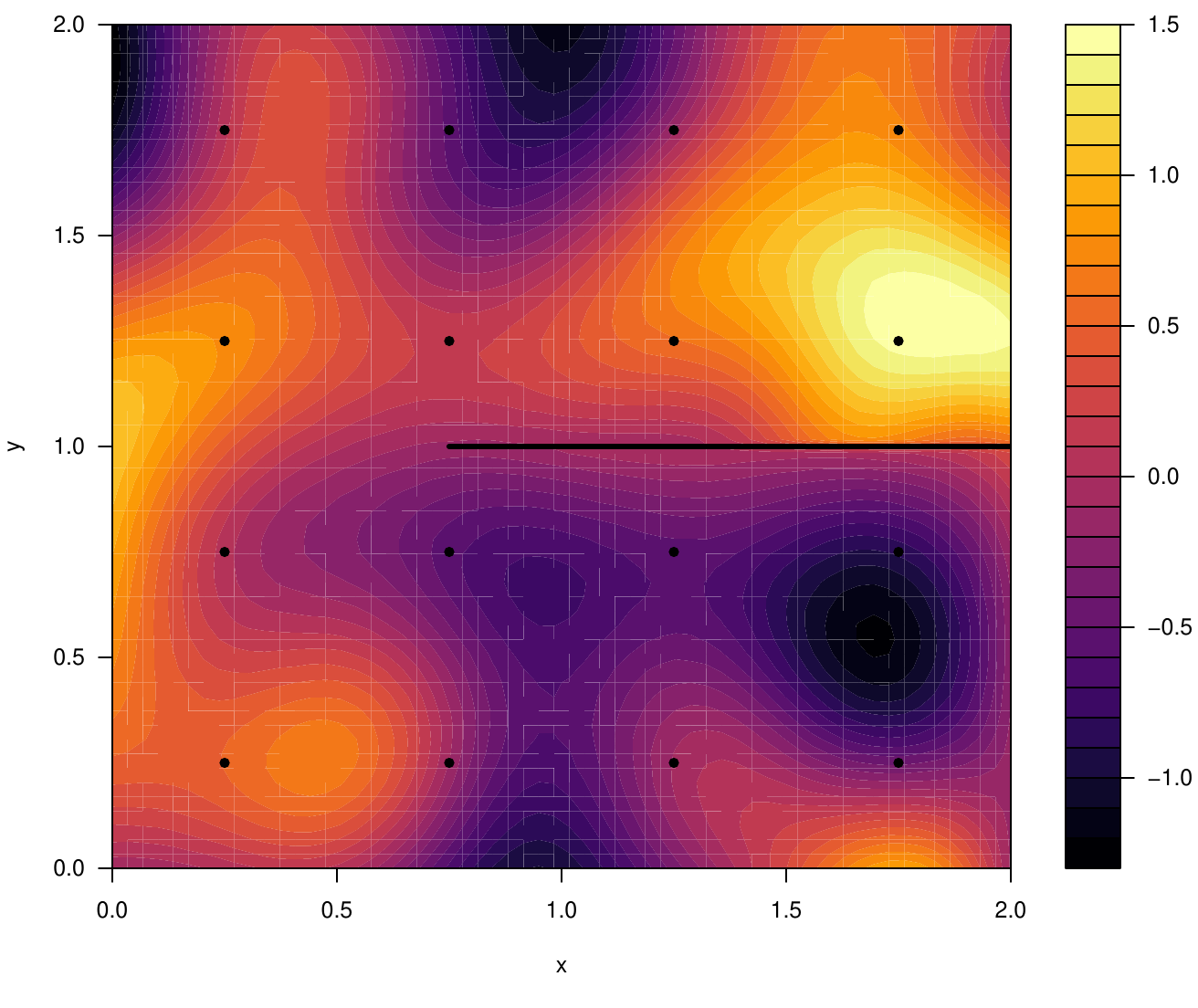}
\vspace{-0.7cm}
\caption{\footnotesize{A single realisation from $f(x)$ adjusted by $D$.}}
\label{fig_toymod1r_c}
\end{subfigure}
\begin{subfigure}{0.49\columnwidth}
\centering
\vspace{0.0cm}
\includegraphics[page=8,scale=0.33,viewport= 0 0 630 530,clip]{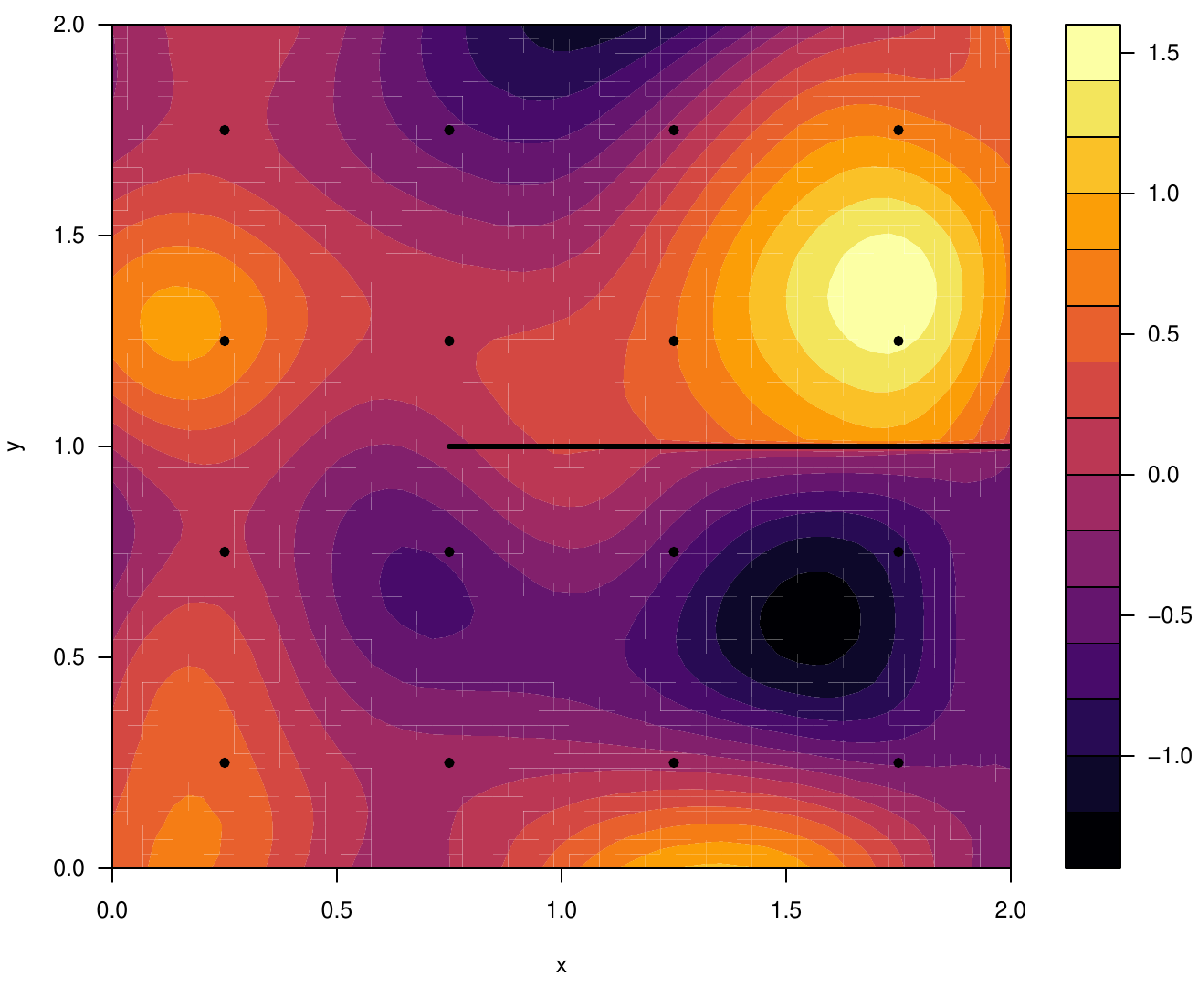}
\vspace{-0.7cm}
\caption{\footnotesize{A single realisation from $f(x)$ adjusted by $D$.}}
\label{fig_toymod1r_d}
\end{subfigure}
\end{center}
\vspace{-0.6cm}
\caption{\footnotesize{Individual realisations from the emulator for $f(x)$ before updating by the runs $D$ (top row) and after 
updating by $D$ (bottom row). The realisations clearly respect the existence of the discontinuity, shown as the horizontal black line. Compare with figure~\ref{fig_toymod1}.}}
\label{fig_toymod1r}
\vspace{-0.8cm}
\end{figure}
}{}

\section{Example Function with non-linear discontinuities}\label{app_D}

Here we define the function $f(x,y)$ with discontinuities situated on non-linear locations, as examined in section~\ref{ssec_nonlinlocations} and shown in figure~\ref{fig_toy3_curved_disc}.
The function $f(x,y)$ is defined over the region $\mathcal{X} = \{\vx \in \mathbb{R}^2: -1 < x < 1, -1 < y < 1\}$ as follows.
We define a region identifier $m$ via the intersection of several circles centred on the points $ \mathbf{a},  \mathbf{b},  \mathbf{c},  \mathbf{d} $ and the origin as:
\ba
m(\vx) &=& \left\{   \begin{matrix} 
0 & \text{if} & |\vx| < 0.4 \\
1 & \text{if}  &  |\vx| >0.4 \; \land \; |\vx - \mathbf{a}|^2< 1 \; \land \;  |\vx - \mathbf{b}|^2 > 1 \\
2 & \text{if}  &   |\vx| >0.4 \; \land \; |\vx - \mathbf{b}|^2 < 1 \; \land \;  |\vx - \mathbf{c}|^2 > 1 \\
3 & \text{if}  &    |\vx| >0.4 \; \land \; |\vx - \mathbf{c}|^2 < 1 \; \land \;  |\vx - \mathbf{d}|^2 > 1 \\
4 & \text{if}  &    |\vx| >0.4 \; \land \; |\vx - \mathbf{d}|^2 < 1 \; \land \; |\vx - \mathbf{a}|^2 > 1 \\
\end{matrix}  \right. \\
\rm{with} && \mathbf{a} = (1,0), \mathbf{b} = (0,1), \mathbf{c} = (-1,0), \mathbf{d} = (0,-1) 
\ea
and define the function to be emulated as
\be
f(x,y) \;\;=\;\; 
0.5(\sin(3x) + \cos(3.5y))  + (-1)^{m(\vx)+1} (|\vx|-0.4)^2  \; \mathbbm{1}_{\{  m(\vx) > 0  \}}
\ee
shown in figure~\ref{fig_toy3_curved_disc_a}.
We choose an embedding surface $v(x,y)$, shown in figure~\ref{fig_toy3_curved_disc_b}, that has suitable jumps over the locations of the discontinuities as:
\ba
v(x,y) &=& \left\{   \begin{matrix} 
0 & \text{if} & m(\vx) =0,2 \text{ or } 4 \\
\frac{1}{2}(m(\vx)-2) (|\vx|-0.4)^2 & \text{if} &  m(\vx) =1\text{ or }3 
\end{matrix}  \right. 
\ea
We then apply the TENSE framework to provide emulator expectations and standard deviations as shown in figures~\ref{fig_toy3_curved_disc_c}
and \ref{fig_toy3_curved_disc_d} respectively, and discussed in section~\ref{ssec_nonlinlocations}.

\section{Definition of Net Present Value (NPV)}\label{app_E}

The Net Present Value or NPV is given by
\be
NPV^{(j)}(\vx) \;\;=\;\; \sum_{i=1}^{N_t} \frac{R_j(\vx,t_i)}{(1+d)^{t_i/\tau}}
\ee
where $R_j(\vx,t_i)$ is the profit for time period $t_i$ (revenue of oil generated, minus expenditure due to water production, injection and other field costs) obtained from the expensive reservoir model, evaluated using the $j$th geological realisation. $d$ is a discounting factor ($8\%$ for the TNO Challenge) with $\tau$ the corresponding discounting time period (typically 365 days). 
%$j$ labels the geological realisation described previously.

\section{The TNO Olympus oil reservoir model}\label{app_F}

Figure~\ref{fig_Olymp_info} shows additional plots of the TNO II Challenge Olympus oil reservoir model, (a) gives the mean oil volume per unit area over the 50 geological realisations while (b) shows the corresponding standard deviation of the oil volume per unit area over the 50 geological realisations. (c) and (d) show the oil volume per unit area from two further examples of the 50 geological realisations.

\ifthenelse{\value{plotstyle}=1}{
\begin{figure}[t]
\begin{center}
\hspace{-0.8cm}\begin{subfigure}{0.52\columnwidth}
\centering
\includegraphics[page=1,scale=0.42,viewport= 0 35 620 550, clip]{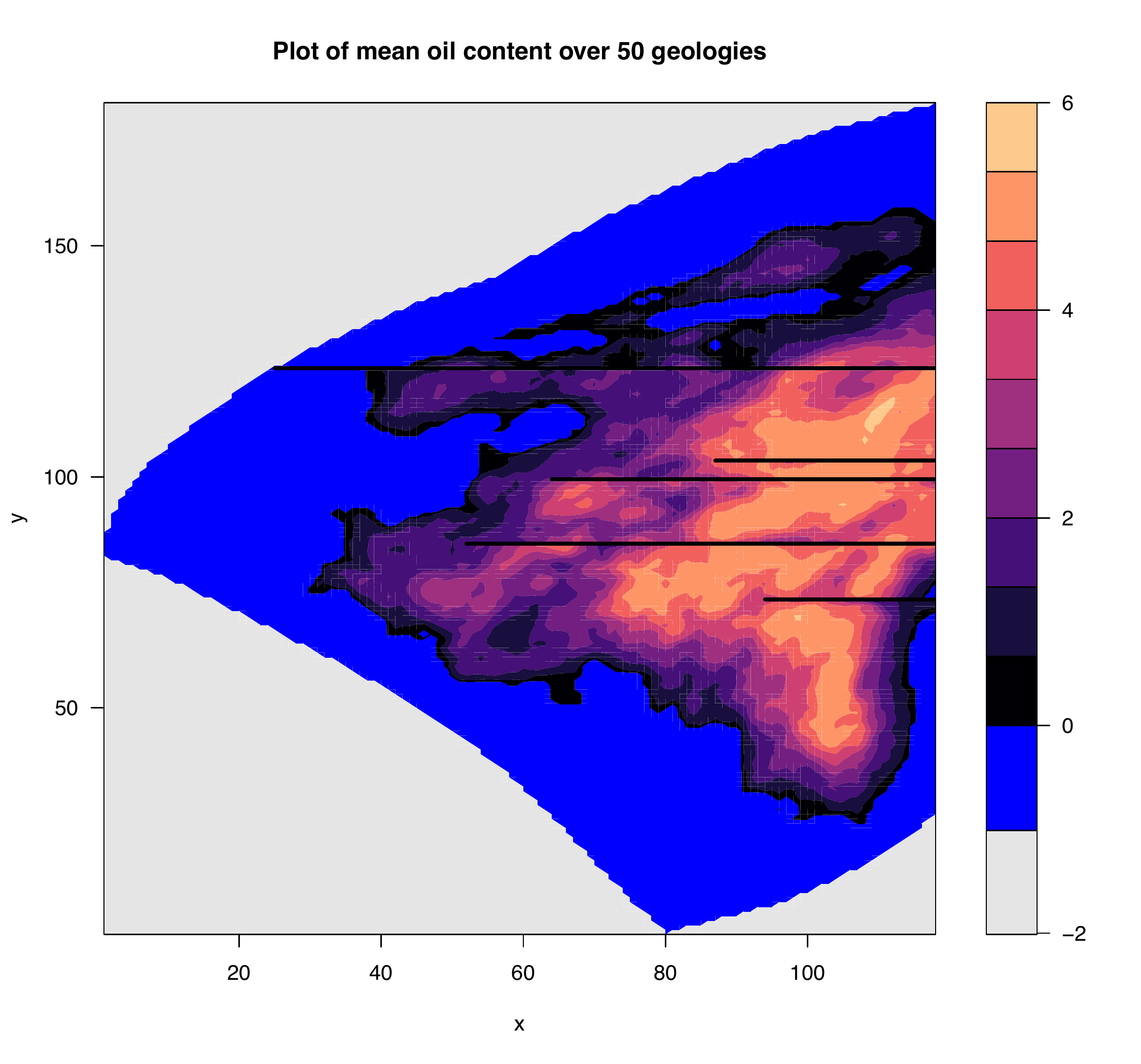}

\vspace{-0.3cm}
\caption{\footnotesize{Mean oil content over 50 geological realisations.}}
\vspace{0.5cm}
\label{fig_Olymp_info_a}
\end{subfigure}
\begin{subfigure}{0.49\columnwidth}
\centering
\includegraphics[page=2,scale=0.42,viewport= 25 35 620 550, clip]{plots_paper/plots_olympus/TNO_map_mean_sd_real_of_5_geologies_with_faults_page2_4_10_14.pdf}

\vspace{-0.3cm}
\caption{\footnotesize{SD of oil content over 50 geological realisations.}}
\vspace{0.5cm}
\label{fig_Olymp_info_b}
\end{subfigure}

\hspace{-0.8cm}\begin{subfigure}{0.52\columnwidth}
\centering
\includegraphics[page=3,scale=0.42,viewport= 0 0 620 550, clip]{plots_paper/plots_olympus/TNO_map_mean_sd_real_of_5_geologies_with_faults_page2_4_10_14.pdf} 

\vspace{-0.3cm}
\caption{\footnotesize{Oil vol. per unit area of a single geological realisation.}}
\label{fig_Olymp_info_c}
\end{subfigure}
\begin{subfigure}{0.49\columnwidth}
\centering
\includegraphics[page=4,scale=0.42,viewport= 25 0 620 550, clip]{plots_paper/plots_olympus/TNO_map_mean_sd_real_of_5_geologies_with_faults_page2_4_10_14.pdf}

\vspace{-0.3cm}
\caption{\footnotesize{Oil vol. per unit area of a single geological realisation.}}
\label{fig_Olymp_info_d}
\end{subfigure}
\end{center}
\caption{\footnotesize{Additional plots of the TNO II Challenge Olympus oil reservoir model. (a) the mean oil volume per unit area over the 50 geological realisations. (b) the standard deviation of the oil volume per unit area over the 50 geological realisations. (c) and (d) oil volume per unit area from two further examples of the 50 geological realisations.}}
\label{fig_Olymp_info}
\end{figure}
}{}
\ifthenelse{\value{plotstyle}=2}{
\begin{figure}[t]
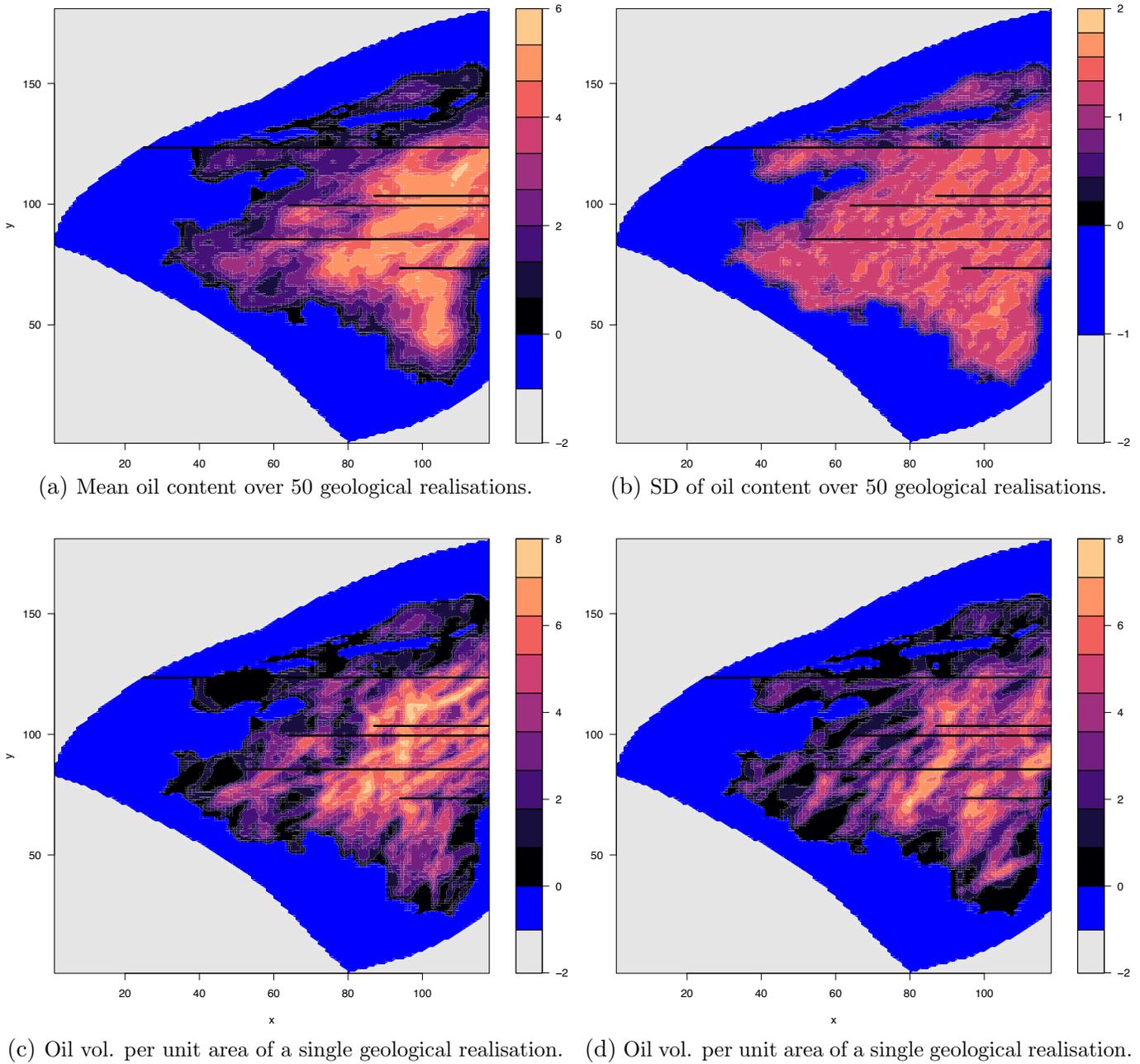

\begin{center}
\hspace{-0.cm}\begin{subfigure}{0.49\columnwidth}
%\centering
\hspace{-1.4cm}\includegraphics[page=2,scale=0.37,viewport= 0 35 620 550, clip]{plots_paper/plots_olympus/TNO_map_mean_sd_real_of_5_geologies_with_faults.pdf}

\vspace{-0.3cm}
\caption{\footnotesize{Mean oil content over 50 geological realisations.}}
\vspace{0.0cm}
\label{fig_Olymp_info_a}
\end{subfigure}
\begin{subfigure}{0.49\columnwidth}
%\centering
\hspace{-0.0cm}\includegraphics[page=4,scale=0.37,viewport= 25 35 620 550, clip]{plots_paper/plots_olympus/TNO_map_mean_sd_real_of_5_geologies_with_faults.pdf}

\vspace{-0.3cm}
\caption{\footnotesize{SD of oil content over 50 geological realisations.}}
\vspace{0.0cm}
\label{fig_Olymp_info_b}
\end{subfigure}
\hspace{-0.0cm}\begin{subfigure}{0.49\columnwidth}
%\centering
\hspace{-1.4cm}\includegraphics[page=10,scale=0.37,viewport= 0 0 620 550, clip]{plots_paper/plots_olympus/TNO_map_mean_sd_real_of_5_geologies_with_faults.pdf} 

\vspace{-0.3cm}
\caption{\footnotesize{Oil vol. per unit area of a single geological realisation.}}
\label{fig_Olymp_info_c}
\end{subfigure}
\begin{subfigure}{0.49\columnwidth}
%\centering
\hspace{-0.0cm}\includegraphics[page=14,scale=0.37,viewport= 25 0 620 550, clip]{plots_paper/plots_olympus/TNO_map_mean_sd_real_of_5_geologies_with_faults.pdf}

\vspace{-0.3cm}
\caption{\footnotesize{Oil vol. per unit area of a single geological realisation.}}
\label{fig_Olymp_info_d}
\end{subfigure}
\end{center}
\vspace{-0.4cm}
\caption{\footnotesize{Additional plots of the TNO II Challenge Olympus oil reservoir model. (a) The mean oil volume per unit area over the 50 geological realisations. (b) The standard deviation of the oil volume per unit area over the 50 geological realisations. (c) and (d) oil volume per unit area from two further examples of the 50 geological realisations.}}
\label{fig_Olymp_info}
\end{figure}
}{}

\section{Torn Embedding Surface $v(x,y)$ for Olympus Model}\label{app_G}

As discussed in section~\ref{ssec_TNO_emulation} we specify an embedding surface $v(\vx)=v(x,y)$ for the Olympus model by tearing along the five discontinuities shown in figure~\ref{fig_olympus_model_b} and bending alternate regions higher and lower into the 3D space using quadratic forms, exploiting a similar strategy to that employed in section~\ref{ssec_torn_embed}.
The embedding surface is shown in figure~\ref{fig_embed_olympus}, with the full definition as follows. Noting that the 
faults/discontinuities occur at $y$ locations $y^{dis} = \{73.5,85.5,99.5,103.5,123.5\}$ which have left end points at $x$ locations
$x^{dis} = \{94,52,64,87,0\}$ and right end points all at $x^{max}=118$ we therefore define:
\be\label{eq_app_def_vxy}
v(x,y) \;=\;  \left\{ \begin{matrix}
\left( \frac{x-x^{dis}_1}{x^{max}-x^{dis}_1} \right)^2  \mathbbm{1}_{\{x>x^{dis}_1\}}     & \text{ for } 0\le y < y^{dis}_1 \\
-1.2 \left( \frac{x-b_1(y)}{x^{max}-b_1(y)} \right)^2  \mathbbm{1}_{\{x>b_1(y)\}}     & \text{ for } y^{dis}_1 \le y < y^{dis}_2 \\
3 \left( \frac{x-b_2(y)}{x^{max}-b_2(y)} \right)^2  \mathbbm{1}_{\{x>b_2(y)\}}     & \text{ for } y^{dis}_2 \le y < y^{dis}_3 \\
0    & \text{ for } y^{dis}_3 \le y < y^{dis}_4 \\
-2 \left( \frac{x-x^{dis}_4}{x^{max}-x^{dis}_4} \right)^2  \mathbbm{1}_{\{x>x^{dis}_4\}}     & \text{ for } y^{dis}_4 \le y < y^{dis}_5 \\
1   & \text{ for } y^{dis}_5 \le y  \\
\end{matrix}
\right.
\ee
%\be
%v(x,y) \;=\;  \left\{ \begin{matrix}
%\left( \frac{x-94}{118-94} \right)^2  \mathbbm{1}_{\{x>94\}}     & \text{ for } 0\le y \le 73.5 \\
%-1.2 \left( \frac{x-b_1(y)}{118-b_1(y)} \right)^2  \mathbbm{1}_{\{x>b_1(y)\}}     & \text{ for } 73.5 \le y \le 85.5 \\
%3 \left( \frac{x-b_2(y)}{118-b_2(y)} \right)^2  \mathbbm{1}_{\{x>b_2(y)\}}     & \text{ for } 85.5 \le y \le 99.5 \\
%0    & \text{ for } 99.5 \le y \le 99.5. \\
%-2 \left( \frac{x-87}{118-87} \right)^2  \mathbbm{1}_{\{x>87\}}     & \text{ for } 103.5 \le y \le 123.5 \\
%1   & \text{ for } 123.5 \le y  \\
%\end{matrix}
%\right.
%\ee
where the lines $b_1(y)$ and $b_2(y)$ that interpolate between fault end points are given by
\ba
b_1(y) &=& x^{dis}_1 \;+\;  \left( \frac{y-y^{dis}_1}{y^{dis}_2 - y^{dis}_1} \right) (x^{dis}_2 - x^{dis}_1),  \quad \quad \text{ for } y^{dis}_1 \le y \le y^{dis}_2 \\
b_2(y) &=& x^{dis}_2 \;+\; \left( \frac{y-y^{dis}_2}{y^{dis}_3 - y^{dis}_2}\right)   (x^{dis}_3 - x^{dis}_2),  \quad \quad \text{ for } y^{dis}_2 \le y \le y^{dis}_3
\ea
%\ba
%b_1(y) &=& 94 + \frac{52-94}{85.5 - 73.5} (y-73.5)  \quad \text{ for } 73.5 \le y \le 85.5 \\
%b_2(y) &=& 52 + \frac{64 - 52}{99.5 - 85.5} (y-85.5)  \quad \text{ for } 85.5 \le y \le 99.5
%\ea
See also figure~\ref{fig_2dzoomed_ind_cor} for a zoomed in view of $v(x,y)$. The above form for $v(x,y)$ was chosen simply to ensure that adjacent regions that are separated by a discontinuity would be suitably distant in the third dimension, to ensure they would therefore be reasonably decorrelated.

\ifthenelse{\value{plotstyle}=2}{\clearpage}{}

\section{TENSE Olympus Details and Further Output}\label{app_H}

As described in section~\ref{ssec_emul_NPV_surface}, the TENSE framework was applied to $D_1$, employing the embedding surface $v(x,y)$, using equations (\ref{eq_NS_squared_cov_struc_3D_full}), (\ref{eq_sigma3D_full_exp1}), (\ref{eq_app_def_vxy}), (\ref{eq_BLm}) and (\ref{eq_BLv}), with $\sigma$ and $m(x)=m$ set to the sample SD and sample mean of the runs $D_1$. Additionally, the 2D correlation length which had been set at $\theta=12$ in the design phase, was subsequently set to the MLE estimate of $\theta=14.6$, using standard normality assumptions. The parameter $\alpha_3$ that features in equation~\ref{eq_sigma3D_full_exp1} was set to $\alpha_3= 0.5$, a choice made in combination with the form of the embedding surface $v(x,y)$ to ensure suitable decorrelation across the five discontinuities, as shown in figure~\ref{fig_2dzoomed_ind_cor}. 
The validity of these settings were checked via leave-one-out emulator diagnostics. The resulting emulator expectation $\ed{D_1}{f(\vx)}$ adjusted by the model evaluations $D_1$ is shown in figure~\ref{fig_Olymp_em1_b} as the coloured contours. 

Figure~\ref{fig_Olymp_em2} shows
further output of the TENSE emulator as applied to the TNO Challenge II Olympus reservoir model. (a) The prior emulator upper credible interval for the NPV, defined as $\e{f(\vx)} + 2\sqrt{\var{f(\vx)}}$ evaluated at each possible well location over the reservoir. (b) The wave 1 emulator upper credible interval for the NPV, defined as 
$\ed{D_1}{f(\vx)} + 2\sqrt{\vard{D_1}{f(\vx)}}$. (c) The wave 2 emulator upper credible interval for the NPV, defined as 
$\ed{D_1 \cup D_2}{f(\vx)} + 2\sqrt{\vard{D_1 \cup D_2}{f(\vx)}}$, which is now similar to the regions highlighted in figure~\ref{fig_Olymp_em1_d}.
(d) The TENSE emulator expectation applied to the standard deviation of the NPV of the 50 geological realisations, showing lower variation in some of the candidate regions 
highlighted in figure~\ref{fig_Olymp_em1_d}.

\ifthenelse{\value{plotstyle}=1}{
\begin{figure}[t]
\begin{center}
\hspace{-0.8cm}\begin{subfigure}{0.52\columnwidth}
\centering
\includegraphics[page=1,scale=0.42,viewport= 25 30 650 550, clip]{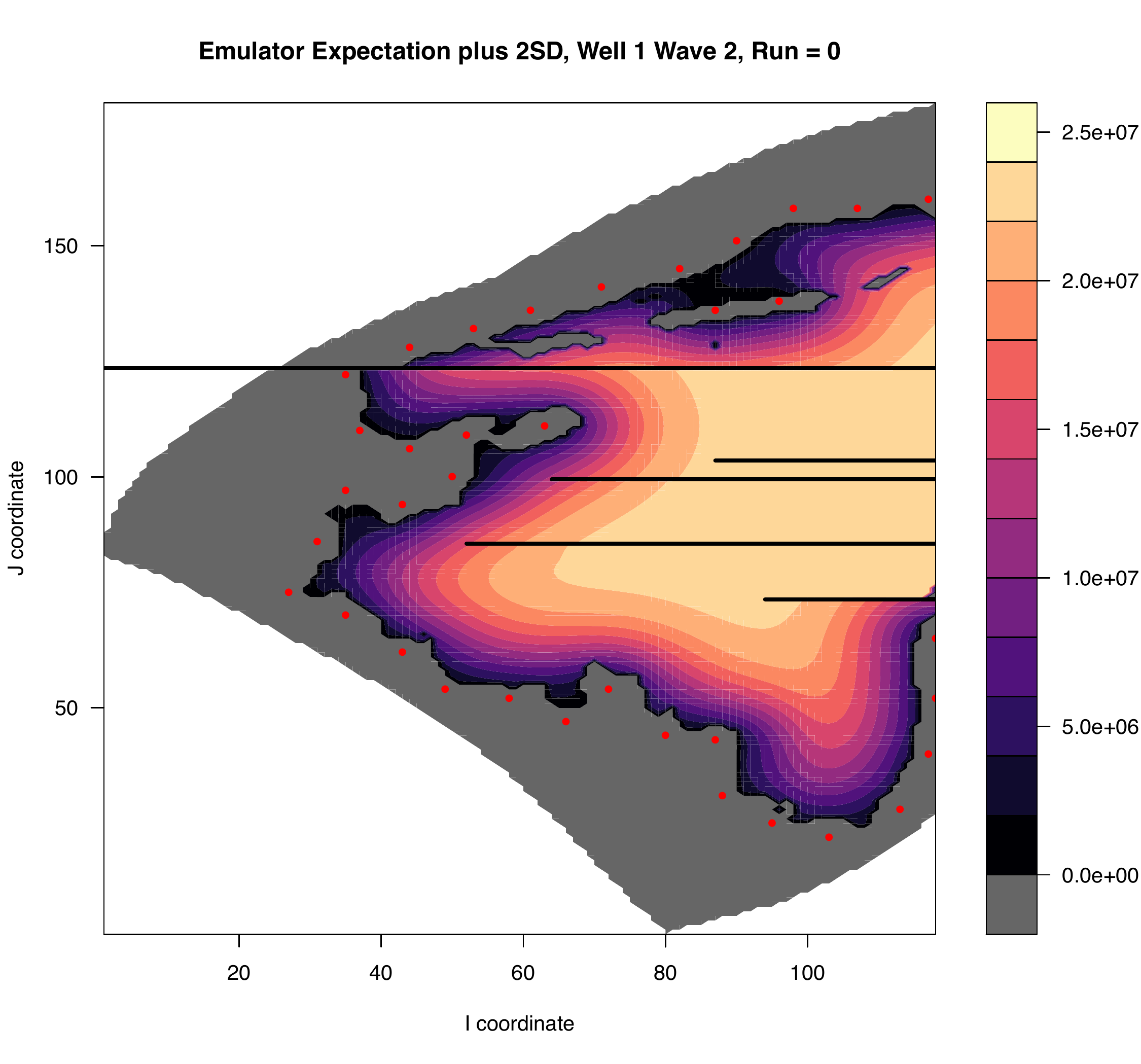}

\vspace{-0.3cm}
\caption{\footnotesize{Prior Upper CI: $\e{f(\vx)} + 2\sqrt{\var{f(\vx)}}$.}}
\vspace{0.5cm}
\label{fig_Olymp_em2_a}
\end{subfigure}
\begin{subfigure}{0.49\columnwidth}
\centering
\includegraphics[page=2,scale=0.42,viewport= 25 30 650 550, clip]{plots_paper/plots_olympus/Emulator_sequential_mean_plus_2sd_w2maxlik_GPmeanw2_para_0to95pts1A_page97_98_99.pdf}

\vspace{-0.3cm}
\caption{\footnotesize{Wave 1 Upper CI: $\ed{D_1}{f(\vx)} + 2\sqrt{\vard{D_1}{f(\vx)}}$.}}
\vspace{0.5cm}
\label{fig_Olymp_em2_b}
\end{subfigure}

\hspace{-0.8cm}\begin{subfigure}{0.52\columnwidth}
\centering
\includegraphics[page=3,scale=0.42,viewport= 25 30 650 550, clip]{plots_paper/plots_olympus/Emulator_sequential_mean_plus_2sd_w2maxlik_GPmeanw2_para_0to95pts1A_page97_98_99.pdf} 

\vspace{-0.3cm}
\caption{\footnotesize{Wave 2 Upper CI: $\ed{D_1 \cup D_2}{f(\vx)} + 2\sqrt{\vard{D_1 \cup D_2}{f(\vx)}}$.}}
\label{fig_Olymp_em2_c}
\end{subfigure}
\begin{subfigure}{0.49\columnwidth}
\centering
\includegraphics[page=10,scale=0.42,viewport= 25 30 650 550, clip]{plots_paper/plots_olympus/Emulator_mean_for_quantiles_etc_at_w2run_95_w2maxlik_GPmeanw2_para1A_with_contours.pdf} 

\vspace{-0.3cm}
\caption{\footnotesize{Emulator expectation of Olympus model SD.}}
\label{fig_Olymp_em2_d}
\end{subfigure}
\end{center}
\caption{\footnotesize{Further output of the TENSE emulator as applied to the TNO Challenge II Olympus reservoir model. (a) The prior emulator upper credible interval for the NPV, defined as $\e{f(\vx)} + 2\sqrt{\var{f(\vx)}}$ evaluated at each possible well location over the reservoir. (b) The wave 1 emulator upper credible interval for the NPV, defined as 
$\ed{D_1}{f(\vx)} + 2\sqrt{\vard{D_1}{f(\vx)}}$. (c) The wave 2 emulator upper credible interval for the NPV, defined as 
$\ed{D_1 \cup D_2}{f(\vx)} + 2\sqrt{\vard{D_1 \cup D_2}{f(\vx)}}$, which is now similar to the regions highlighted in figure~\ref{fig_Olymp_em1_d}.
(d) The TENSE emulator expectation applied to the standard deviation of the NPV of the 50 geological realisations, showing lower variation in some of the candidate regions 
highlighted in figure~\ref{fig_Olymp_em1_d}.}}
\label{fig_Olymp_em2}
\end{figure}
}{}
\ifthenelse{\value{plotstyle}=2}{
\begin{figure}[t]
\begin{center}
\hspace{-0.0cm}\begin{subfigure}{0.49\columnwidth}
%\centering
\hspace{-1.4cm}\includegraphics[page=97,scale=0.37,viewport= 25 30 650 550, clip]{plots_paper/plots_olympus/Emulator_sequential_mean_plus_2sd_w2maxlik_GPmeanw2_para_0to95pts1A.pdf}

\vspace{-0.3cm}
\caption{\footnotesize{Prior Upper CI: $\e{f(\vx)} + 2\sqrt{\var{f(\vx)}}$.}}
\vspace{0.0cm}
\label{fig_Olymp_em2_a}
\end{subfigure}
\begin{subfigure}{0.49\columnwidth}
%\centering
\hspace{0.2cm}\includegraphics[page=98,scale=0.37,viewport= 25 30 650 550, clip]{plots_paper/plots_olympus/Emulator_sequential_mean_plus_2sd_w2maxlik_GPmeanw2_para_0to95pts1A.pdf}

\vspace{-0.3cm}
\caption{\footnotesize{Wave 1 Upper CI: $\ed{D_1}{f(\vx)} + 2\sqrt{\vard{D_1}{f(\vx)}}$.}}
\vspace{0.0cm}
\label{fig_Olymp_em2_b}
\end{subfigure}

\hspace{-0.0cm}\begin{subfigure}{0.49\columnwidth}
%\centering
\hspace{-1.4cm}\includegraphics[page=99,scale=0.37,viewport= 25 30 650 550, clip]{plots_paper/plots_olympus/Emulator_sequential_mean_plus_2sd_w2maxlik_GPmeanw2_para_0to95pts1A.pdf} 

\vspace{-0.3cm}
\caption{\footnotesize{W2 UCI: $\ed{D_1 \cup D_2}{f(\vx)} + 2\sqrt{\vard{D_1 \cup D_2}{f(\vx)}}$.}}
\label{fig_Olymp_em2_c}
\end{subfigure}
\begin{subfigure}{0.49\columnwidth}
%\centering
\hspace{0.2cm}\includegraphics[page=10,scale=0.37,viewport= 25 30 650 550, clip]{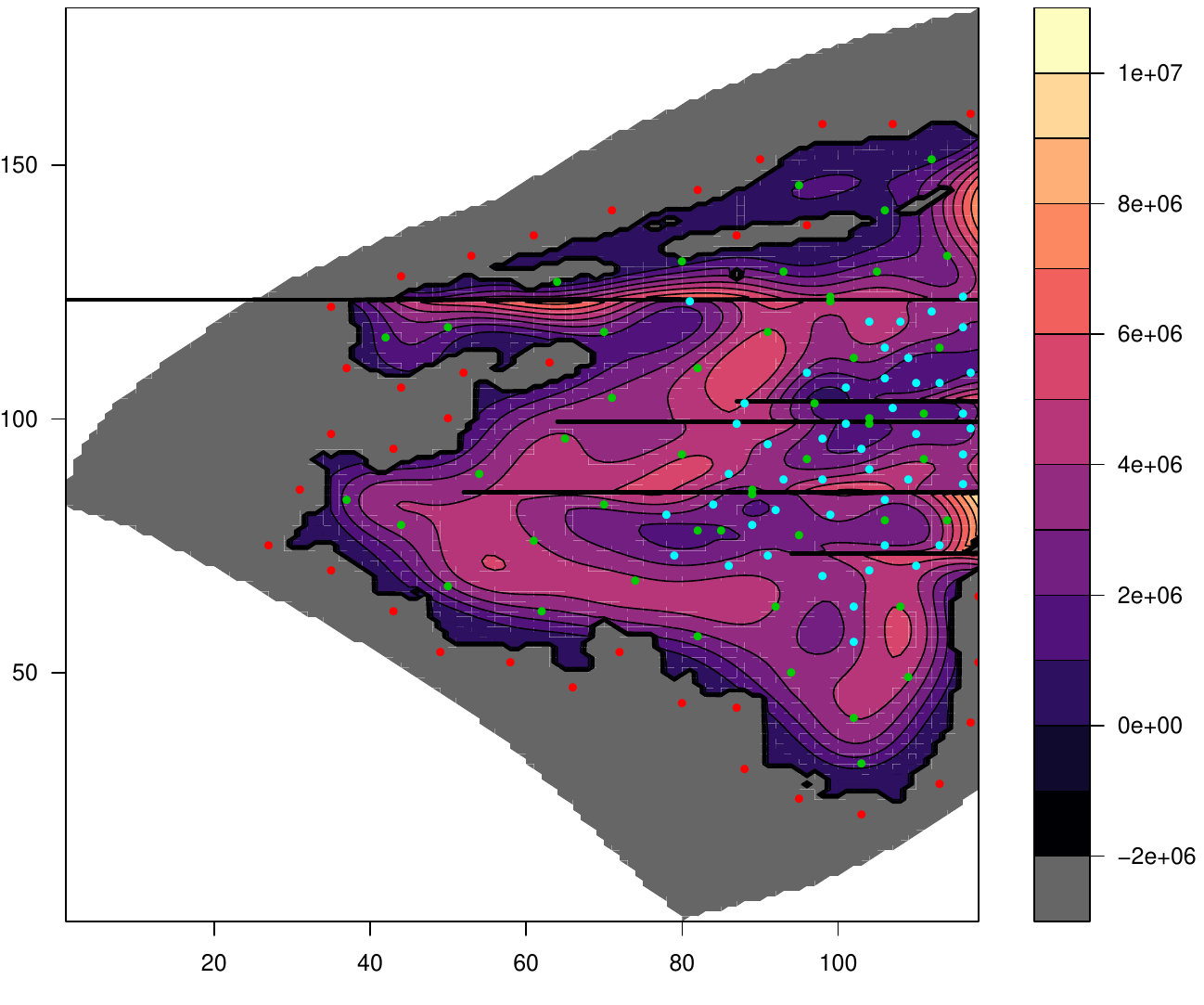} 

\vspace{-0.3cm}
\caption{\footnotesize{Emulator expectation of Olympus model SD.}}
\label{fig_Olymp_em2_d}
\end{subfigure}
\end{center}
\vspace{-0.4cm}
\caption{\footnotesize{Further output of the TENSE emulator as applied to the TNO Challenge II Olympus reservoir model. (a) The prior emulator upper credible interval for the NPV, defined as $\e{f(\vx)} + 2\sqrt{\var{f(\vx)}}$ evaluated at each possible well location over the reservoir. (b) The wave 1 emulator upper credible interval for the NPV, defined as 
$\ed{D_1}{f(\vx)} + 2\sqrt{\vard{D_1}{f(\vx)}}$. (c) The wave 2 emulator upper credible interval for the NPV, defined as 
$\ed{D_1 \cup D_2}{f(\vx)} + 2\sqrt{\vard{D_1 \cup D_2}{f(\vx)}}$, which is now similar to the regions highlighted in figure~\ref{fig_Olymp_em1_d}.
(d) The TENSE emulator expectation applied to the standard deviation of the NPV of the 50 geological realisations, showing lower variation in some of the candidate regions 
highlighted in figure~\ref{fig_Olymp_em1_d}.}}
\label{fig_Olymp_em2}
\end{figure}
}{}

\ifthenelse{\value{plotstyle}=1}{\clearpage}{}

\section{TENSE Quantile Emulation of Olympus}\label{app_I}

Here we detail further analysis performed on the Olympus model using the TENSE framework. 

While the full optimisation of the Olympus model with respect to multiple well configurations is not the focus of this work, we do make the following observations. 
As discussed in~\cite{eage:/content/papers/10.3997/2214-4609.202035109}, due to the imperfection of the simulator, the notion of finding the optimum decision (in this case well placement) is somewhat misleading. Instead, when providing decision support it is more informative to provide classes of good decisions, such as shown in figure~\ref{fig_Olymp_em1_d}, for further consideration by the decision maker who may, as is common in the oil industry, have a set of additional preferences unknown to the statistician/reservoir analyst. Examples of these may include unknown risk preferences, political, financial or environmental considerations, or other corporate logistical issues. 
Anticipation of these issues by the analyst can partially inform the $\delta$ parameter used in the definition of the region of interest (see equation~(\ref{eq_non_imp_contraint})).

Concerning risk preferences, the TENSE framework can be used to perform quantile emulation, that is to emulate various quantiles of the stochastic NPV output (where the stochasticity is induced by the geological uncertainty), instead of just emulating the mean over the 50 geological realisations (see equation~(\ref{eq_NPVmean1})) as instructed by the TNO challenge. Plots of the emulator expectation for the $10\%, 25\%, 50\%, 75\%$ and $90\%$ NPV quantiles are shown in figure~\ref{fig_Olymp_qem1} which can be used to identify more risk averse locations for the first producer well. For example, consideration of the $25\%$ NPV quantile may suggest the region between the top two faults is preferable to the other regions with equally high mean NPV. Similarly, the TENSE emulation of the standard deviation of the NPV output induced by the 50 geological realisations is also shown in figure~\ref{fig_Olymp_em2_d}, which shows, slightly counterintuitively, that the standard deviation due to the geological uncertainty is generally lower in regions with higher expected mean NPV (as well as, trivially, for regions with very low NPV). This behaviour is confirmed by examining box-plots of the run data, given in figure~\ref{fig_boxplots_Olymp}. All of this provides the decision maker with a rich set of easily accessible additional information.

\ifthenelse{\value{plotstyle}=1}{
\begin{figure}[t]
\vspace{-0.5cm}
\begin{center}
\begin{subfigure}{0.41\columnwidth}
\centering
\hspace{-1.2cm}\includegraphics[page=3,scale=0.39,viewport= 25 35 535 550, clip]{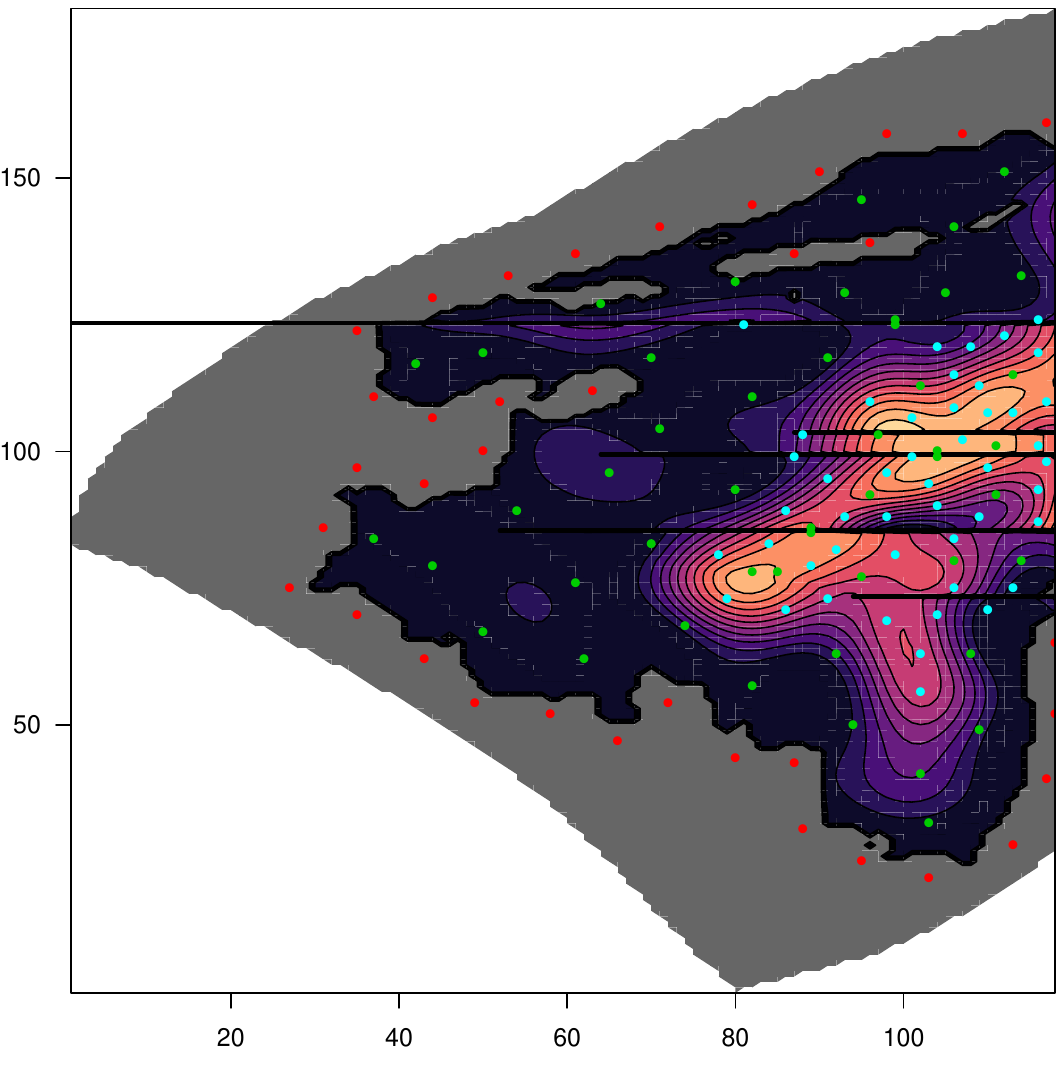}

\vspace{-0.3cm}
\caption{\footnotesize{TENSE expectation for 10\% NPV quantile.}}
\vspace{0.2cm}
\label{fig_Olymp_qem1_a}
\end{subfigure}
\begin{subfigure}{0.49\columnwidth}
\centering
\includegraphics[page=4,scale=0.39,viewport= 25 35 650 550, clip]{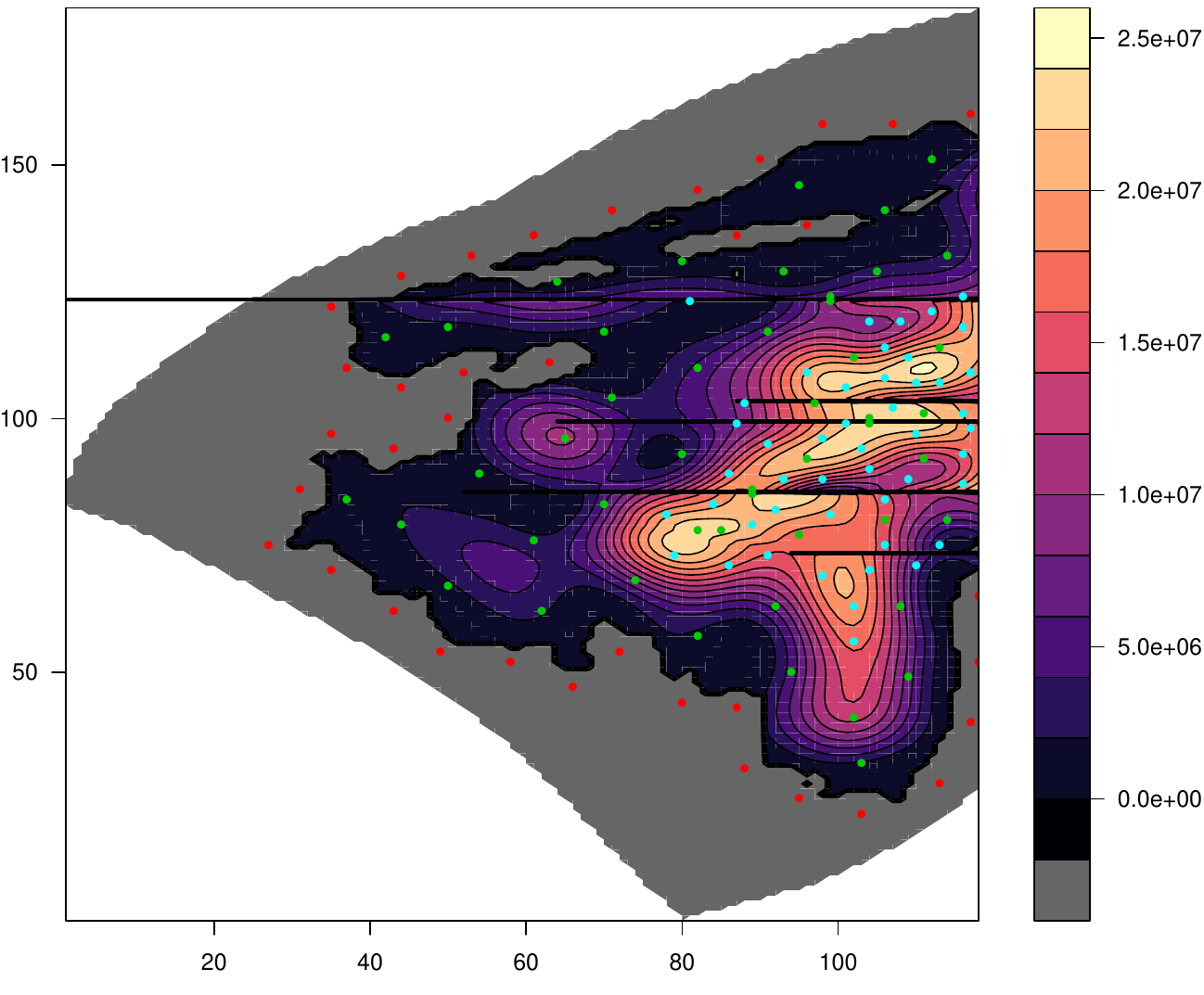} 

\vspace{-0.3cm}
\caption{\footnotesize{TENSE expectation for 25\% NPV quantile.}}
\vspace{0.2cm}
\label{fig_Olymp_qem1_b}
\end{subfigure}
\begin{subfigure}{0.41\columnwidth}
\centering
\hspace{-1.2cm}\includegraphics[page=5,scale=0.39,viewport= 25 35 535 550, clip]{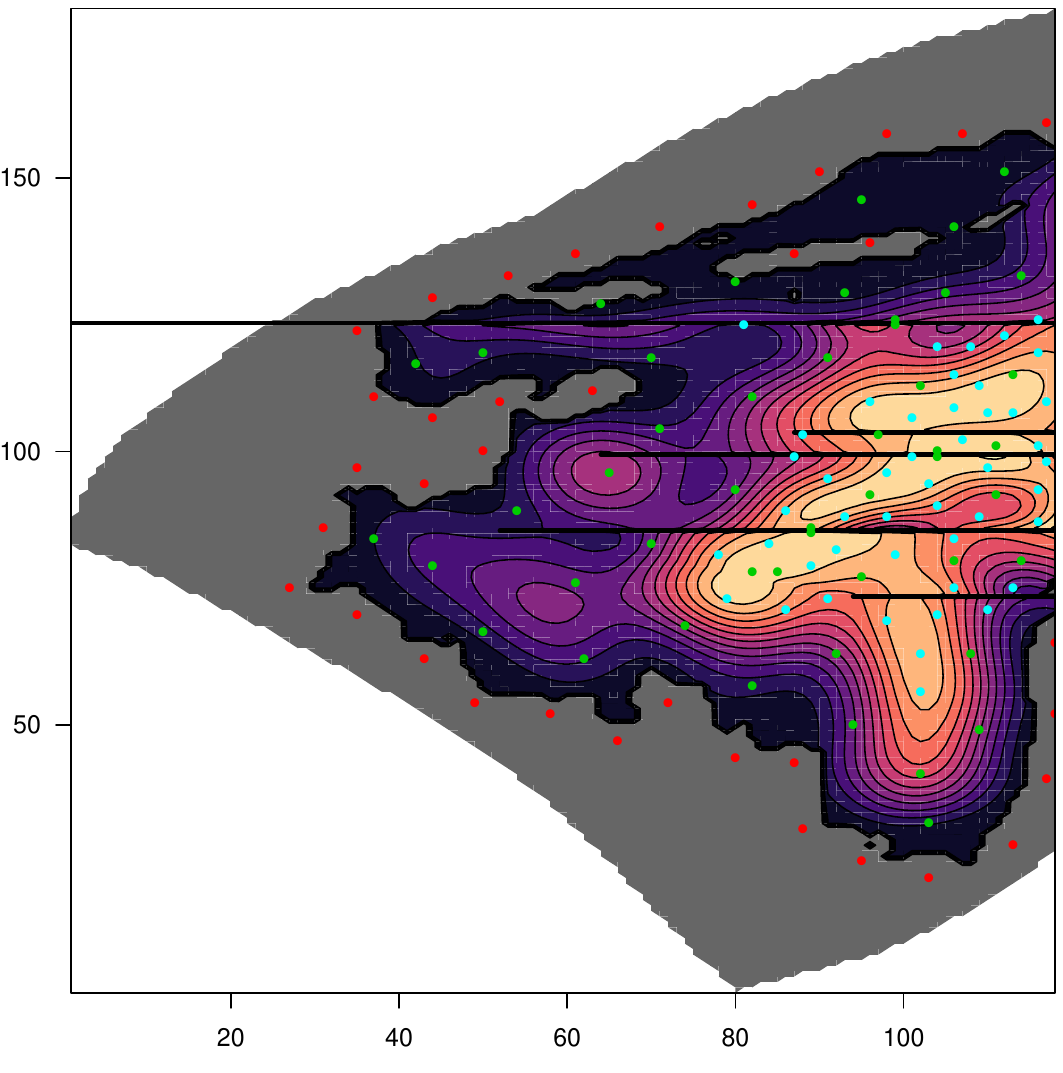} 

\vspace{-0.3cm}
\caption{\footnotesize{TENSE expectation for 50\% NPV quantile.}}
\vspace{0.2cm}
\label{fig_Olymp_qem1_c}
\end{subfigure}
\begin{subfigure}{0.49\columnwidth}
\centering
\includegraphics[page=6,scale=0.39,viewport= 25 35 650 550, clip]{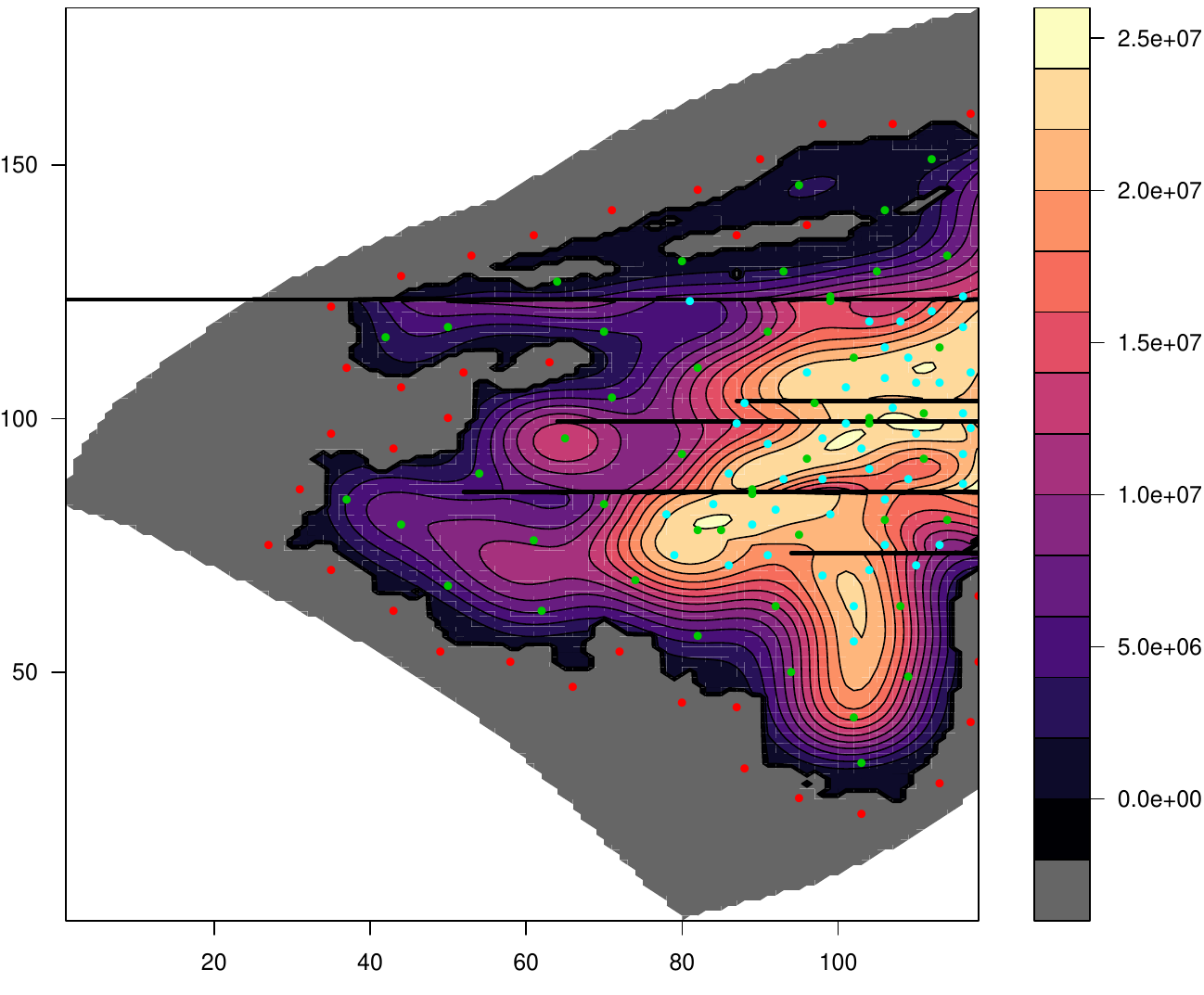}

\vspace{-0.3cm}
\caption{\footnotesize{TENSE expectation for 75\% NPV quantile.}}
\vspace{0.2cm}
\label{fig_Olymp_qem1_d}
\end{subfigure}
\begin{subfigure}{0.49\columnwidth}
\centering
\includegraphics[page=7,scale=0.39,viewport= 25 35 650 550, clip]{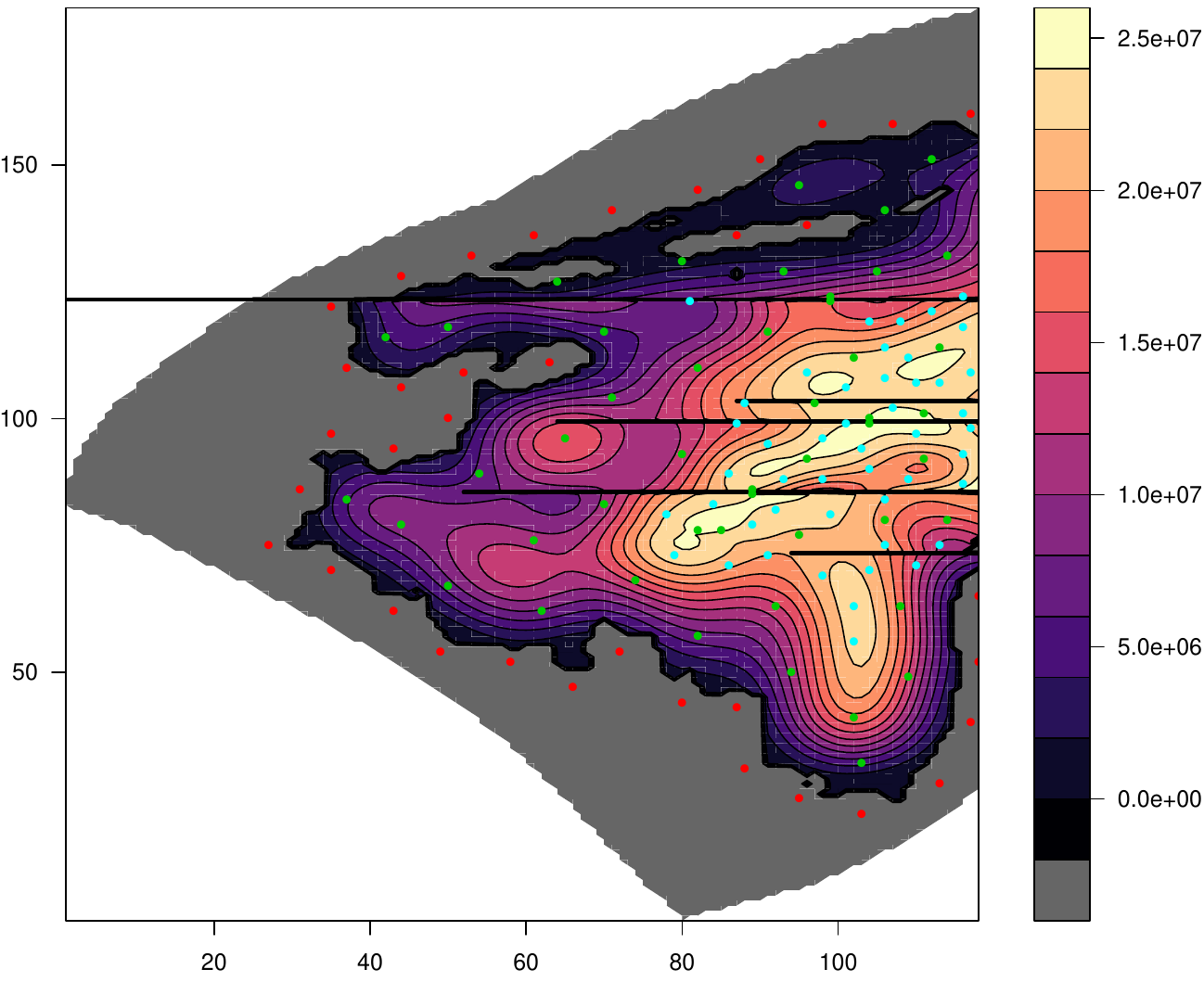}

\vspace{-0.3cm}
\caption{\footnotesize{TENSE expectation for 90\% NPV quantile.}}
\label{fig_Olymp_qem1_e}
\end{subfigure}
\end{center}
\vspace{-0.4cm}
\caption{\footnotesize{Quantile emulation using the TENSE framework, applied to the 50 geological realisations of NPV.}}
\label{fig_Olymp_qem1}
\end{figure}
}{}
\ifthenelse{\value{plotstyle}=2}{
\begin{figure}[t]
\vspace{-0.2cm}
\begin{center}
\begin{subfigure}{0.49\columnwidth}
\centering
\hspace{-0.3cm}\includegraphics[page=3,scale=0.33,viewport= 25 35 535 550, clip]{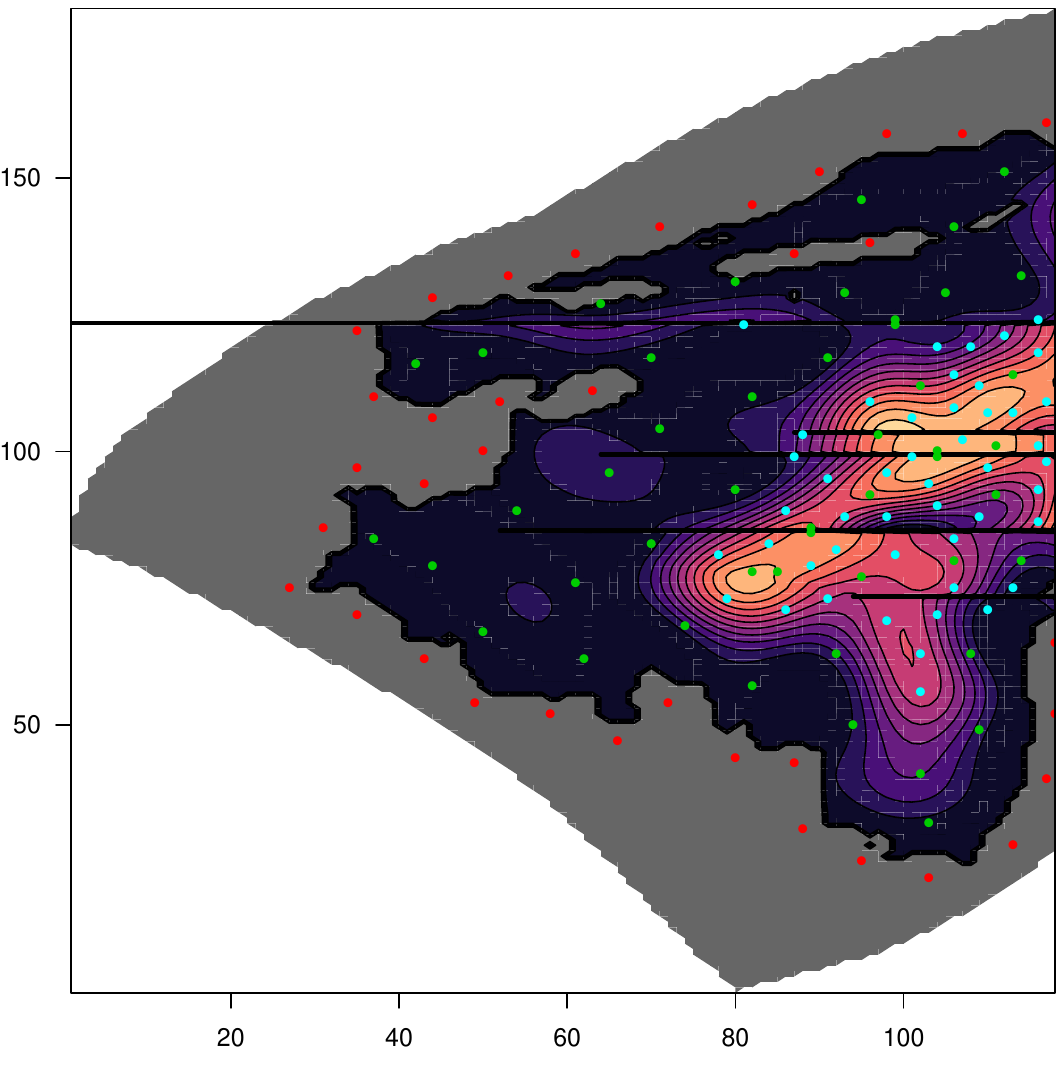}

\vspace{-0.3cm}
\caption{\footnotesize{TENSE expectation for 10\% NPV quantile.}}
\vspace{0.2cm}
\label{fig_Olymp_qem1_a}
\end{subfigure}
\begin{subfigure}{0.49\columnwidth}
\centering
\hspace{0.2cm}\includegraphics[page=4,scale=0.33,viewport= 25 35 650 550, clip]{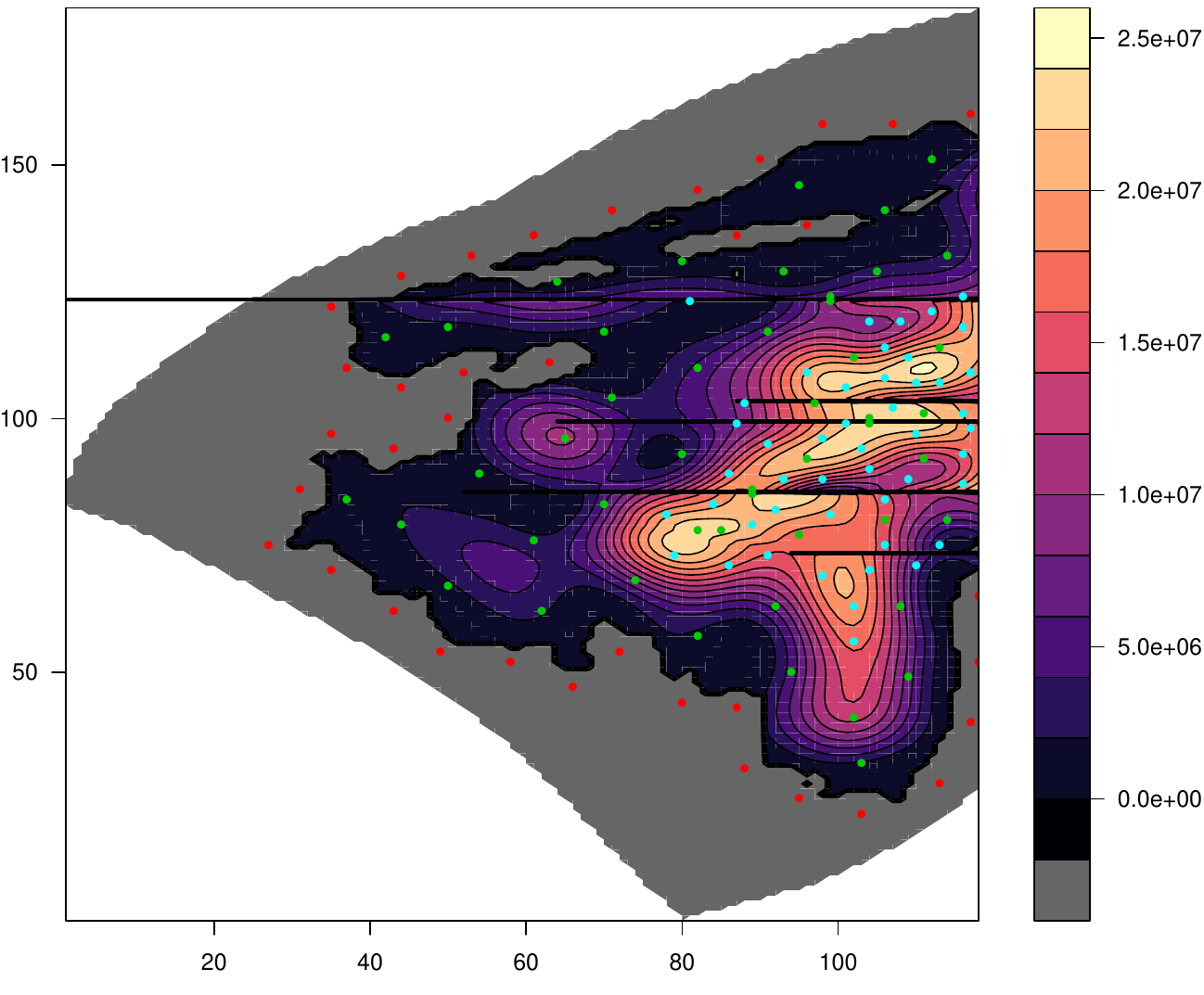} 

\vspace{-0.3cm}
\caption{\footnotesize{TENSE expectation for 25\% NPV quantile.}}
\vspace{0.2cm}
\label{fig_Olymp_qem1_b}
\end{subfigure}
\begin{subfigure}{0.49\columnwidth}
\centering
\hspace{-0.3cm}\includegraphics[page=5,scale=0.33,viewport= 25 35 535 550, clip]{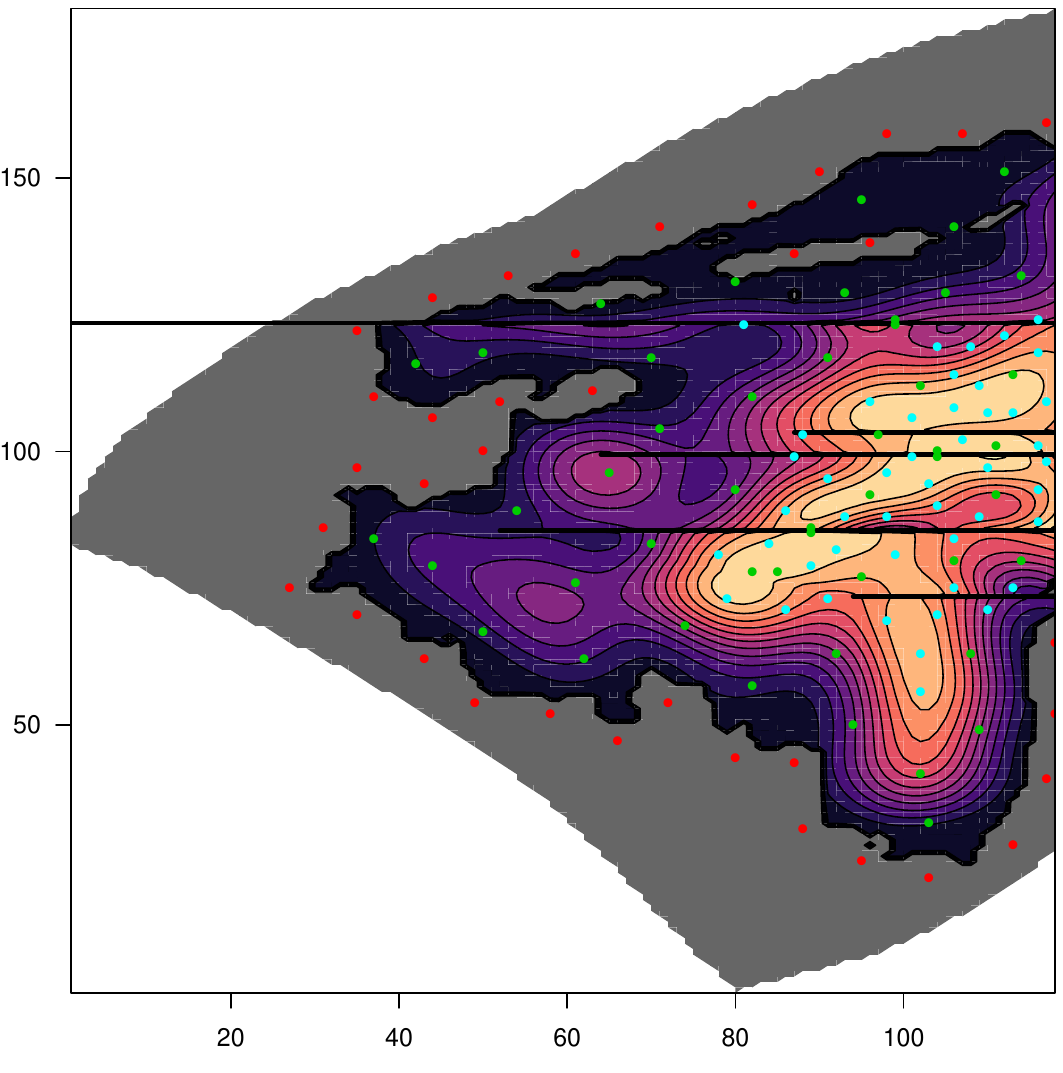} 

\vspace{-0.3cm}
\caption{\footnotesize{TENSE expectation for 50\% NPV quantile.}}
\vspace{0.2cm}
\label{fig_Olymp_qem1_c}
\end{subfigure}
\begin{subfigure}{0.49\columnwidth}
\centering
\hspace{0.2cm}\includegraphics[page=6,scale=0.33,viewport= 25 35 650 550, clip]{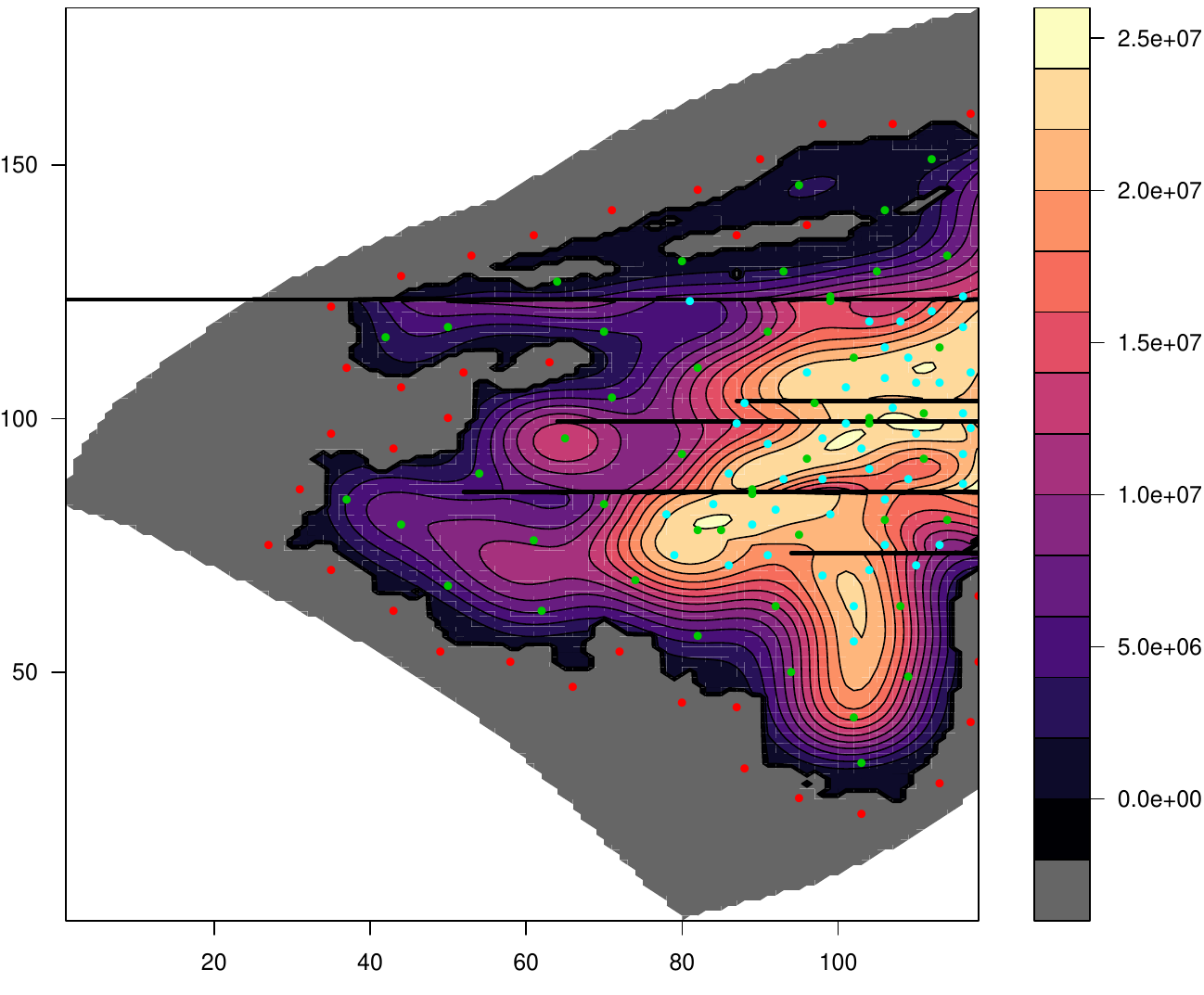}

\vspace{-0.3cm}
\caption{\footnotesize{TENSE expectation for 75\% NPV quantile.}}
\vspace{0.2cm}
\label{fig_Olymp_qem1_d}
\end{subfigure}
\begin{subfigure}{0.49\columnwidth}
\centering
\includegraphics[page=7,scale=0.33,viewport= 25 35 650 550, clip]{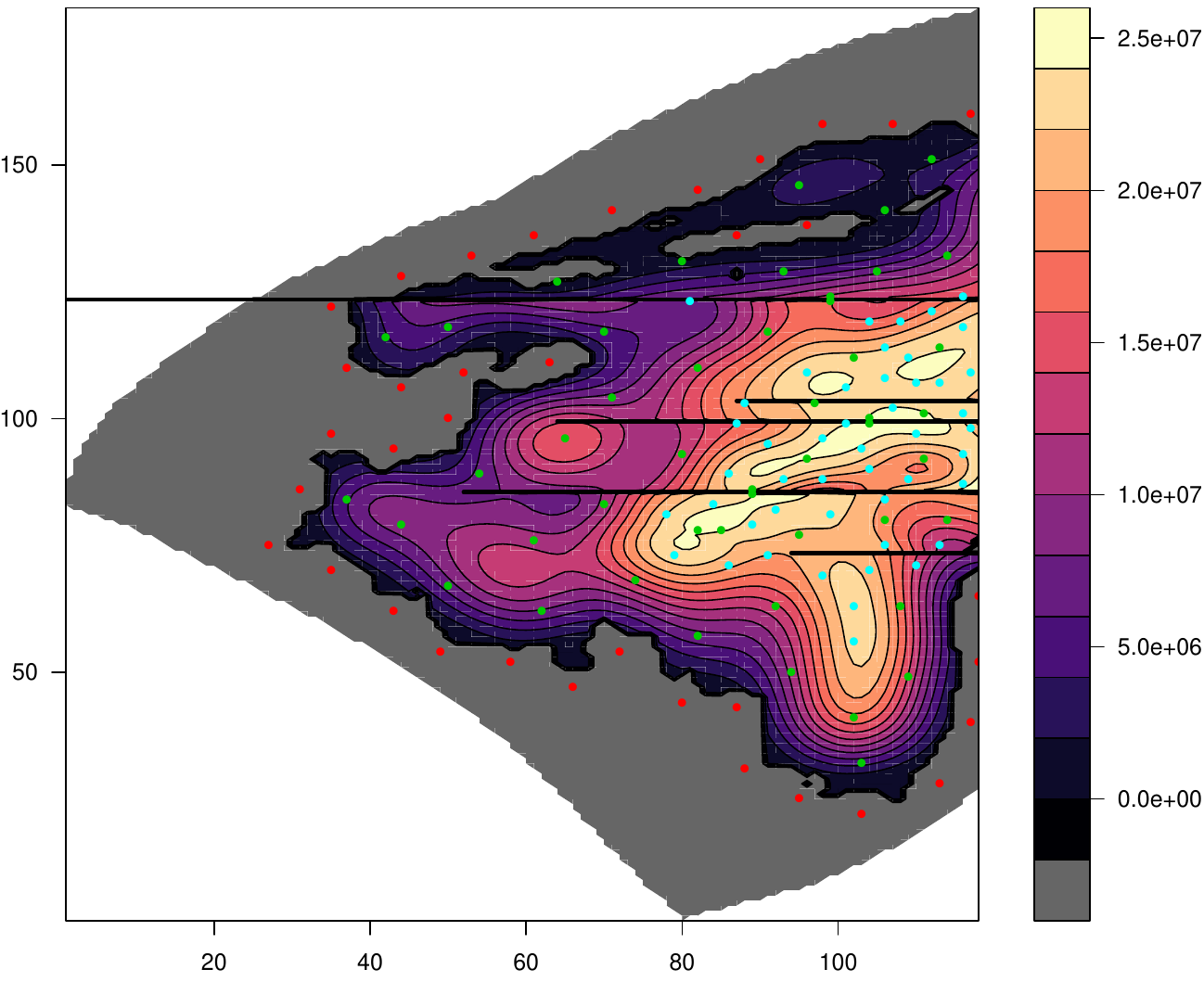}

\vspace{-0.3cm}
\caption{\footnotesize{TENSE expectation for 90\% NPV quantile.}}
\label{fig_Olymp_qem1_e}
\end{subfigure}
\end{center}
\vspace{-0.4cm}
\caption{\footnotesize{Quantile emulation using the TENSE framework, applied to the 50 geological realisations of NPV.}}
\label{fig_Olymp_qem1}
\end{figure}
}{}

\ifthenelse{\value{plotstyle}=1}{
\begin{figure}
\begin{center}
\includegraphics[page=2,scale=0.44,viewport= 0 10 1130 675, clip]{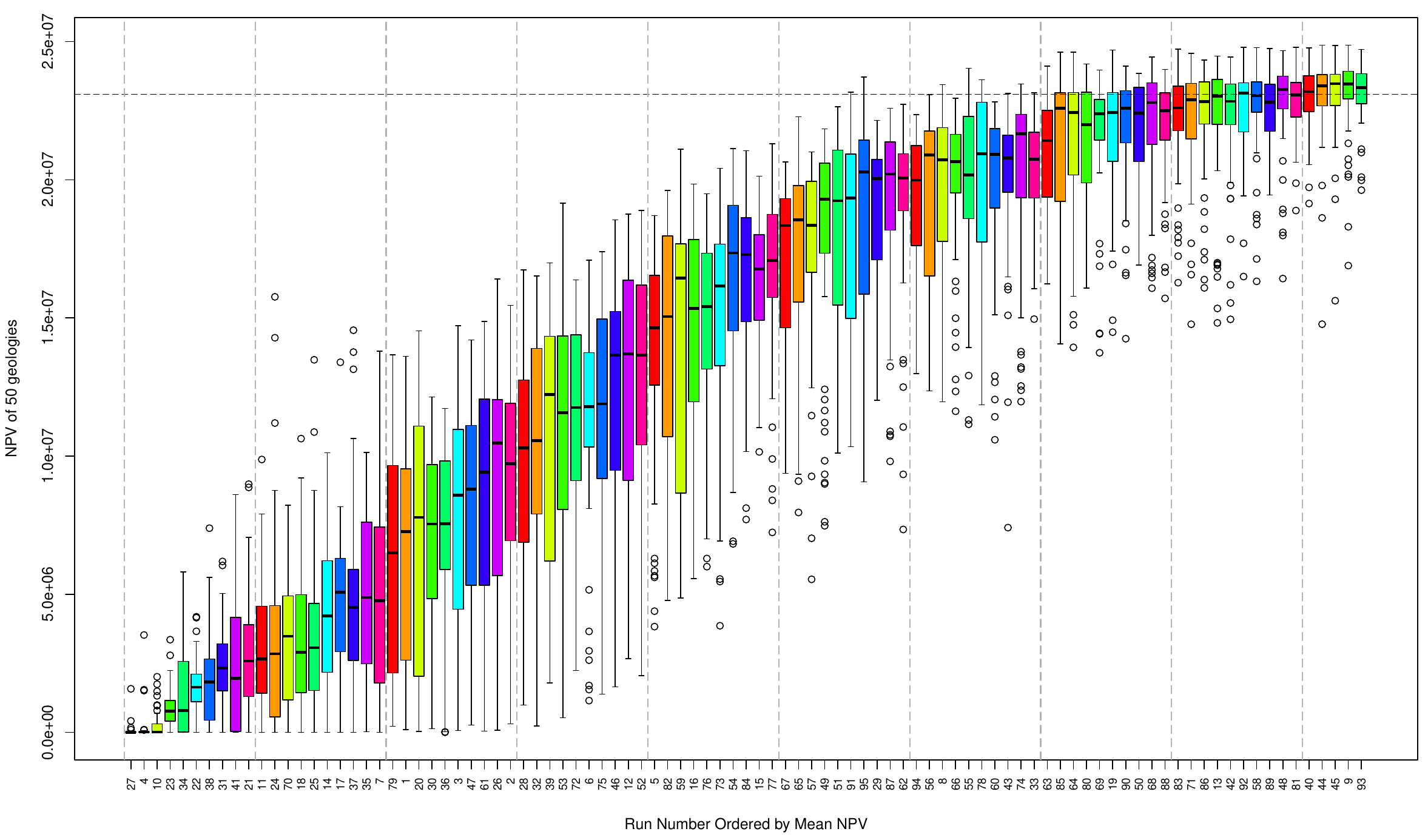} 
\end{center}
\vspace{-0.6cm}
\caption{\footnotesize{Boxplot summaries of the 50 NPV outputs corresponding to the 50 geological realisations, for each of the 95 runs composing wave 1 and wave 2. Runs are ordered according to increasing mean NPV. Note that for runs with higher mean NPV the spread in the 50 realisations is lower than runs with intermediate mean NPV, in accordance with figures~\ref{fig_Olymp_em1_d} and \ref{fig_Olymp_em2_d}. The highest mean NPV seen in any run so far $f^+$ is shown as the horizontal dashed line.}}
\label{fig_boxplots_Olymp}
\end{figure}
}{}
\ifthenelse{\value{plotstyle}=2}{
\begin{figure}
\hspace{-0.8cm}\includegraphics[angle=0,page=2,scale=0.35,viewport= 0 10 1130 675, clip]{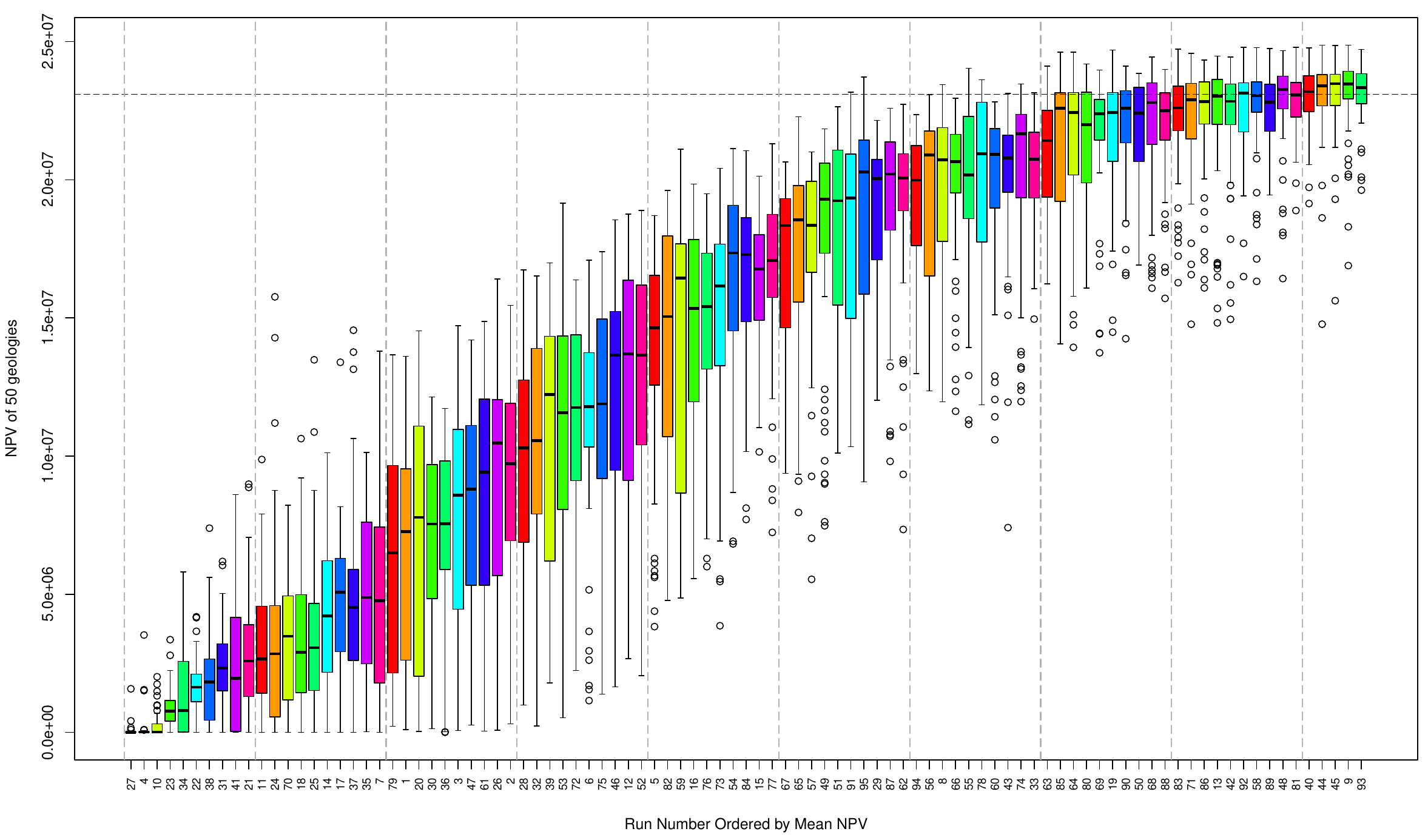} 
\vspace{-0.2cm}
\caption{\footnotesize{Boxplot summaries of the 50 NPV outputs corresponding to the 50 geological realisations, for each of the 95 runs composing wave 1 and wave 2. Runs are ordered according to increasing mean NPV. Note that for runs with higher mean NPV the spread in the 50 realisations is lower than runs with intermediate mean NPV, in accordance with figures~\ref{fig_Olymp_em1_d} and \ref{fig_Olymp_em2_d}. The highest mean NPV seen in any run so far $f^+$ is shown as the horizontal dashed line.}}
\label{fig_boxplots_Olymp}
\end{figure}
}{}